\documentclass[12pt,letterpaper]{article}
\usepackage{amssymb}
\usepackage{eqsection,latexsym,epsf}
\usepackage{epsfig,amsfonts,cite}

\newtheorem{df}{Definition}
\newtheorem{thm}{Theorem}
\newtheorem{coro}{Corollary}
\newtheorem{lemma}{Lemma}
\newtheorem{propo}{Proposition}
\newtheorem{conj}{Conjecture}

\def\soc{{\sf d}}
\def\phb{\hat{\varphi}}
\def\hw{\hat{w}}
\def\ind{{\mathbb I}}
\def\Z{{\mathbb Z}}
\def\hOmega{\widehat{\Omega}}

\def\cX{{\cal X}}
\def\cY{{\cal Y}}

\def\ed{\stackrel{\rm d}{=}}

\def\Fbe{F_{\rm B}}
\def\Ube{U_{\rm B}}
\def\Hbe{H_{\rm B}}

\def\e{z}
\def\eh{\hat{z}}

\def\Ad{{\sf A}}
\def\ad{{\sf a}}
\def\Jd{{\sf J}}
\def\jd{{\sf j}}
\def\Hd{{\sf H}}
\def\hd{{\sf h}}
\def\Ai{\overline{\sf A}}
\def\Ji{\overline{\sf J}}
\def\Hi{\overline{\sf H}}
\def\Ud{{\sf U}}
\def\ud{{\sf u}}
\def\Vd{{\sf V}}
\def\vd{{\sf v}}
\def\Ui{\overline{\sf U}}
\def\Vi{\overline{\sf V}}
\def\Di{\overline{\sf D}}
\def\invdistr{\overline{\cal A}}

\def\Rho{P}
\def\rdes{r_{\rm des}}
\def\Vnod{{\cal V}}
\def\Cnod{{\cal C}}
\def\ha{\hat{a}}
\def\Edges{{\cal E}}
\def\Symm{{\cal S}}
\def\partf{{\cal Z}}

\def\graph{{\cal G}}
\def\E{{\mathbb E}}
\def\Et{{\mathbb E}_s}
\def\Eg{{\mathbb E}_{\cal G}}
\def\Ec{{\mathbb E}_{C}}
\def\Ey{{\mathbb E}_{y}}
\def\Eyz{{\mathbb E}_{y,z}}
\def\Pc{{\mathbb P}_C}
\def\R{{\mathbb R}}

\def\hRho{\hat{\Rho}}
\def\hLambda{\hat{\Lambda}}
\def\GMG{{\rm LDGM}(\cdots)}            
\def\PCG{{\rm LDPC}(\cdots)}
\def\GMS{{\rm LDGM}(n,\Lambda,\Rho)}    
\def\PCS{{\rm LDPC}(n,\Lambda,\Rho)} 
\def\GMP{{\rm LDGM}(n,\gamma,\Rho)}     
\def\PCP{{\rm LDPC}(n,\gamma,\Rho)}
\def\GMM{{\rm LDGM}(n,\gamma,\Lambda,\Rho)}     
\def\PCM{{\rm LDPC}(n,\gamma,\Lambda,\Rho)}
\def\H{{\mathbb H}}

\def\ub{\underline{b}}
\def\ui{\underline{i}}
\def\ux{\underline{x}}
\def\uhx{\underline{\hat{x}}}
\def\hx{\hat{x}}
\def\uy{\underline{y}}
\def\uhy{\underline{\hat{y}}}
\def\uz{\underline{z}}
\def\hy{\hat{y}}
\def\h{\hat{h}}
\def\Rh{\widehat{R}}

\def\uX{\underline{X}}

\def\uY{\underline{Y}}
\def\uhY{\underline{\hat{Y}}}

\def\uZ{\underline{Z}}
\def\us{\underline{\sigma}}
\def\uu{\underline{u}}
\def\uv{\underline{v}}

\def\Qv{Q_{\Vnod}}
\def\Qc{Q_{\Cnod}}
\def\Qeff{Q_{\rm eff}}
\def\di{\partial i}
\def\da{\partial a}

\def\sign{{\rm sign}}
\def\atanh{{\rm arctanh}}
\def\l2{\log_2}

\def\PB{P_{\rm B}}
\def\Pb{P_{\rm b}}
\def\Bin{P}
\def\Cap{{\tt C}}
\def\Prob{{\mathbb P}}

\def\bentro{{\tt h}}             

\def\ve{\varepsilon}

\def\om{\overline{m}}

\def\proof{\hspace{1.cm}{\bf Proof:}\hspace{0.1cm}}
\def\prooft{\hspace{0.5cm}{\bf Proof:}\hspace{0.1cm}}
\def\endproof{\hfill$\Box$\vspace{0.4cm}}

\newcommand{\<}{\langle}
\renewcommand{\>}{\rangle}

\begin{document}

\title{Tight bounds for LDPC and LDGM codes under MAP decoding}

\author{
  { Andrea Montanari\footnote{  Laboratoire de Physique Th\'{e}orique de l'Ecole Normale
  Sup\'{e}rieure, 24, rue Lhomond, 75231 Paris CEDEX 05, FRANCE
Internet: {\tt Andrea.Montanari@lpt.ens.fr}
(UMR 8549, Unit{\'e}   Mixte de Recherche du
  CNRS et de l'ENS). Financial support has been provided by the 
European Community under contracts  STIPCO and EVERGROW
(IP No. 1935 in the
complex systems initiative of the Future and Emerging Technologies
directorate of the IST Priority, EU Sixth Framework).}}
              \\
}

\maketitle

\thispagestyle{empty} 

\abstract{A new method for analyzing low density parity check (LDPC) codes
and low density generator matrix (LDGM) codes under bit maximum a posteriori 
probability (MAP) decoding is introduced. The method is based
on a rigorous approach to spin glasses developed by Francesco Guerra. It
allows to construct lower bounds on the entropy of the transmitted message
conditional to the received one. Based on heuristic statistical mechanics
calculations, we conjecture such bounds to be tight.
The result holds for standard irregular {\it ensembles} when used over binary 
input output symmetric channels.

The method is first developed for Tanner graph ensembles with Poisson 
left degree distribution. It is then generalized to
`multi-Poisson' graphs, and, by a completion procedure, to 
arbitrary degree distribution.}

\vspace{1cm}


\clearpage

%
%
\section{Introduction}

Codes based on random graphs are of huge practical and theoretical 
relevance. The analysis of such communication schemes is currently 
in a mixed status. From a practical point of view, the most
relevant issue is the analysis of linear-time decoding algorithms. 
As far as message-passing algorithms are concerned, our understanding
is rather satisfactory. Density evolution 
\cite{GallagerThesis,Tornado,RichardsonUrbanke,RichardsonUrbankeIntroduction} 
allows to compute exact 
thresholds for vanishing bit error probability, 
at least in the large blocklength
limit. These results have been successfully employed for designing 
capacity-approaching code ensembles \cite{ImprovedLDPCC,Chung}.

A more classical problem is the evaluation of decoding schemes which are 
optimal with respect to some fidelity criterion, such as word MAP
(minimizing the block error probability) or symbol MAP decoding
(minimizing the bit error probability). Presently, this issue has
smaller practical relevance than the previous one, and nonetheless
its theoretical interest is great. Any progress in this direction
would improve our understanding of the effectiveness of belief propagation
(and similar message-passing algorithms) in general inference
problems. Unhappily, the status of this research 
area~\cite{TypicalPairs,Bounds} is not as advanced
as the analysis of iterative techniques. In most cases one is
able to provide only upper and lower bounds on the thresholds
for vanishing bit error probability in the large blocklength
limit. Moreover, the existing techniques seems completely unrelated
from the ones employed in the analysis of iterative decoding.
This is puzzling. We know that, at least for some code
constructions and some channel models, belief propagation 
has performances which are close to optimal. The same has been observed
empirically in inference problems. Such a convergence in behavior 
hints at a desirable convergence in the analysis techniques.

This paper aims at bridging this gap. We introduce a new technique 
which allows to derive lower bounds on the entropy of the
transmitted message conditional to the received one. We conjecture
that the lower bound provided by this approach is indeed tight. 
Interestingly enough, the basic objects involved in the new bounds
are probability densities over ${\mathbb R}$, as in the density
evolution analysis of iterative decoding. These densities 
are required moreover to satisfy the same `symmetry' condition 
(see Sec. \ref{VariablesSection} for a definition) as the messages 
distributions in density evolution. The bound can be optimized 
with respect to the densities. A necessary condition for the densities 
to be optimal is that they correspond to a fixed point of density evolution
for belief propagation decoding.

The method presented in this paper is based on recent developments
in the rigorous theory of mean field spin glasses. 
Mean field spin glasses are theoretical models for
disordered magnetic alloys, displaying an extremely rich
probabilistic structure \cite{SpinGlass,TalagrandBook}. As shown by
Nicolas Sourlas \cite{Sourlas1,Sourlas2,Sourlas4}, there exists 
a precise mathematical correspondence between such models
and error correcting codes. Exploiting this correspondence,
a number of heuristic techniques from statistical mechanics have been 
applied to the analysis of coding systems, including
LDPC codes \cite{Saad1,Saad2,Mio,NostroDynamical} and turbo codes 
\cite{Turbo1,Turbo2}. Unhappily, the results obtained through this 
approach were non-rigorous, although, most of the times, they were 
expected to be exact. 

Recently, Francesco Guerra and Fabio Lucio 
Toninelli~\cite{Guerra,GuerraToninelli} succeeded in developing
a general technique for constructing bounds on the free energy 
of mean field spin glasses. The technique, initially applied
to the Sherrington-Kirkpatrick model, was later extended
by Franz and Leone \cite{FranzLeone} to deal with Ising systems on 
random graph with Poisson distributed degrees. Finally,  Franz, Leone
and Toninelli \cite{FranzLeoneToninelli} adapted it to systems on graphs 
with general degree distributions. This paper adds two improvements to this 
line of research. It generalizes it to Ising systems with (some classes of)
biased coupling distributions\footnote{The reader unfamiliar with the 
statistical physics jargon may skip this statement.}. 
Furthermore, it introduces a new way of dealing with general degree 
distributions  which (in our view) is considerably simpler than the
approach of Ref.~\cite{FranzLeoneToninelli}.
Using the new technique, we are able to prove that the asymptotic
expression for the conditional entropy of irregular LDPC
ensembles derived in \cite{NostroDynamical}
is indeed a lower bound on the real conditional entropy. This gives
further credit to the expectation that the results of \cite{NostroDynamical}
are exact, and we formalize this expectation as a conjecture in 
Sec.~\ref{GeneralizationSection}.

The new technique is based upon an interpolation procedure which progressively
eliminates right (parity check) nodes from the Tanner graph. 
This procedure is considerably simpler for graph ensembles with Poisson left 
(variable) degree distribution. Such graph can be in fact constructed
by adding a uniformly random right node at a time, 
independently from the others.
We shall therefore adopt a three steps strategy. We first prove 
our bound for Poisson ensembles. This allows to explain the important
ideas of the interpolation technique in the simplest possible context.
Unhappily Poisson ensembles may have quite bad error-floor properties due to
low degree node and are not very interesting for practical 
purposes~\footnote{One exception to this statement is provided by
Luby Transform codes \cite{LTCodes}. These can be regarded as Poisson 
ensembles, and, due to the large average right degree, have an arbitrary small
error floor}.
Next, we generalize the bound to `multi-Poisson' ensembles. 
These can be constructed by a sequence of rounds such that, within each 
round, right nodes are added independently of each other. In other words 
multi-Poisson graphs are obtained as the superposition of several 
Poisson graphs. 
Finally, we show that a general degree distribution can be 
approximated arbitrarily well using a `multi-Poisson' construction.
Together with 
continuity of the bound, this implies our general result.

In Section \ref{DefinitionsSection} we introduce the code ensembles
to be considered. Symbol-MAP decoding scheme is defined in 
Sec.~\ref{DecodingSection}, together with some basic probabilistic notations.
Section \ref{VariablesSection} collects some remarks on symmetric random
variables to be used in the proof of our main results. We then prove
that the per-bit conditional entropy of the transmitted message
concentrates in probability with respect to the code
realization. This serves as a justification for considering its 
ensemble average.
Our main
result, i.e. a lower bound on the average conditional entropy is stated
in Sec.~\ref{MainSection}. This Section also contains the proof for Poisson
ensembles. The proof for multi-Poisson and standard ensembles is 
provided (respectively) in Sections~\ref{sec:MultiProof} and 
\ref{StandardProofSection}.
Section \ref{ExampleSection} presents 
several applications of the new bound together with a general strategy 
for optimizing it. Finally, we draw our conclusion and discuss extensions 
of our work in Section \ref{GeneralizationSection}.
Several technical calculations are deferred to the Appendices.
%
%
\section{Code ensembles}
\label{DefinitionsSection}

In this Section we define the code ensembles to be analyzed in the 
rest of the paper. By `standard ensembles' we refer to the irregular ensembles
considered, e.g., in Refs.~\cite{Tornado,RichardsonUrbanke}. Poisson
ensembles are characterized by Poisson left degree distribution. Finally 
multi-Poisson codes can be thought as `combinations' of Poisson codes, 
and are mainly a theoretical device for approximating standard 
ensembles.

For each of the three families, we shall proceed by introducing a family of 
Tanner graph ensembles. In order to specify a Tanner graph, we need to
exhibit a set of left (variable) nodes $\Vnod$, of size
$|\Vnod|=n$, a set of right (check) nodes $\Cnod$, with $|\Cnod| = m$
and a set of edges $\Edges$, each edge joining a left
and a right node. If $i\in\Vnod$ denotes a generic left node and $a\in\Cnod$ 
a generic right node, an edge joining them will be given as  $(i,a)\in \Edges$.
Multiple edges are allowed (although only their parity matters
for the code definition). Furthermore, two graphs obtained through 
a permutation of the variable or of the check nodes are regarded as distinct
(nodes are `labeled').
The neighborhood of the variable  (check) node $i\in\Vnod$ ($a\in\Cnod$) 
is denoted by $\di$ ($\da$). In formulae $\di = \{ a\in\Cnod \, :\; 
(i,a)\in\Edges\}$ and $\da = \{ i\in\Vnod\, :\; (i,a)\in\Edges\}$.

A Tanner graph ensemble will be generically indicated as
$(\cdots)$, where `$\cdots$' is a set of relevant parameters. 
Expectation with respect to a Tanner graph ensemble will be denoted
by $\Eg$. 

Next, we define
$\PCG$ and $\GMG$ codes as the LDPC and LDGM codes associated to a random 
Tanner graph from the $(\cdots)$ ensemble. Since this construction does
not depend upon the particular family of Tanner graphs to be considered, 
we formalize it here.
\begin{df}
\label{DefCodes}
Let $\H = \{ H_{ai} : a\in\Cnod; i\in\Vnod\}$ be the adjacency matrix
of a Tanner graph from the  ensemble  $(\cdots)$:
$H_{ai}=1$ if $(i,a)$ appears in $\Edges$ an odd number of times,
and $H_{ai}=0$ otherwise. Then
\begin{enumerate}
\item A code from the $\GMG$ {\it ensemble} is  the linear
code on $GF[2]$ having $\H$ as generator matrix. The {\it design rate}
of this {\it ensemble} is defined as $\rdes = n/\Eg m$.
\item A code from the $\PCG$ {\it ensemble} is the linear code 
on $GF[2]$ having $\H$ as parity check matrix. The {\it design rate}
of this {\it ensemble} is $\rdes = 1-\Eg m/n$.
\end{enumerate}
\end{df}

Before actual definitions of graph ensembles, it is convenient to introduce 
some notations for describing them. For a given graph we define
the degree profile $(\hLambda,\hRho)$ as a couple of polynomials
\begin{eqnarray}
\hLambda(x) = \sum_{l=2}^{l_{\rm max}}\hLambda_lx^l\, ,\;\;\;\;\;\;\;
\hRho(x) = \sum_{k=2}^{k_{\rm max}}\hRho_k x^k\, ,\label{eq:SequenceDefinition}
\end{eqnarray}
such that $\hLambda_i$ ($\hRho_i$) is the fraction of left (right) 
nodes of degree $i$. The degree profile  $(\hLambda,\hRho)$,
will be in general a random variable depending on the particular graph 
realization. On the other hand, each ensemble will be assigned 
a non-random `design degree sequence'  $(\Lambda,\Rho)$. This is the degree
profile that the ensemble is designed to achieve (and in some cases
achieves with probability approaching one in the large blocklength limit).
Both $\Lambda(x)$ and $\Rho(x)$ will have non-negative coefficients and
satisfy the normalization condition $\Lambda(1) = \Rho(1) =1$.
Finally, it is useful to introduce the `edge perspective'
degree sequences: $\lambda(x) = \sum_{l}\lambda_lx^{l-1}\equiv
\Lambda'(x)/\Lambda'(1)$, and  $\rho(x) = \sum_{k}\rho_kx^{k-1}\equiv
\Rho'(x)/\Rho'(1)$.

In the following Sections, `with high probability' (w.h.p.) and similar 
expressions will refer to the large blocklength limit, with the other
code parameters kept fixed.
%
%
\subsection{Standard ensembles}
\label{StandardDefinitionSection}

Standard ensembles are discussed in several papers
\cite{Tornado0,Tornado,RichardsonUrbanke,RichardsonUrbankeIntroduction,ImprovedLDPCC,RichardsonUrbankeBook}.
Their performances under iterative decoding have been thoroughly investigated 
allowing for ensemble optimization~\cite{Tornado,Chung}. A standard
ensemble of factor graphs is defined by  assigning the blocklength 
$n$ and the design degree sequence $(\Lambda,\Rho)$. We shall assume the 
maximum left and right degrees $l_{\rm max}$ and $k_{\rm max}$ to be finite.
\begin{df}
A graph from the standard ensemble $(n,\Lambda,\Rho)$ includes $n$ 
left nodes and $m \equiv n\Lambda'(1)/\Rho'(1)$ right nodes
(i.e. $\Vnod = [n]$ and $\Cnod = [m]$), and is constructed as follows.
Partition the set of left nodes uniformly at random into $l_{\rm max}$ subsets
$\{\Vnod_l\}$ with  $|\Vnod_l| = n\Lambda_l$. For any 
$l=2,\dots,l_{\rm max}$, associate $l$ `sockets' to each $i\in\Vnod_l$.
Analogously, partition the right nodes into sets $\{\Cnod_k\}$ with  
$|\Cnod_k| = n\Rho_k$, and associate $k$ sockets to the nodes in
$\Cnod_k$. Notice that the total number of sockets on each side is
$n\Lambda'(1)$.

Choose a uniformly random permutation over $n\Lambda'(1)$ objects and connect
the sockets accordingly (two connected sockets form an edge). 
\end{df}
The ensemble $(n,\Lambda,\Rho)$ is non-empty only if the numbers
$n\Lambda'(1)/\Rho'(1)$ and $\{n\Lambda_l,m\Rho_k\}$  are integers. 
The design rate of the $\PCS$ ensemble is $\rdes = 1-\Lambda'(1)/\Rho'(1)$,
while for the $\GMS$ ensemble $\rdes = \Rho'(1)/\Lambda'(1)$.
It is clear that the degree profile $(\hLambda,\hRho)$
concentrates around the design degree sequences  $(\Lambda,\Rho)$

An equivalent construction of a graph in the standard ensemble 
is the following. As before, we shall partition nodes and associate them  
sockets. Furthermore we shall keep track of the number of `free'
sockets at variable node $i$ after $t$ steps in the procedure,
through an integer $\soc_i(t)$. Therefore, at the beginning set 
$\soc_i(0) = l$ for any $i\in\Vnod_l$.
Next, for any $a=1,\dots, m$ consider the $a$-th
check node, and assume that $a\in\Cnod_k$. For $r=1,\dots,k$ do the following
operations: $(i)$ choose $i^a_r$ in $\Vnod$ with probability distribution
$w_i(t) = \soc_i(t)/(\sum_j \soc_j(t))$; $(ii)$ Set $\soc_{i^a_r}(t+1) = 
\soc_{i^a_r}(t)-1$ and $\soc_i(t+1) = \soc_i(t)$ for any $i\neq i^a_r$;
$(iii)$ increment $t$ by 1. Finally $a$ is connected to 
$i^a_1,\dots i^a_k$. The graph $\graph$ obtained after the last right node 
$a=m$ is connected to the left side, is distributed according to the 
standard ensemble as defined above.
%
%
\subsection{Poisson ensembles}
\label{PoissonDefinitionSection}

A Poisson ensemble is specified by the blocklength 
$n$, a real number $\gamma>0$, and a right degree design sequence
$\Rho(x)$. Again, we require the maximum right degree $k_{\rm max}$
to be finite.
\begin{df}
A Tanner  graph from the Poisson ensemble $(n,\gamma,\Rho)$ is constructed
as follows. The graph has $n$ variable nodes $i\in \Vnod\equiv[n]$.
For any $2\le k\le k_{\rm max}$, choose $m_k$ from a Poisson 
distribution with parameter $n\gamma\Rho_k/\Rho'(1)$. The graph has
$m = \sum_k m_k$ check nodes $a\in\Cnod\equiv\{ 
(\ha,k) : \ha\in[m_k]\; , 2\le k \le k_{\rm max}\}$. 

For each parity check node $a=(\ha,k)$ choose $k$ variable nodes 
$i^a_1,\dots,i^a_k$ uniformly at random in $\Vnod$, and
connect $a$ with $i^a_1,\dots,i^a_k$.
\end{df}
A few remarks are in order: they are understood to hold in the large 
blocklength limit $n\to\infty$ with $\gamma$ and $\Rho$ fixed.
$(i)$ The number of check nodes is a Poisson random variable with mean 
$\Eg m = n\gamma/\Rho'(1)$. Moreover, $m$ concentrates
in probability around its expectation. $(ii)$ The right degree profile  
$\hRho$ concentrates around its expectation $\Eg \hRho_k =\Rho_k+O(1/n)$. 
$(iii)$ The left degree profile
$\hLambda$ has expectation
 $\Eg \hLambda_l = \gamma^l e^{-\gamma}/l !+O(1/n)$, and
concentrates around its average. 
In view of these remarks, we define the design left degree sequence 
$\Lambda$ of a Poisson ensemble to be  given by $\Lambda(x)=e^{\gamma (x-1)}$. 

The design rate\footnote{In the first case, the actual rate $r$  is
equal to $n/m$, and therefore, because of the observation 
$(i)$ above, concentrates around the design rate.
In the $\PCP$ ensemble, the actual rate $r$ is always larger
or equal to $1-m/n$, because the rank of the parity check matrix
$\H$ is not larger than $m$. Notice that $1-m/n$ concentrates around 
the design rate.
It is not hard to show that the actual rate
is, in fact, strictly larger than $r_{\rm des}$ with high probability. 
A lower bound on the number of codewords $N(\H)$ can in fact
be obtained by counting all the codewords such that  $x_i = 0$
for all the variable nodes $i$ with degree larger than $0$.
This implies $r\ge \hLambda_0$. Take $\gamma$ and $\Rho(x)$ such that
$e^{-\gamma}>1-\gamma/\Rho'(1)+\delta$ for some $\delta>0$
(this can be always done
by chosing $\gamma$ large enough). Since $\hLambda_0$ is closely
concentrated around $e^{-\gamma}$ in the $n\to\infty$ limit,
we have $r>r_{\rm des}+\delta $ with high probability.} 
of the $\GMP$ ensemble is $\rdes = \Rho'(1)/\gamma$,
while, for a $\PCP$ ensemble, $\rdes = 1-\gamma/\Rho'(1) = 
1-\Eg m/n$.

The important simplification arising for Poisson ensembles is that their rate
can be easily changed in a continuous way. Consider for instance the problem
of sampling a graph from the $(n,\gamma+\Delta\gamma,\Rho)$
ensemble. To the first order in $\Delta\gamma$ this can be done
as follows. First generate a graph from the $(n,\gamma,\Rho)$.
Then, for each $k\in\{3,\dots,k_{\rm max}\}$, add a check node
of degree $k$ with probability $\Delta\gamma\,\Rho_k/\Rho'(1)$, and
connect it to $k$ uniformly random variable nodes $i_1,\dots,i_k$.
Technically, this property will allow to  compute derivatives 
with respect to the code rate, cf. App.~\ref{StraightforwardApp}.
%
%
\subsection{Multi-Poisson ensembles}
\label{MultiPoissonDefinitionSection}

We introduce multi-Poisson ensembles
them in order to `approximate' graphs in standard ensembles as the union
of several Poisson sub-graphs. The construction proceeds by  
a finite number of rounds. During each round, we add a certain number of right 
nodes to the graph. The adjacent left nodes are drawn
independently using a biased
distribution. The bias drives the procedure towards the design 
left degree distribution, and is most effective in the limit of a large 
number of `small' stages.
A multi-Poisson ensemble is fully specified by the blocklength
$n$, a design degree sequence $(\Lambda,\Rho)$ (with
$\Lambda(x)$ and $\Rho(x)$ having maximum degree, respectively, 
$l_{\rm max}$, and $k_{\rm max}$), and a  real
number $\gamma> 0$ describing the number of checks to be added at each 
round. 
The number of rounds is defined to be $t_{\rm max}\equiv \lfloor \Lambda'(1)/
\gamma\rfloor-1$. Below we adopt the notation $[x]_+ = x$ if $x\ge 0$ and $=0$
otherwise.
\begin{df}\label{def:MultiPoisson}
A Tanner  graph $\graph$ from the multi-Poisson ensemble 
$(n,\Lambda,\Rho, \gamma)$ is defined by the following procedure. 
The graph has $n$ variable nodes $i\in \Vnod\equiv[n]$, which are partitioned 
uniformly at random into $l_{\rm max}$ subsets $\{\Vnod_l\}$, with
$|\Vnod_l|= n\Lambda_l$. For each $l=2,\dots,l_{\rm max}$ and each 
$i\in\Vnod_l$, let $\soc_i(0) = l$. Let $\graph_0$ be the graph with 
variable nodes $\Vnod$ and without any check 
node. We shall define a sequence $\graph_0,\dots,\graph_{t_{\rm max}}$,
and set $\graph = \graph_{t_{\rm max}}$.

For any $t=0,\dots,t_{\rm max}-1$, $\graph_{t+1}$ is constructed from
$\graph_t$ as follows. For any 
$2\le k\le k_{\rm max}$, choose $m^{(t)}_k$ from a Poisson distribution 
with parameter $n\gamma\Rho_k/\Rho'(1)$.  Add 
$m^{(t)} = \sum_{k} m_k^{(t)}$ check nodes to $\graph_t$ and denote them
by $\Cnod_t \equiv\{ (\hat{a},k,t) : \hat{a}\in [m_k^{(t)}],\, 
2\le k\le k_{\rm max}  \}$. For each node $a=(\ha,k,t)\in\Cnod_t$, 
choose $k$ variable nodes $i^a_1, \dots,i^a_k$ independently 
in $\Vnod$, with distribution 
$w_i(t) = [\soc_i(t)]_+/(\sum_{j}[\soc_j(t)]_+)$, 
and connect $a$ with $i^a_1,\dots,i^a_k$. Finally, set 
$\soc_i(t+1) = \soc_i(t) -\Delta_i(t)$, where $\Delta_i(t)$ is the 
number of times node $i$ has been chosen during  round $t$. 
\end{df}
Notice that the above procedure may fail if $\sum_{j}[\soc_j(t)]_+$ 
vanishes for some $t<t_{\rm max}-1$. However, it is easy 
to understand that, in the large blocklength limit, the procedure will 
succeed with high probability (see the proof of Lemma~\ref{MultiLemma} below).

The motivation for the above definition is that, as $\gamma\to 0$ at
$n$ fixed, it reproduces the definition of standard ensembles
(see the formulation at the end of Sec.~\ref{StandardDefinitionSection}). 
At non-zero $\gamma$, the multi-Poisson ensemble differ from the 
standard one  in that the probabilities
$w_i(t)$ are changed only every about $n\gamma$ edge connections.
On the other hand, we shall be able to analyze multi-Poisson ensembles in the 
asymptotic limit $n\to\infty$ with the other 
parameters --the design distributions $(\Lambda,\Rho)$ and the 
stage `step-size' $\gamma$-- kept fixed.
It is therefore crucial to estimate the `distance' between 
multi-Poisson and standard ensembles for large $n$ at fixed $\gamma$.

We formalize this idea using a coupling argument. Let us recall here
that, given two random variables $X\in\cX$ and $Y\in\cY$, a coupling among 
them is a random variable $(X',Y')\in\cX\times\cY$ such that 
the marginal distributions of $X'$ and  $Y'$ are the same as (respectively)
those of $X$ and $Y$. Furthermore, we define a `rewiring' as the 
elementary operation of either adding or removing  a function node from 
a Tanner graph.
The following Lemma is proved in Appendix~\ref{CouplingApp}.
\begin{lemma}\label{MultiLemma}
Let $0<\gamma<1$ and $(\Lambda,\Rho)$ be a degree sequence pair. 
Then there exist two $n$--independent positive numbers 
$A(\Lambda,\Rho),b(\Lambda,\Rho)>0$ and a coupling 
$(\graph_{\rm s},\graph_{\rm mP})$, 
between the standard ensemble $(n,\Lambda,\Rho)$ and the
multi-Poisson ensemble $(n,\Lambda,\Rho,\gamma)$, 
such that w.h.p. $\graph_{\rm s}$ is obtained from $\graph_{\rm mP}$ with a
 number of rewirings smaller than
 $A(\Lambda,\Rho)n\gamma^{b(\Lambda,\Rho)}$.
\end{lemma}
In other words, we can obtain a random Tanner graph from the standard ensemble
by first generating a multi-Poisson graph with the same design degree 
sequences and small $\gamma$, and then changing a small fraction of edges.

Although it is convenient to define multi-Poisson ensembles in terms of
the design degree distribution $\Lambda(x)$, such a distribution
{\em does not} coincide with the actual degree profile achieved by 
the above construction, even in the large blocklength limit. 
In order to clarify this statement, let us define the expected degree 
distribution $\Lambda^{(n,\gamma)}(x) \equiv \Ec \hLambda(x)$ for blocklength 
$n$ and step parameter $\gamma$. Furthermore, 
let\footnote{In App.~\ref{app:DegreeMultiPoisson} we shall prove the 
existence (in an appropriate sense)
of this limit and provide an efficient way of computing 
$\Lambda^{(\gamma)}(x)$.} 
$\Lambda^{(\gamma)}(x)\equiv\lim_{n\to\infty}\Lambda^{(n,\gamma)}(x)$.
We claim that, in general, $\Lambda^{(\gamma)}(x)\neq \Lambda(x)$.
This can be verified explicitly using the 
characterization of $\Lambda^{(\gamma)}(x)$ provided in 
Appendix \ref{app:DegreeMultiPoisson}. However, as a consequence of
Lemma \ref{MultiLemma}, for small $\gamma$, $\Lambda^{(\gamma)}(x)$ is 
`close' to $\Lambda(x)$.
\begin{coro}\label{coro:Degree}
Let $0<\gamma<1$ and $(\Lambda,\Rho)$ be a degree sequence pair. 
Then there exist two $n$--independent positive numbers 
$A(\Lambda,\Rho),b(\Lambda,\Rho)>0$ such that 
$|\Lambda^{(\gamma)}_l-\Lambda_l|\le 
A(\Lambda,\Rho)n\gamma^{b(\Lambda,\Rho)}$ for each $l\in\{2,\dots,
l_{\rm max}\}$. Moreover 
\begin{eqnarray}
\lim_{\gamma\to 0} ||\Lambda^{(\gamma)}-\Lambda|| = 0\, ,\;\;\;\;\;\;\;\;\;\;
\mbox{where} \;\;\;\;  ||\Lambda^{(\gamma)}-\Lambda||\equiv
\frac{1}{2}\sum_{l} |\Lambda^{(\gamma)}_l-\Lambda_l|\, .
\end{eqnarray}
\end{coro}
The distance $||\mu-\nu||$ defined above is often called the `total
variation distance': we refer to App.~\ref{CouplingApp} for some properties. 

{\bf Proof:}\hspace{0.1cm}
Let $(\graph_{\rm s},\graph_{\rm mP})$ be a pair of Tanner graphs distributed
as in the coupling of Lemma \ref{MultiLemma}. Their marginal distributions
are, respectively, the standard  $(n,\Lambda,\Rho)$ and the
multi-Poisson $(n,\Lambda,\Rho,\gamma)$ ones. Denote by 
$\hLambda_l(\graph_{\cdot})$ the fraction of degree-$l$ variable nodes
in graph $\graph_{\cdot}$. Then, there exist $A$ and $b$ such that
\begin{eqnarray}
\lim_{n\to\infty}\Prob[\exists \, l\; {\rm s.t.}\; |\hLambda_l(\graph_{\rm s})-
\hLambda_l(\graph_{\rm mP})|>A\gamma^b] = 0\, .
\end{eqnarray}
This follows from Lemma \ref{MultiLemma} together with the fact that each 
rewiring induces a change bounded by $k_{\rm max}/n$ in 
the degree profile. Therefore, using the notation 
$\Lambda^{(n)}_l = \E\hLambda_l(\graph_{\rm s})$ for the expected
degree profile in the standard ensemble at finite blocklength, we get
\begin{eqnarray}
|\Lambda^{(\gamma,n)}_l-\Lambda^{(n)}_l| = |\E[\hLambda_l(\graph_{\rm s})-
\hLambda_l(\graph_{\rm mP})]|\le \E| \hLambda_l(\graph_{\rm s})-
\hLambda_l(\graph_{\rm mP})|\le A\gamma^b+o_n(1)\, .
\end{eqnarray}
The first thesis follows by taking the $n\to\infty$ limit.

Convergence in total variation distance follows immediately:
\begin{eqnarray}
||\Lambda^{(\gamma)}-\Lambda|| & = &\frac{1}{2}\sum_{l}
|\Lambda_l^{(\gamma)}-\Lambda_l|
\le  \frac{1}{2}\sum_{l\le l_*}
|\Lambda_l^{(\gamma)}-\Lambda_l|+ \frac{1}{2}\sum_{l>l_*}\Lambda_l+
\frac{1}{2}\sum_{l>l_*}\Lambda_l^{(\gamma)} =\nonumber\\
&=&\frac{1}{2}\sum_{l\le l_*}
|\Lambda_l^{(\gamma)}-\Lambda_l|+ \frac{1}{2}(1-\sum_{l\le l_*}\Lambda_l)+
\frac{1}{2}(1-\sum_{l>l_*}\Lambda_l^{(\gamma)})\, .
\end{eqnarray}
The last expression can be made arbitrarily small by chosing
$\gamma$ and $l_*$ appropriately. For instance, one can choose $l_*$ 
in such a way that  $\sum_{l\le l_*}\Lambda_{l}\ge 1-\ve$, and then
$\gamma$ such that $A\gamma^b\le \ve/l_*$. This implies
$||\Lambda^{(\gamma)}-\Lambda|| \le 3\ve$.
\endproof
%
%
\section{Decoding Schemes}
\label{DecodingSection}

In the LDGM case the codeword bits are naturally associated to
the check nodes. A string $\uhx =\{\hx_a:a\in\Cnod\}\in\{0,1\}^m$
is a codeword if and only if there exists an information message
$\ux =\{ x_i: i\in\Vnod\}\in\{0,1\}^n$ such that
\begin{eqnarray}
\hx_a = x_{i^a_1}\oplus\cdots\oplus x_{i^a_k}\, ,\label{EncodingLDGM}
\end{eqnarray}
for each $a=(\ha,k)\in\Cnod$. Here $\oplus$ denotes the sum modulo 2.
Encoding consists in choosing an information message 
$\ux$ with uniform probability distribution and constructing the corresponding 
codeword $\uhx$ using the equations (\ref{EncodingLDGM}).
Notice that, because the code is linear, each codeword is the image of
the same number of information messages.
Therefore choosing an information message uniformly at random
probability is equivalent to choosing a codeword uniformly at random.

In the LDPC case the codeword bits can be associated to variable nodes.
A string $\ux =\{ x_i: i\in\Vnod\}\in\{0,1\}^n$ 
is a codeword if and only if it satisfies the parity check equations
\begin{eqnarray}
x_{i^a_1}\oplus\cdots\oplus x_{i^a_k} = 0\, ,\label{ParityChecks}
\end{eqnarray}
for each $a=(\ha,k)\in\Cnod$. In the encoding process we pick a 
codeword with uniform distribution.

The codeword, chosen according to the above encoding process,
is transmitted on a binary-input output symmetric channel (BIOS) 
with output alphabet ${\cal A}$ and 
transition probability density $Q(y|x)$. In the following we shall
use a discrete notation for ${\cal A}$. It is straightforward
to adapt the formulas below to the continuous case. If, for
instance ${\cal A} = {\mathbb R}$, sums should be replaced by
Lebesgue integrals: $\sum_{y\in{\cal A}}\cdot\to\int\! dy\;\; \cdot\;$.

The channel output has the form 
$\uhy = \{ y_a: a\in\Cnod\}\in {\cal A}^m$ in the LDGM case or
$\uy = \{ y_i: i\in\Vnod\}\in {\cal A}^n$ LDPC case. In order to
keep unified notation for the two cases, we shall introduce a simple
convention which introduces a fictitious output $\uy$, associated
to the variable nodes, (in the LDGM case) or $\uhy$, associated
to the check nodes (in the LDPC case). 
If an LDGM is used, $\uy$ takes by definition a standard
value $\uy_*=(*,\dots,*)$, while of course $\uhy$ is determined 
by the transmitted codeword and the channel realization. 
The character $*$ should be thought as an erasure.
If we are considering a LDPC, 
$\uy$ is the channel output, while $\uhy$ takes a standard value 
$\uhy_0 = (0,\dots,0)$.

We will focus on the probability distribution $P(\ux|\uy,\uhy)$
of the vector $\ux$, conditional to the channel output $(\uy,\uhy)$. 
Depending upon the family of codes employed (whether LDGM or
LDPC), this distribution has different meanings. It is the distribution
of the information message in the LDGM case, and the distribution of the
transmitted codeword in the LDPC case. It can be always written in the form
\begin{eqnarray}
P(\ux|\uy,\uhy) = \frac{1}{\partf(\uy,\uhy)}\, \prod_{a\in\Cnod}
\Qc(\hy_a|x_{i^a_1}\oplus\cdots\oplus x_{i^a_k})
\prod_{i\in\Vnod} \Qv (y_i|x_i)\, .
\label{ProbabilityDistribution}
\end{eqnarray}
The precise form of the functions $\Qc(\cdot|\cdot)$ and $\Qv(\cdot|\cdot)$
depends upon the family of codes:
\begin{enumerate}
\item[] For LDGM's:
\begin{eqnarray}
\Qc(\hy|\hx) = Q(\hy|\hx)\, ,
\;\;\;\;\;\; \Qv(y|x) = \left\{\begin{array}{rl}
1& \mbox{if $y=*$,}\\
0& \mbox{otherwise}\end{array}\right.\, .
\end{eqnarray}
\item[] For LDPC's
\begin{eqnarray}
\Qc(\hy|\hx) =  \left\{\begin{array}{rl}
1& \mbox{if $\hy=\hx$,}\\
0& \mbox{otherwise}\end{array}\right.\, ,
\;\;\;\;\;\; \Qv(y|x) = Q(y|x)\, .
\end{eqnarray}
\end{enumerate}
\begin{figure}
\centerline{\epsfig{figure=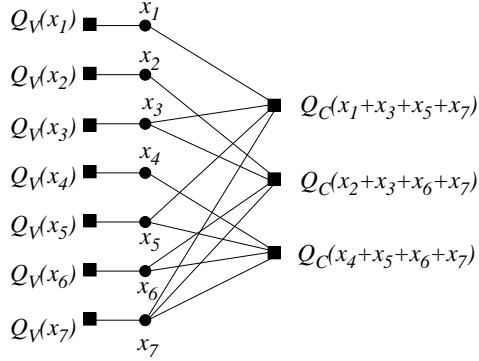,angle=0,width=0.35\linewidth}}
\caption{\small Factor graph representation of the probability distribution
of the transmitted codeword (information message for LDGM codes)
conditional to the channel output, cf. Eq. (\ref{ProbabilityDistribution}). 
Notice that, for the sake of compactness the $y$ arguments have been omitted.}
\label{FactorFig}
\end{figure}
The probability distribution (\ref{ProbabilityDistribution}) can be 
conveniently represented in the factor graphs language \cite{Factor}, see 
Fig. \ref{FactorFig}. 
There are two type of factor nodes in such a graph: nodes corresponding
to $\Qv(\cdot|\cdot)$ terms on the left, and nodes corresponding
to $\Qc(\cdot|\cdot)$ on the right.
It is also useful to introduce a specific notation for
the expectation with respect to the distribution 
(\ref{ProbabilityDistribution}). If $F:\{0,1\}^{\Vnod}\to\R$ is a function
of the codeword (information message for LDGM codes), we define:
\begin{eqnarray}
\<F(\ux)\> \equiv \sum_{\ux} F(\ux)\, P(\ux|\uy,\uhy)\, .
\end{eqnarray}
In the proof of our main result, see Sec. \ref{MainSection}, it will be 
also useful to consider several i.i.d. copies of $\ux$, each one
having a distribution of the form (\ref{ProbabilityDistribution}).
If  $F:\{0,1\}^{\Vnod}\times\cdots\times\{0,1\}^{\Vnod} \to\R$,
$(\ux^{(1)},\dots,\ux^{(q)})\mapsto F(\ux^{(1)},\dots,\ux^{(q)})$ 
is a real function of $q$ such copies, we denote its expectation value by
\begin{eqnarray}
\< F(\ux^{(1)},\dots,\ux^{(q)})\>_* \equiv \sum_{\ux^{(1)}\dots\ux^{(q)}} 
F(\ux^{(1)},\dots,\ux^{(q)})\; P(\ux^{(1)}|\uy,\uhy)\cdots P(\ux^{(q)}|
\uy,\uhy)\, .\label{ReplicaDef}
\end{eqnarray}

We are interested in two different decoding schemes: bit MAP (for
short MAP hereafter) and iterative belief propagation (BP) decoding.
In MAP decoding we follow the rule:
\begin{eqnarray}
x^{\rm MAP}_i = \arg\max_{x_i} P(x_i|\uy,\uhy)\, ,
\end{eqnarray}
where $P(x_i|\uy,\uhy)$ is obtained by marginalizing the probability 
distribution (\ref{ProbabilityDistribution}) over $\{x_j; j\neq i\}$. 

BP decoding~\cite{Pearl} 
is a message passing algorithm. The received message is encoded
in terms of log-likelihoods $\{h_i:i\in\Vnod\}$  and $\{J_a:a\in\Cnod\}$
as follows (notice the unusual normalization):
\begin{eqnarray}
h_i  = \frac{1}{2}\log \frac{\Qv(y_i|0)}{\Qv(y_i|1)}\, ,\;\;\;\;\;\;
\;\;\;\;\;\;
J_a  = \frac{1}{2}\log \frac{\Qc(\hy_a|0)}{\Qc(\hy_a|1)}\, .
\label{LogLikelihoodDef}
\end{eqnarray}
The messages $\{ u_{a\to i}, v_{i\to a} \}$ are 
updated recursively following the rule
\begin{eqnarray}
u_{a\to i} &:= & \atanh\{\tanh J_a\!\!
\prod_{j\in \partial a\backslash i}\!\!\tanh v_{j\to a}\}\, ,
\label{BP1}\\
v_{i\to a}  &:= & h_i+\sum_{b\in\partial i\backslash a} u_{b\to i}\, .
\label{BP2}
\end{eqnarray}
After iterating Eqs. (\ref{BP1}), (\ref{BP2}) a certain fixed number
of times, all the incoming messages to a certain bit $x_i$ are used to estimate
its value.

In the following, we shall denote by $\Ec$ the expectation with 
respect to one of the code  ensembles defined in this Section. 
Which one of the  ensembles
defined above will be  clear from the context.  
We will denote by $\Ey$ the expectation with respect to the 
received message $(\uy,\uhy)$, assuming the transmitted one to
be the all-zero codeword (or, in some cases, with respect to one of 
the received symbols).
%
%
\section{Random variables}
\label{VariablesSection}

It is convenient to introduce a few notations and simple results 
for handling some particular classes of random variables. Since these
random variables  will appear several times in the following,
this introduction will help to keep the presentation more compact. 
Here, as in the other Sections, we shall sometimes use the standard 
convention of denoting random variables by upper case letters 
(e.g. $X$) and deterministic values by lower case letters (e.g. $x$).
Moreover, we use the symbol $\ed$ to denote identity in 
distribution (i.e. $X\ed Y$ if $X$ and $Y$ have the same distribution).

The most important class of random variables in the present
paper is the following.
\begin{df}
A random variable $X$ is said to be {\it symmetric} (and 
we write $X\in \Symm$) if it takes values in $(-\infty,+\infty]$ and
\begin{eqnarray}
\E_X[ g(x)] = \E_X[ e^{-2x}g(-x)]\, ,\label{SymmetricDefinition}
\end{eqnarray}
for any real function $g$  such that at least one of the expectation 
values exists.
\end{df} 
This definition was already given in \cite{RichardsonDesign}. 
Notice however that 
the two definitions differ by a factor $2$ in the normalization of $X$.
The introduction of symmetric random variables is motivated by the following
observations  \cite{RichardsonDesign}:
\begin{enumerate}
\item   If $Q(y|x)$ is the transition probability of a BIOS and 
$y$ is distributed according to $Q(y|0)$, then the log-likelihood
\begin{eqnarray}
\ell(y) = \frac{1}{2}\log\frac{Q(y|0)}{Q(y|1)}\, ,\label{Apriori}
\end{eqnarray}
is a symmetric random variable. In particular the log-likelihoods
$\{h_i:i\in {\cal V}\}$ and $\{J_a:a\in{\cal C}\}$ defined in 
(\ref{LogLikelihoodDef}) are symmetric random variables.
\item Two important symmetric variables are: $(i)$ $X=0$ with probability 1;
$(ii)$ $X=+\infty$ with probability $1$. Both can be considered as particular 
examples of the previous observation. Just take an erasure channel 
with erasure probability $1$ (case $i$) or $0$ (case $ii$). 
\item If $X$ and $Y$ are symmetric then $Z=X+Y$ is symmetric. 
Here the sum is extended to the domain $(-\infty,+\infty]$ by imposing the
rule $x+\infty= +\infty$.
\item If $X$ and $Y$ are symmetric then $Z=\atanh(\tanh X\tanh Y)$ 
is symmetric. 
The functions $x\mapsto\tanh x$ and $x\mapsto\atanh\, x$ 
are extended to the domains (respectively) $(-\infty,+\infty]$ 
and $(-1,+1]$ by imposing the  rules $\tanh(+\infty) = +1$
and $\atanh(+1)=+\infty$.
\end{enumerate}
As a consequence of the above observations, if the messages 
$\{ u_{a\to i}, v_{i\to a} \}$ in BP decoding are initialized to
some symmetric random variable, remain symmetric at each step of the
decoding procedure. This follows directly from the update equations
(\ref{BP1}) and (\ref{BP2}).

Remarkably, the family of symmetric variables is `stable'
also under MAP decoding. This is stated more precisely in the
result below \cite{RichardsonUrbankeBook}.
\begin{lemma}
Let $P(\ux|\uy,\uhy)$ be the probability distribution of the channel input
(information message) $\ux = (x_1\dots x_n)$, conditional to 
the channel output $\uy = (y_1\dots y_n)$, $\uhy = (\hy_1\dots \hy_n)$
as given in Eq. (\ref{ProbabilityDistribution}). Assume the channel is BIOS 
and $\ux$ is the codeword of (is coded using a) linear code. 
Let $\ui \equiv \{i_1\dots i_k\}\subseteq [n]$ and define
\begin{eqnarray}
\ell_{\ui}(\uy) = \frac{1}{2}\, 
\log\frac{P(x_{i_1}\oplus\dots\oplus x_{i_k}=0|\uy)}
{P(x_{i_1}\oplus\dots\oplus x_{i_k}=1|\uy)}\, .\label{Aposteriori}
\end{eqnarray}
If $\uy$ is distributed according to the channel, conditional to the all-zero
codeword being transmitted then $\ell_{\ui}(\uy)$ is a symmetric random
variable.
\end{lemma}

To complete our brief review of properties of symmetric random variables, it
is useful to collect a few identities to be used several times in the 
following (throughout the paper we use $\log$ and $\log_2$
to denote, respectively, natural and base 2 logarithms). 
\begin{lemma}
\label{Tricks}
Let $X$ be a symmetric random variable. Then the following identities hold
\begin{eqnarray}
\E_X \tanh^{2k-1}X = \E_X\tanh^{2k}X = \E_X\frac{\tanh^{2k}X}{1+\tanh X}\, ,
\;\;\;\;\;\mbox{for $k\ge 1$,} \label{Identity1}\\
\E_X\log(1+\tanh X) = \sum_{k=1}^{\infty} \left(
\frac{1}{2k-1}-\frac{1}{2k}\right)\E_X\tanh^{2k}X\, .\label{Identity2}
\end{eqnarray}
\end{lemma}
\proof The identities (\ref{Identity1}) follow from the observation
that, because of Eq. (\ref{SymmetricDefinition}), we have 
\begin{eqnarray}
\E\, g(X) = \frac{1}{2}\E\, [g(X)+e^{-2X}g(-X)]\, .\label{eq:Game}
\end{eqnarray}
The desired result is obtained by substituting either $g(X) = \tanh^{2k}X$
or $g(X) = \tanh^{2k-1}X$.

In order to get (\ref{Identity2}), apply the identity (\ref{eq:Game})
to $g(X) =\log(1+\tanh X)$ and get
\begin{eqnarray}
\E\log(1+t) &=& \frac{1}{2}\E\frac{1}{1+t}[(1+t)
\log (1+t) + (1-t)\log(1-t)] =\\
&=&\E \sum_{k=1}^{\infty} \left(
\frac{1}{2k-1}-\frac{1}{2k}\right) \frac{t^{2k}}{1+t}\, ,
\end{eqnarray}
where we introduced the shorthand $t = \tanh X$.
At this point you can switch sum and expectation because of the monotone 
convergence theorem. \endproof

The space of symmetric random variables is useful because log-likelihoods
(for our type of problems) naturally belong to this space. It is also useful
to have a compact notation for the distribution of a binary variable $x$ whose 
log-likelihood is  $u\in (-\infty,+\infty]$. We therefore define
\begin{eqnarray}
\Bin_u(x) =\left\{\begin{array}{ll}
(1+e^{-2u})^{-1}, & \mbox{if $x=0$,}\\
(1+e^{-2u})^{-1}e^{-2u}, & \mbox{if $x=1$.}\end{array}\right.\label{BinDef}
\end{eqnarray}
%
%
\section{Concentration}
\label{ConcentrationSection}

Our main object of interest will be the entropy per input symbol
of the transmitted message conditional to the received one
(we shall generically measure entropies in bits).
We take the average of this quantity with respect to the 
code  ensemble to define
\begin{eqnarray}
h_n & = &\frac{1}{n}\Ec H_n(\uX|\uY,\uhY) = \label{EntropyDensity1}\\ 
& = & \frac{1}{n}\Ec \Ey 
\log_2 \partf(\uy,\uhy) - \frac{\gamma}{\Rho'(1)}\sum_{y} \Qc(y|0)\log_2 \Qc(y|0)\nonumber\\
&& -\sum_{y} \Qv(y|0)\log_2 \Qv(y|0)\, .
\label{EntropyDensity2}
\end{eqnarray}
In passing from Eq. (\ref{EntropyDensity1}) to (\ref{EntropyDensity2}),
we exploited the symmetry of the channel, and fixed the transmitted message 
to be the all-zero codeword. 

Intuitively speaking, the conditional entropy appearing in 
Eq.~(\ref{EntropyDensity1}) allows to estimate the typical 
number of inputs with non-negligible probability for a given channel output. 
The most straightforward rigorous justification for looking at
the conditional entropy is provided by Fano's inequality~\cite{Cover}
which we recall here.
\begin{lemma}
\label{FanoLemma}
Let $\PB(C)$ ($\Pb(C)$) be the block (bit) error probability for a code
$C$ having blocklength $n$ and rate $r$. Then
\begin{enumerate}
\item $\PB(C)\ge [H_n(\uX|\uY,\uhY)-1]/(n r) $.
\item $\bentro(\Pb(C))\ge  H_n(\uX|\uY,\uhY)/n$.
\end{enumerate}
Here $\bentro(x) \equiv -x\log_2 x-(1-x)\log_2(1-x)$ denotes the binary entropy 
function.
\end{lemma}

The rationale for taking the expectation with respect to the code 
in the definition (\ref{EntropyDensity1}) is the following
concentration result
\begin{thm}
Let $H_n(\uX|\uY,\uhY)$ be the relative entropy for a code drawn 
from any of the  ensembles $\GMG$ or $\PCG$ defined in 
Section \ref{DefinitionsSection}, when used to communicate over a 
binary memoryless channel.
Then there exist two ($n$-independent) constants $A,B>0$ 
such that, for any $\ve>0$
\begin{eqnarray}
\Pc \{ |H_n(\uX|\uY,\uhY)-n h_n|>n\ve\} \le A\, e^{-nB\ve^2}\, .
\end{eqnarray}
Here $\Pc$ denotes the probability with respect to the code realization.
\label{TheoremConcentration}
\end{thm} 

In particular this result implies that, if $h_n$ is bounded away from zero
in the $n\to\infty$ limit, then, with high probability with respect to
the code realization, the bit error rate is bounded away from 0.
The converse  (namely that $\lim_{n\to\infty}h_n= 0$ implies
$\lim_{n\to\infty} \Pc[\Pb(C)>\delta] =0$) is in general false.
However, for many cases of interests (in particular for LDPC  ensembles) 
we expect this to be the case.
We refer to Sec. \ref{GeneralizationSection}
for further discussion of this point.
\vspace{0.1cm}

\prooft We use an Azuma inequality~\cite{Azuma} argument similar to the
one adopted by Richardson and Urbanke to prove concentration
under message passing decoding \cite{RichardsonUrbanke}.

Notice that the code-dependent contribution to $H_n(\uX|\uY,\uhY)$
is $\Ey \log_2 \partf(\uy,\uhy)$, cf. Eq. (\ref{EntropyDensity2}). 
We are therefore led to construct a Doob martingale as follows.
First of all fix $m_{\rm max} = (1+\delta)\Eg m$, and
condition to $m\le m_{\rm max}$ ($m$ being the number of right nodes). 
A code $C$ in this
`constrained'  ensemble can be thought as
a sequence of $m_{\rm max}$ random variables $c_1,\dots, c_{m_{\rm max}}$.
The variable $c_t\equiv (k,\underline{i})$ consists in the 
degree $k$ of the $t$-th check node, plus the list 
$\underline{i} =\{ i^a_1\dots i^a_k\}$ of adjacent variable nodes on the 
Tanner graph. We adopt the convention $c_t=*$ if $t>m$.
Let $C_t\equiv (c_1,\dots,c_t)$ and define the random variables
\begin{eqnarray}
X_t = \Ec[\Ey\log_2 \partf(\uy,\uhy)|C_t, m\le m_{\rm max}]\, ,\;\;\;\;\;\;\;\;
t=0,1,\dots,m_{\rm max}\, .
\end{eqnarray}
It is obvious that the sequence $X_0,\dots X_{m_{\rm max}}$ form a martingale.
In particular $\E[X_{t}\, |\, X_0\dots X_{t-1}] = X_{t-1}$.
In order to apply Azuma inequality we need an upper bound on the 
differences $|X_t-X_{t-1}|$. 
Consider two Tanner graphs which differ in a unique check node and let
$\Delta$ be a uniform upper bound on the difference in
$\Ey\log_2 \partf(\uy,\uhy)$ among these two graphs. 
Since $X_{t-1}$ is the expectation of $X_t$ with respect to the
choice of the $t$-th check node, $|X_t-X_{t-1}|\le \Delta$ as well. 
In order to derive such an upper bound $\Delta$, we shall compare two 
graphs differing in a unique check node, with the graph
obtained by erasing this node. 

More precisely, consider a Tanner graph in the  ensemble having 
$m$ check nodes,
and channel output $\uy$, $\uhy$. Now add a single check node,
to be denoted as $0$. Let $\hy_0$ be the corresponding observed value,
and $i^0_1\dots i^0_k$ the positions of the adjacent bits. 
In the LDGM case $\hy_0$ will be drawn from the channel distribution,
while it is fixed to $0$ if the code is a LDPC. Evaluate the
difference of the corresponding partition functions. We claim that
\begin{eqnarray}
\E_{\hy_0}\Ey\log_2 \partf_{m+1}(\uy,\uhy,\hy_0) -\Ey\log \partf_{m}(\uy,\uhy) = 
\E_{\hy_0}\Ey \log_2 \< Q_{\Cnod}(\hy_0|x_{i_1}\oplus\cdots\oplus x_{i_k})\>\, ,
\label{eq:DifferencePartf}
\end{eqnarray}
where $\<\cdot\>$ denotes the expectation with respect to the distribution
(\ref{ProbabilityDistribution}) for the $m$-check nodes code. 
In order to prove such a formula, we write explicitely 
 $\log_2 \< Q_{\Cnod}(\hy_0|x_{i_1}\cdots x_{i_k})\>$
 using Eq.~(\ref{ProbabilityDistribution}). We get
\begin{eqnarray}
\log_2\left\{\frac{1}{\partf_{m}(\uy,\uhy)}\,\sum_{\ux} \prod_{a\in[m]}
\Qc(\hy_a|x_{i^a_1}\cdots x_{i^a_k})
\prod_{i\in [n]} \Qv (y_i|x_i) \cdot 
Q_{\Cnod}(\hy_0|x_{i_1}\cdots x_{i_k})\right\}\, .
\end{eqnarray}
Next we apply the definition of  $\partf_{m+1}(\uy,\uhy,\hy_0)$,
and take expectation with respect to $\uy$, $\uhy$ and $\hy_0$.

Using the definitions (\ref{Apriori}) and (\ref{Aposteriori}) we obtain
\begin{eqnarray}
\< Q_{\Cnod}(\hy_0|x_{i_1}\oplus\cdots\oplus x_{i_k})\> = 
\frac{1}{2}\left[1+\tanh l(y_0)\, \tanh \ell_{\ui}(\uy)\right]\, ,
\end{eqnarray}
with $\ui = (i_0\dots i_k)$.
It follows therefore from Lemma \ref{Tricks} that 
\begin{eqnarray}
-1 \le \E_{\hy_0}\Ey
\log \< Q_{\Cnod}(\hy_0|x_{i_1}\oplus\cdots\oplus x_{i_k})\>\le 0\, .
\label{eq:BoundDiffPartf}
\end{eqnarray}
Therefore the difference in conditional entropy among two Tanner graph 
which differ in a unique check node is at most $2$ bits (one bit for removing
the check node plus one bit for adding it in a different position).
Arguing as above, this yields $|X_{t+1}-X_t|\le 2$ and
Azuma inequality implies
\begin{eqnarray}
\Pc \{ |H_n(\uX|\uY,\uhY)-n h_n|>n\ve\, |\, m\le m_{\rm max}\} 
\le A_1\, e^{-n B_1\ve^2}\, .\label{eq:Concentration1}
\end{eqnarray}
with $A_1 = 2$ and $B_1 = P'(1)/[8\gamma(1+\delta)]$. 

In the case of standard ensembles, we are done, because the condition
$m\le m_{\rm max} = (1+\delta)\Eg m$ holds surely ($m$ is a deterministic
quantity). For Poisson and multi-Poisson ensembles, we have still to show that
the event $m>m_{\rm max}$ does not modify significantly the estimate
(\ref{eq:Concentration1}). 
From elementary theory of Poisson random variables we know that, 
for any $\delta >0$ there exist $A_2,B_2>0$ such that 
$\Pc [m> m_{\rm max}]\le A_2 \, e^{-n B_2\delta^2}$. Notice that 
\begin{eqnarray}
\Pc \{ |H_n(\uX|\uY,\uhY)-n h_n|>n\ve\} &\le & 
\Pc \{ |H_n(\uX|\uY,\uhY)-n h_n|>n\ve\, |\, m\le m_{\rm max}\} +\\ 
&&+\Pc [m> m_{\rm max}]\le\nonumber\\
&\le& A_1\, e^{-n B_1\ve^2}+A_2 \, e^{-n B_2\delta^2}\, .
\end{eqnarray}
The thesis is obtained by choosing $\delta=\ve$, $B=\min(B_1,B_2)$,
and $A = A_1+A_2$.
\endproof
%
%
\section{Main result and proof for Poisson ensembles}
\label{MainSection}

As briefly mentioned in the Introduction, the main ideas in the proof
are most clearly described in the context of simple Poisson ensembles.
We shall therefore discuss them in detail here, will be more succinct 
when using similar arguments for multi-Poisson ensembles.
Some of the  calculations and technical  details are deferred to 
Appendices \ref{StraightforwardApp} and \ref{PositivityApp}.

In order to state our main result in compact form, it is useful to introduce
two infinite families $\{ U_A\}$, $\{ V_B \}$ of i.i.d. random 
variables. The indices $A$, and $B$ run over whatever set including the cases 
occurring in the paper. We adopt the convention that any two variables 
of these families carrying distinct subscripts are independent. 
The distribution of the $U$ and $V$ variables shall moreover satisfy 
a couple of requirements specified in the definition below. 
\begin{df}
Fix a degree sequence pair $(\Lambda,\Rho)$, and let
$\rho_k$ be the edge perspective right degree distribution
$\rho_k = \Rho_k/\Rho'(1)$.
Let $V_1, V_2, \dots\ed V$ be a family of i.i.d. symmetric random variables, 
$k$ an integer with distribution $\rho_{k}$, $J$ a symmetric random variable 
distributed as the log-likelihoods $\{J_a\}$ of the parity check bits, 
cf. Eq. (\ref{LogLikelihoodDef}), and
\begin{eqnarray}
U^V \equiv \atanh\left[\tanh J\prod_{i=1}^{k-1}\tanh V_i\right]\, .
\label{CnodeFunction}
\end{eqnarray}
The random variables $U$, $V$ are said to be admissible if they
are independent, symmetric and $U\ed U^V$.

For any couple of admissible random variables $U$, $V$, we define
the associated trial entropy as follows
\begin{eqnarray}
\phi_V(\Lambda,\Rho) & \equiv & -\Lambda'(1)\,\E_{u,v}\l2\left[ \sum_x \Bin_u(x)\Bin_v(x) \right]+
\E_l\E_y\E_{\{u_i\} }\l2\left[ \sum_x \frac{Q_{\Vnod}(y|x)}{Q_{\Vnod}(y|0)}
\prod_{i=1}^l \Bin_{u_i}(x)\right]+
\nonumber\\
&&+\frac{\Lambda'(1)}{P'(1)}\,\E_k\E_{\hy} \E_{\{v_i\} }
\l2\left[\sum_{x_1\dots x_k}
\frac{Q_{\Cnod}(\hy|x_1\oplus \cdots\oplus x_k)}{Q_{\Cnod}(\hy|0)}\prod_{i=1}^k \Bin_{v_i}(x_i)\right]\,  ,\label{FreeEnergy}
\end{eqnarray}
where $l$ and $k$ are two integer random variables with distribution 
(respectively) $\Lambda_l$ and $\Rho_k$. Hereafter we shall drop 
the reference to the degree distributions in $\phi_V(\Lambda,\Rho)$
whenever this is clear from the context.
\end{df}
Notice that, in the notation for the trial entropy, we put in evidence 
its dependence just on the distribution of the $V$-variables. In fact we 
shall think of the distribution of the $U$-variables
as being determined by the relation $U\ed U^V$, see Eq. (\ref{CnodeFunction}).

The main result of this paper is stated below.
\begin{thm}
\label{MainTheorem}
Let $\Rho(x)$ be a polynomial with non-negative coefficients such
that $P(0) = P'(0) = 0$, and assume that $P(x)$ is convex for 
$x\in[-x_0,x_0]$.
\begin{enumerate}
\item Let $h_n$ be the expected conditional entropy per bit for either
of the Poisson ensembles $\GMP$ or  $\PCP$. If $x_0\ge 1$, then
\begin{eqnarray}
h_n\ge\sup_{V}\phi_V(\Lambda,\Rho)\, .\label{TheoremPoisson}
\end{eqnarray}
\item Let $h_n$ be the expected conditional entropy per bit for either
of the multi-Poisson ensembles $\GMM$ or  $\PCM$. If $x_0>e$, then
\begin{eqnarray}
\lim\inf_{n\to\infty}h_n\ge\sup_{V}\phi_V(\Lambda^{(\gamma)},\Rho)\, .\label{eq:MultiMain}
\end{eqnarray}
\item Let $h_n$ be the expected conditional entropy per bit for either
of the standard ensembles $\GMS$ or  $\PCS$. If $x_0>e$, then
\begin{eqnarray}
\lim\inf_{n\to\infty}h_n\ge\sup_{V}\phi_V(\Lambda,\Rho)\, .\label{eq:StdMain}
\end{eqnarray}
\end{enumerate}
Here the $\sup$ has to be taken over the space of admissible random variables.
\end{thm} 
\begin{figure}
\centerline{\epsfig{figure=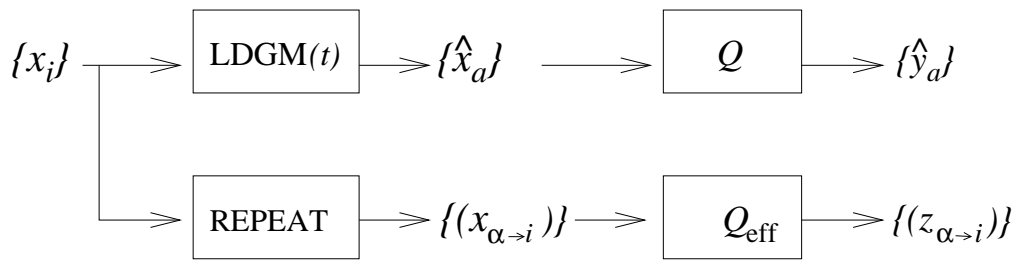,angle=0,width=0.75\linewidth}}
\caption{\small A communication scheme interpolating between an LDGM 
code and an irregular repetition code. Notice that the repetition
part is transmitted through a different (effective) channel.}
\label{EffChFig}
\end{figure}
\hspace{0.5cm}{\bf Proof [Poisson ensembles]:}\hspace{0.1cm}
Computing the conditional entropy (\ref{EntropyDensity2}) is difficult 
because the probability distribution (\ref{ProbabilityDistribution})
does not factorize over the bits $\{x_i\}$.
Guerra's idea \cite{Guerra}
consists in interpolating between the original distribution 
(\ref{ProbabilityDistribution}) and a much simpler one which factorizes 
over the bits.
In the LDPC case, this corresponds to progressively removing parity check 
conditions (\ref{ParityChecks}), from the code definition. 
For LDGM's, it amounts to removing bits from the codewords 
(\ref{EncodingLDGM}). In both cases the design rate is augmented.
In order to compensate the increase in conditional entropy 
implied by this transformation we imagine to re-transmit some of the
bits $\{x_i\}$ through an `effective' channel.  It turns out that the
transition probability of this effective channel can be chosen in such a
way to `match' the transformation of the original code.

In practice, for any $s\in [0,\gamma]$ we define the interpolating system
as follows. We construct a code from the same  ensemble family
as the original one, but with modified parameters $(n,s,\Rho)$. 
Notice that both the block length and the right degree distribution 
remained unchanged.
The design rate, on the other hand has been increased to 
$\rdes(s) = \Rho'(1)/s$ (in the LDGM case) or $\rdes(t) = 1-s/\Rho'(1)$
(in the LDPC case).
A codeword  from this code is transmitted through the original channel,
with transition matrix $Q(\,\cdot\,|\,\cdot\,)$.
It is useful to denote by $\Cnod_s$ the set of factor nodes  
for a given value of $t$. Of course $\Cnod_s$ is a random variable.

As anticipated, we must compensate for the rate loss 
in the above procedure. We therefore repeat each bit $x_i$ 
$l_i$ times and transmit it through an auxiliary channel with  
transition probability $\Qeff(\,\cdot\,|\,\cdot\,)$. 
The $l_i$'s are taken to be 
independent Poisson random variables with parameter $(\gamma-s)$.
We can therefore think this construction as a code formed of two
parts: a code from the ${\rm LDGM}(n,s,\Rho)$ or ${\rm LDPC}(n,s,\Rho)$
{\it ensemble}, plus an irregular repetition code.
Each part of the code is transmitted through a different channel.
In Fig.~\ref{EffChFig} we present a schematic description of this two--parts 
coding system (the scheme refers to the LDGM case).

\begin{figure}
\centerline{\epsfig{figure=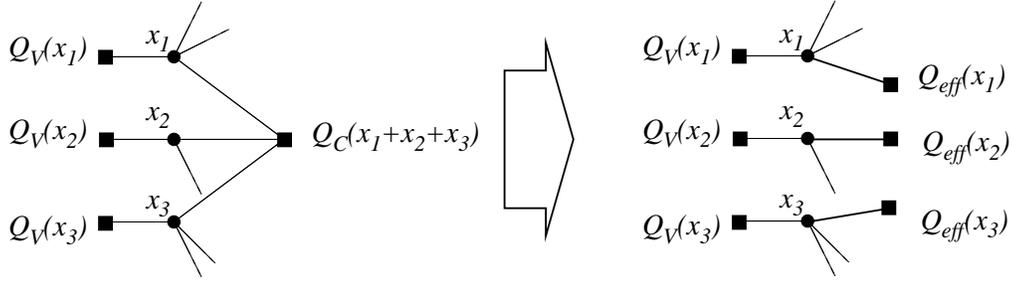,angle=0,width=0.75\linewidth}}
\caption{\small Action of the interpolation procedure on the factor graph
representing the probability distribution of the channel input (or, in 
LDGM case, the information message) conditional to the channel output.
For the sake of simplicity we dropped the arguments $y_i$ (for $\Qv$),
$\hy_a$ (for $\Qc$) and $z_{\alpha\to i}$ (for $\Qeff$).}
\label{InterGraph}
\end{figure}
The received message will have the form $(\uy,\uhy,\uz)$, with
$\uz = \{ z_{\alpha_i\to i}\}$, $\alpha_i\in [l_i]$. 
We can write the conditional probability 
for $\ux = (x_1\dots x_n)$ to be the transmitted codeword conditional to 
the output $(\uy,\uhy,\uz)$ of the channel as follows:
\begin{eqnarray}
P(\ux|\uy,\uhy,\uz) = \frac{1}{\partf_s}\, \prod_{a\in\Cnod_s}
\Qc(\hy_a|x_{i^a_1}\oplus\cdots\oplus x_{i^a_k})\,
\prod_{i\in\Vnod} \Qv (y_i|x_i)\,\prod_{i\in\Vnod}
\prod_{\alpha_i=1}^{l_i} \Qeff (z_{\alpha_i\to i}|x_i)\, ,
\label{InterProbabilityDistribution}
\end{eqnarray}
with $\partf_s=\partf(\uy,\uhy,\uz)$ fixed by the normalization condition
of $P(\ux|\uy,\uhy,\uz)$. Notice that, for $s=\gamma$ the original
distribution (\ref{ProbabilityDistribution}) is recovered since $l_i=0$
for any $i\in \Vnod$. On the other hand, if $s=0$, then  $m=0$
and the Tanner graph contains no check nodes: $\Cnod_0 = \emptyset$.
We are left with a simple irregular repetition code. The action
of the interpolation procedure on the factor graph is depicted in 
Fig.~\ref{InterGraph}.

The following steps in the proof will depend upon the effective channel
transition probability $\Qeff(\,\cdot\,|\,\cdot\,)$ only through its 
log-likelihood distribution. We therefore define the random variable $U$ as
\begin{eqnarray}
U\ed \frac{1}{2}\log \frac{\Qeff(Z|0)}{\Qeff(Z|1)}\, .\label{UChannelRelation}
\end{eqnarray}
Where $Z$ is distributed according to $\Qeff(z|0)$.
Notice that $U$ is symmetric and that, for any symmetric $U$ we can 
construct at least one BIOS channel whose log-likelihood is
distributed as $U$ \cite{RichardsonUrbanke}. Hereafter we shall consider
$\Qeff(\,\cdot\,|\,\cdot\,)$ to be determined  by such a construction.
When necessary, we shall adopt a discrete notation for the
output alphabet of the effective channel $\Qeff(\,\cdot\,|\,\cdot\,)$. 
However, our derivation is equally valid for a continuous output
alphabet.

It is natural to consider the conditional entropy-per-bit of the 
interpolating model. With an abuse of notation we denote by $\Ec$ 
the expectation both with respect to the code  ensemble and 
with respect to the $l_i$'s. We define
\begin{eqnarray}
h_{n}(s) & \equiv & \frac{1}{n}\, \Ec H_n(\uX|\uY,\uhY,\uZ) = 
\label{EntropyInterpolation}\\
&=&\frac{1}{n}\,\Ec\Eyz\log_2 \partf(\uy,\uhy,\uz)-\frac{s}{\Rho'(1)}
\sum_{y}\Qc(y|0)\log_2 \Qc(y|0)-
\nonumber\\
&&-\sum_{y}\Qv(y|0)\log_2 \Qv(y|0)-(\gamma-s)\sum_z \Qeff(z|0)\log_2 \Qeff(z|0)\, .
\label{EntropyInterpolation2}
\end{eqnarray}
Notice that $h_n(s)$ depends implicitly upon the 
distribution of $U$. In passing from Eq. (\ref{EntropyInterpolation})
to Eq. (\ref{EntropyInterpolation2}) we used  the symmetry 
condition for the two channels and assumed the transmitted message to be
the all-zero codeword.

It is easy to compute the conditional entropy (\ref{EntropyInterpolation})
at the extremes of the interpolation interval. For $s = \gamma$ we recover
the original probability distribution (\ref{ProbabilityDistribution})
and the conditional entropy $h_n$.  When $s=0$, on the other hand,
the factor graph no longer contains function nodes of degree larger than
one and the partition function can be computed explicitly. Therefore we have
\begin{eqnarray}
h_n(\gamma) & = & h_n\, \\
h_n(0) & = & \Eyz \E_l \log_2\left[\sum_x\frac{\Qv(y|x)}{\Qv(y|0)}
\prod_{\alpha=1}^l\Qeff(z_{\alpha}|x)\right]-
\gamma\sum_z\Qeff(z|0)\log_2 \Qeff(z|0)\, .\phantom{AA}
\end{eqnarray}
where $l$ is a poissonian variable  with parameter $\gamma$.
As anticipated, $h_n(0)$ can be expressed uniquely in terms of the
distribution of the $U$ variables. Using Eq. (\ref{UChannelRelation}), we get
\begin{eqnarray}
h_n(0) & = & 
\E_y\E_l\E_{\{u_i\} }\log_2\left[ \sum_x \frac{Q_{\Vnod}(y|x)}{Q_{\Vnod}(y|0)}
\prod_{\alpha=1}^l P_{u_i}(x)\right]\, + \, \gamma\, \E_u \log_2 (1+e^{-2u})\, .
\label{eq:InitialEntropy}
\end{eqnarray}
The next step is also very natural. Since we are interested in estimating 
$h_n$, we write
\begin{eqnarray}
h_n = h_n(0)+\int_0^\gamma \frac{dh_n}{ds}(s) \;\; ds \, .\label{Integral}
\end{eqnarray}
A straightforward calculation yields:
\begin{eqnarray}
\frac{dh_n}{ds}(s) &=& \sum_k \frac{\Rho_k}{\Rho'(1)}\, \frac{1}{n^k}\,
\sum_{i_1\dots i_k} \E_{\hy}\Et \log_2\left\<
\frac{\Qc(\hy|x_{i_1}\oplus\cdots\oplus x_{i_k})}
{\Qc(y|0)}\right\>_s-\nonumber\\
&&-\frac{1}{n}\sum_{i\in{\cal V}}
\E_z\Et\log_2\left\< \frac{\Qeff(z|x_i)}{\Qeff(z|0)}\right\>_s\, ,
\label{EntropyDerivative}
\end{eqnarray}
where the average $\<\cdot\>_s$ is taken with respect to the 
interpolating probability distribution (\ref{InterProbabilityDistribution}).
The details of this computation are  reported in 
App.~\ref{StraightforwardApp}. Of course the right-hand side of 
Eq. (\ref{EntropyDerivative}) is still quite hard to evaluate
because the averages $\<\cdot\>_s$ are complicated objects.
We shall therefore approximate them by much simpler averages
under which the $x_i$'s are independent and have log-likelihoods
which are distributed as $V$--variables. More precisely we define
\begin{eqnarray}
\frac{d\h_n}{dt}(s) & = & \sum_k \frac{\Rho_k}{\Rho'(1)}\, \frac{1}{n^k}\,
\sum_{i_1\dots i_k} \E_{\hy}\E_v \log_2\left\{\sum_{x_{i_1}\dots x_{i_k}}
\frac{\Qc(\hy|x_{i_1}\oplus\cdots\oplus x_{i_k})}
{\Qc(y|0)}\prod_{j=1}^kP_{v_j}(x_{i_j})\right\}-\nonumber\\
&&-\frac{1}{n}\sum_{i\in{\cal V}}
\E_z\E_v\log_2\left\{\sum_{x_i} \frac{\Qeff(z|x_i)}{\Qeff(z|0)} P_v(x_i)
\right\}\, .
\label{EntropyDerivativeApprox}
\end{eqnarray}
Notice that in fact this expression does not depend  upon $s$.
Summing and subtracting it to Eq. (\ref{Integral}),
and after a few algebraic manipulations, we obtain
\begin{eqnarray}
h_n = \phi_V + \int_0^{\gamma} \left[\frac{dh_n}{dt}(s)-
\frac{d\h_n}{dt}(s)\right] ds \;\;
\equiv \;\; \phi_V + \int_0^{\gamma} R_n(s)
\;\; ds\, .\label{RemainderDef}
\end{eqnarray}
The proof is completed by showing that $R_n(s)\ge 0$ for any 
$s\in[0,\gamma]$. This calculation is reported in App.~\ref{PositivityApp}.
\endproof
%
%
\section{Proof for multi-Poisson ensembles}
\label{sec:MultiProof}

The proof for multi-Poisson ensembles follows the same strategy as for 
Poisson ensembles. The only difference is that the interpolating system is 
obviously more complex.

Given the ensemble parameters $(n,\gamma,\Lambda,\Rho)$, and defining as 
in Sec.~\ref{MultiPoissonDefinitionSection},
$t_{\rm max}\equiv\lfloor\Lambda'(1)/\gamma\rfloor-1$, we introduce an 
interpolating ensemble for each pair $(t_*,s)$, with 
$t_*\in\{0,\dots, t_{\rm max}-1\}$ and $s\in[0,\gamma]$ as follows.
The first $t_*$ rounds (i.e. $t=0,\dots,t_*-1$ in Definition
\ref{def:MultiPoisson}) in the graph construction are the same as for the 
original multi-Poisson graph. 

Next, during round $t_*$, 
$m^{(t_*)}_k$ is drawn from a Poisson distribution 
with  mean $ns\Rho_k/\Rho'(1)$ (instead of $n\gamma\Rho_k/\Rho'(1)$), and 
$m^{(t_*)}_k$ right (check)
nodes of degree $k$ are added for each $k=2,\dots,k_{\rm max}$.
The neighbors of each of this check node are i.i.d. random variables
in $\Vnod$ with distribution 
$w_i(t) = [\soc_i(t_*)]_+/(\sum_{j}[\soc_j(t_*)]_+)$.
In order to compensate for the smaller number of right
nodes, an integer $l_i(t_*)$ is drawn for each $i\in\Vnod$ from 
a Poisson distribution with mean $n(\gamma-s)w_i(t)$. 
As in the previous Section, this means that $l_i(t_*)$ repetitions
of the bit $x_i$ will be transmitted through an effective channel
with transition probability $\Qeff(\,\cdot\,|\,\cdot\,)$. Finally,
the number of free sockets is updated by setting 
$\soc_i(t_*+1) = \soc_i(t_*)-\Delta_i(t_*)-l_i(t_*)$, where $\Delta_i(t_*)$
is the number of times the node $i$ has been chosen as neighbor of a
right node in this round.

During rounds $t=t_*+1,\dots, t_{\rm max}-1$, no right node is 
added to the factor graph. On the other hand, for each $i\in\Vnod$ a
random integer $l_i(t)$ is drawn from a Poisson distribution with 
mean $n\gamma w_i(t_*)$. This means that the bit $x_i$ will be further
retransmitted $l_i$ times through the effective channel.
Furthermore the number of free sockets 
is updated at the end of each round by setting $\soc_i(t+1) = 
\soc_i(t)-l_i(t)$.

This completes the definition of the Tanner graph ensemble. 
We denote by $\Cnod_{(t_*,s)}$ the set of function nodes and by 
$l_i=\sum_{t=t_*}^{t_{\rm max}-1}l_i(t)$ the total number of times the bit
$x_i$ is transmitted through the effective channel. The overall conditional
distribution of the channel input given the output has the 
same form (\ref{InterProbabilityDistribution}) as for the simple Poisson 
ensemble.

The overall number of function nodes in the $(t_*,s)$ ensemble
is easily seen to be a Poisson variable with mean $n(\gamma t_*+s)/\Rho'(1)$.
In fact we add Poisson$(n\gamma/\Rho'(1))$ function nodes during each of 
the first $t_*$ rounds, and  Poisson$(ns/\Rho'(1))$
during the $(t_*+1)$-th round. Analogously, the overall number of
repetitions  $\sum_i l_i$ is a Poisson variable with mean 
$n[\gamma(t_{\rm max}-t_*-1)+\gamma-s]$. Using these remarks, we get the 
expression
\begin{eqnarray}
h_{n}(t_*,s) & = & \frac{1}{n}\, \Ec H_n(\uX|\uY,\uhY,\uZ) = 
\label{EntropyInterpolationMP}\\
&=&\frac{1}{n}\,\Ec\Eyz\log_2 \partf(\uy,\uhy,\uz)-\frac{1}{\Rho'(1)}
(\gamma t_*+s)
\sum_{y}\Qc(y|0)\log_2 \Qc(y|0)-\label{EntropyInterpolation2MP}\\
&&-\sum_{y}\Qv(y|0)\log_2 \Qv(y|0)-(\gamma(t_{\rm max}\!-\!t_*)-s)
\sum_z \Qeff(z|0)\log_2 \Qeff(z|0)\, .\nonumber
\end{eqnarray}

Let us now look at a few particular cases of the interpolating system.
For $(t_*=t_{\rm max}-1,s=\gamma)$ we recover the original multi-Poisson 
ensemble. Moreover, for any $t_*\in\{0,\dots,t_{\rm max}-2\}$
the ensembles $(t_*,s=\gamma)$ and  $(t_*+1,s=0)$ are identically distributed.
Finally, when $(t_*=0,s=0)$ the set of factor nodes is empty with probability
one, and the resulting coding scheme is just an irregular repetition 
code (bit $x_i$ being repeated $l_i$ times) used over the channel
$\Qeff(\,\cdot\,|\,\cdot\,)$. If we denote by $h_n(t_*,s)$ the expected
entropy per bit in the interpolating ensemble, we get 
\begin{eqnarray}
h_n(t_{\rm max}\!-\!1,\gamma) & = & h_n\, \\
h_n(0,0) & = & \E_{y,z}\E_l\, \log_2\left[\sum_x\frac{\Qv(y|x)}{\Qv(y|0)}
\prod_{\alpha=1}^l\Qeff(z_{\alpha}|x)\right]-
\gamma t_{\rm max}\sum_z\Qeff(z|0)\log_2 \Qeff(z|0)\, .\, \nonumber\\
\end{eqnarray}
Here the expectations on $y$, $\{z_{\alpha}\}$ are taken with respect to 
the distributions $\Qv(y|0)$, $\Qeff(z|0)$, while $l$ is distributed 
according to the expected degree profile $\Lambda^{(n,\gamma)}_l$.
Moreover, we used the fact that $\sum_l \Lambda^{(n,\gamma)}_l\, l=
n\gamma t_{\rm max}$. Finally, as in the previous Section
(cf. Eq.~(\ref{eq:InitialEntropy})) the conditional entropy
$h_n(0,0)$ depends upon the effective channel transition probability 
only through the distribution of the log-likelihood $U$.

The next step consist in writing
\begin{eqnarray}
h_n(t_*,\gamma) = h_n(t_*,0) +\int_0^{\gamma}\!\frac{dh_n}{ds}(t_*,s) \;\;
ds\, ,
\end{eqnarray}
for $t_*\in\{0,\dots,t_{\rm max}-1\}$. Using the fact that 
$h_n(t_*,\gamma) =h_n(t_*+1,0)$, this implies
\begin{eqnarray}
h_n = h_n(0,0)+\sum_{t_*=0}^{t_{\rm max}-1}\int_0^{\gamma}\!
\frac{dh_n}{ds}(t_*,s) \;\; ds\, .\label{eq:IntegralMP}
\end{eqnarray}
The derivative with respect to the interpolation parameter
is similar the simple Poisson case:
\begin{eqnarray}
\frac{dh_n}{ds}(t_*,s) &=& \sum_k \frac{\Rho_k}{\Rho'(1)}\, \,
\sum_{i_1\dots i_k} \E_{\hy}
\E^{\{i_1\dots i_k\}} w_{i_1}(t_*)\cdots w_{i_k}(t_*)\log_2\left\<
\frac{\Qc(\hy|x_{i_1}\oplus\cdots\oplus x_{i_k})}
{\Qc(y|0)}\right\>_{(t_*,s)}-\nonumber\\
&&-\sum_{i\in{\cal V}}
\E_z\E^{\{i\}}w_{i}(t_*)
\log_2\left\< \frac{\Qeff(z|x_i)}{\Qeff(z|0)}\right\>_{(t_*,s)}
\, + \, \varphi(n;t_*,s)\, ,
\label{EntropyDerivativeMP}
\end{eqnarray}
where $\<\,\cdot\,\>_{(t_*,s)}$ denotes expectation with respect to the 
conditional distribution (\ref{InterProbabilityDistribution}) 
appropriate for the $(t_*,s)$ ensemble and the expectations
$\E^{\{i_1\dots i_k\}}$ are defined in App.~\ref{app:multi-Derivative}.
Moreover, the term $\varphi(n;t_*,s)$ has the following properties.
\begin{enumerate}
\item[A.] $|\varphi(n;t_*,s)|\le C_1$ for some constant $C_1$ which depends 
uniquely upon the ensemble parameters $\Lambda$, $\Rho$ and $\gamma$.
\item[B.] $\varphi(n;t_*,s)\le C_2(t_*,s)\sqrt{(\log n)^3/n}$ for some function
$C_2(t_*,s)$ which does not depend upon $n$.
\end{enumerate}
We  refer to App.~\ref{app:multi-Derivative} for the details 
of this computation.
Notice that the equivalent  expression for Poisson codes,
cf. Eq.~(\ref{EntropyDerivative}), is recovered by setting $w_i(t_*) = 1/n$
and dropping the correction $\varphi(\cdot)$.

Finally, we introduce an `approximation' $\frac{d\h_n}{ds}(t_*,s)$ to
Eq.~(\ref{EntropyDerivativeMP}) analogous to 
Eq.~(\ref{EntropyDerivativeApprox}). More precisely, we replace
the expectations $\<\,\cdot\,\>_{(t_*,s)}$ with expectations
over product measures of the form $\prod_{i}P_{v_i}(x_{i})$,  the
ensemble averages $\E^{\{i_1\dots i_k\}}$ with averages over i.i.d. $V_i$'s,
and we drop the remainder $\varphi(\cdot)$.
Using Eq.~(\ref{eq:IntegralMP}) and after rearranging various terms
(the relation $\Lambda^{(\gamma,n)}{}'(1) = \gamma t_{\rm max}$ is useful here)
we end up with
\begin{eqnarray}
h_n& = &\phi_V(\Lambda^{(\gamma,n)},\Rho) + \sum_{t_*=0}^{t_{\rm max}-1}
\int_0^{\gamma} \left[\frac{dh_n}{dt}(t_*;s)-
\frac{d\h_n}{dt}(t_*;s)\right] ds =\nonumber\\
&=& \;\; \phi_V(\Lambda^{(\gamma,n)},\Rho)  + \sum_{t_*=0}^{t_{\rm max}-1}\int_0^{\gamma} R_n(t_*,s)
\;\; ds\;\;+o_n(1)\, ,
\label{RemainderDefMP}
\end{eqnarray}
where the second inequality follows by applying  the dominated convergence
theorem to $\varphi(\cdot)$.
The proof is finished by showing that $R_n(t_*,s)$ is non-negative
for any $t_*\in\{0,\dots,t_{\rm max}-1\}$ and $s\in[0,\gamma]$.
this calculation is very similar to the simple Poisson case,
and is discussed in App.~\ref{MultiPositivityApp}.\endproof
%
%
\section{Proof for standard ensembles}
\label{StandardProofSection}

We proved points 1 and 2 of Theorem \ref{MainTheorem} directly. We 
will now show that 2 implies the lower bound for standard ensembles
(point 3) which is the practically more relevant case.

The idea is that the standard ensemble $(n,\Lambda,\Rho)$ is indeed
`very close' to the multi-Poisson ensemble $(n,\gamma,\Lambda,\Rho)$ 
for small $\gamma$. In order to implement this idea, we state a
preliminary result here.
\begin{lemma}\label{RewiringLemma}
Let $C_1$ and $C_2$ be two codes with the same blocklength
$n$ from any of the ensembles defined in Section \ref{DefinitionsSection} 
(the ensemble does not need to be the same). Assume they are used to 
communicate through the same noisy channel and let
$H_1(X|Y_1)$, $H_2(X|Y_2)$ be the corresponding conditional entropies. 
If  $C_1$ can be obtained from $C_2$ with $n_{\rm r}$ rewirings,
then $|H_1(X|Y_1)-H_2(X|Y_2)|\le  n_{\rm r}$.
\end{lemma}
\prooft Recall that a rewiring was defined as either the removal
or the addition of a function node to the Tanner graph. 
We already proved (cf. Eqs.~(\ref{eq:DifferencePartf}) to 
(\ref{eq:BoundDiffPartf}) and relative discussion) that the introduction
of a single function node induces a change in the conditional entropy
which is smaller than 1 bit.  
\endproof

Let now $0<\gamma<1$, and
consider a pair of Tanner graphs $(\graph_{\rm s},\graph_{\rm mP})$
distributed according to the coupling in Lemma \ref{MultiLemma}.
In particular, the marginal distribution of 
$\graph_{\rm s}\ed (n,\Lambda,\Rho)$ and $\graph_{\rm mP}\ed
(n,\Lambda,\Rho,\gamma)$. With an abuse of notation, denote by 
$h_n(\graph_{\rm s})$ and $h_n(\graph_{\rm mP})$ the corresponding 
conditional entropy {\em per bit} and by $h_n$, $h_n^{(\gamma)}$ their 
ensemble averages. From Lemmas \ref{MultiLemma} and \ref{RewiringLemma} 
it follows that
\begin{eqnarray}
\lim_{n\to\infty}
\Prob[|h_n(\graph_{\rm mP})-h_n(\graph_{\rm s})|\ge A\gamma^b] = 0\, ,
\label{eq:LimitProbEntroDiff}
\end{eqnarray}
where the constants $A$ and $b$ are as in Lemma \ref{MultiLemma}.
Therefore
\begin{eqnarray}
|h^{(\gamma)}_n-h_n| = |\E[h_n(\graph_{\rm mP})-h_n(\graph_{\rm s})]|\le
\E | h_n(\graph_{\rm mP})-h_n(\graph_{\rm s}) |\le A\gamma^b +o_n(1)\, ,
\end{eqnarray}
where the last inequality follows from Eq.~(\ref{eq:LimitProbEntroDiff})
together with the remark that $h_n(\graph)\le 1$ for any code.
By taking the large blocklength limit, we get
\begin{eqnarray}
\lim\inf_{n\to\infty}h_n\ge \lim\inf_{n\to\infty}h_n^{(\gamma)}-A\gamma^b
\ge \phi_V(\Lambda^{(\gamma)},\Rho)-A\gamma^b\, ,
\end{eqnarray}
where the last inequality follows from our lower bound on multi-Poisson 
ensembles, Eq.~(\ref{eq:MultiMain}). Next we notice that 
$\phi_V(\Lambda,\Rho)$ is a continuous function of $\Lambda$
(with respect to the total variation distance, see App.~\ref{CouplingApp} 
for a definition), 
once the degree distribution $\Rho$, and the $V$--variables distribution 
have been fixed. Moreover by Corollary \ref{coro:Degree} 
$\lim_{\gamma\to 0}\Lambda^{(\gamma)} = \Lambda$ in total variation distance 
sense. 
Therefore we can obtain the 
thesis (\ref{eq:StdMain}) by taking the $\gamma\to 0$ limit in the 
last expression.\endproof
%
%
\section{Examples and applications}
\label{ExampleSection}

The optimization problem in Eq. (\ref{TheoremPoisson}) is, in general,
rather difficult. Nevertheless, one can easily obtain sub-optimal bounds
on the entropy $h_n$, by cleverly chosing the distribution of the 
$V$--variables to be used in Eq. (\ref{FreeEnergy}).
Moreover, bounds can be optimized through density evolution.
Although a complete discussion of the optimization problem is beyond the 
scope of this paper, a rather simple approach, cf. Sec.~\ref{sec:BoundOpt}
and Tab.~\ref{TableLDPCBSC},
already gives very good results (indeed we believe them to be optimal).

Our main focus will be here on standard $(n,\Lambda,\Rho)$ ensembles.
As in our original definition, cf. Eq.~(\ref{eq:SequenceDefinition}),
we shall generically consider the case in which $\Lambda_0=\Lambda_1=0$ 
(no degree  0 or degree 1 variable nodes).
However, most of the arguments can be adapted to Poisson
ensembles too. On the other hand, we shall {\em always} assume
$\Rho_0=\Rho_1= 0$ (no degree  0 or degree 1 check nodes).

Throughout this Section, we shall use the notation $h=\lim\inf_{n\to\infty}
h_n$ for the asymptotic conditional entropy per symbol.
%
%
\subsection{Shannon threshold}
\label{ShannonSection}
Assume  $V=0$ with probability 1,
and therefore, from Eq. (\ref{CnodeFunction}),
$U=0$ with probability 1. Plugging these distributions
into Eq. (\ref{FreeEnergy}) we get
\begin{eqnarray}
\phi_V = -\frac{\Lambda'(1)}{P'(1)}+\,\E_y
\,\l2\left[\sum_x\frac{Q_{\Vnod}(y|x)}{Q_{\Vnod}(y|0)}\right]
+\frac{\Lambda'(1)}{P'(1)}\,\E_{\hy}\,\l2\left[
\sum_x\frac{Q_{\Cnod}(\hy|x)}{Q_{\Cnod}(\hy|0)}\right]\, .
\end{eqnarray}
Using Theorem \ref{MainTheorem}, and after a few manipulations, we get
\begin{eqnarray}
h& \ge & 1-\frac{\Cap(Q)}{\rdes},\;\;\;\;\;\;\;\;\;\;\mbox{for the $\GMS$ 
  ensemble} ,\\
h& \ge & \rdes-\Cap(Q),\;\;\;\;\;\;\;\mbox{for the $\PCS$ 
ensemble} .\label{LDPCCapacity}
\end{eqnarray}
Where we denoted $\Cap(Q)$ the capacity of the BIOS channel with transition 
probability $Q(y|x)$:
\begin{eqnarray}
\Cap(Q) = 1-\sum_{y\in{\cal A}}Q(y|0)\l2\left[\sum_x\frac{Q(y|x)}
{Q(y|0)}\right]\, .
\end{eqnarray}
In other words reliable communication (which requires $h_n\to 0$ as 
$n\to\infty$) can be achieved only if the design rate is smaller than
channel capacity. For the LDGM ensemble this statement is
equivalent to the converse of channel coding theorem, because $\rdes$
is concentrated around the actual rate. 
This is the case also for standard LDPC ensembles with no degree 0 or 
degree 1 variable nodes~\cite{DiRichardsonUrbanke}.

For general LDPC ensembles, Eq.~(\ref{LDPCCapacity}) is slightly
weaker than the channel coding theorem because the actual rate can
be larger than the design rate. However, as shown in the next Sections,
the bound can be easily improved changing the 
distribution of $V$.

Of course this results could have been derived from information-theoretic
arguments. However it is nice to see that it is indeed contained in
Theorem \ref{MainTheorem}.
%
%
%
%
\subsection{Non-negativity of the entropy}
Let us consider the opposite limit: $V=+\infty$  
with probability 1, and distinguish two cases:

\underline{$\PCS$  ensemble}. 
As a consequence of Eq. (\ref{CnodeFunction}),
also $U=+\infty$ with probability 1. Using  Theorem \ref{MainTheorem}
and Eq.~(\ref{FreeEnergy}), it is easy to obtain
\begin{eqnarray}
h\ge \Lambda_0\, H_Q(X|Y)\, ,\label{PositiveLDPC}
\end{eqnarray}
where $H_Q(X|Y) = 1-\Cap(Q)$ is the relative entropy for a single bit
transmitted across the channel $Q(y|x)$. The interpretation of 
Eq. (\ref{PositiveLDPC}) is straightforward. Typically, a fraction 
$ \Lambda_0$ of the variable nodes have degree zero. The relative
entropy (\ref{EntropyDensity1}) is lower-bounded by the entropy of
these variables.

\underline{$\GMS$  ensemble}. Equation (\ref{CnodeFunction}) implies
that $U$ is distributed as the log-likelihoods $J_a=(1/2)\log Q(y|0)/Q(y|1)$,
see Eq.~(\ref{LogLikelihoodDef}). It is easy then to evaluate the bound:
\begin{eqnarray}
h\ge \E_l\E_y\, \l2\left[\sum_x\prod_{i=1}^l\frac{Q(y_i|x)}
{\sum_{x'}Q(y_i|x')}\right]\, .\label{PositiveLDGM}
\end{eqnarray}
The meaning of this inequality is, once again, quite clear:
\begin{eqnarray}
H_n(\uX|\uhY)\ge \sum_{i=1}^n H_n(X_i|\uX^{(i)},\uhY)\ge
\sum_{i=1}^n H_n(X_i|\{x_j\!=\!0\,\,\forall j\neq i\};\uhY)\, ,
\end{eqnarray}
where we introduced $\uX^{(i)}\equiv \{X_j\}_{j\neq i}$. The above inequalities
are consequences of the entropy chain rule and of the fact that 
conditioning reduces entropy.
Taking the expectation with respect to the code  ensemble and letting 
$n\to\infty$ yields (\ref{PositiveLDGM}).
%

%
%
%
\subsection{Binary Erasure Channel}
\label{BECSection}

Let us define the binary erasure channel ${\rm BEC}(\epsilon)$. We have 
${\cal A} = \{0,1,*\}$ and $Q(0|0) = Q(1|1) = 1-\epsilon$,
$Q(*|0) = Q(*|1) = \epsilon$. Since the log-likelihoods 
(\ref{LogLikelihoodDef})
take values in $\{0,+\infty\}$ it is natural to assume the same property
to hold for the variables $U$ and $V$. We denote by $\e$ ($\eh$) the 
probability for $V$ ($U$) to be 0. As in the previous example we
distinguish two cases

\underline{$\PCS$  ensemble}. Equation (\ref{CnodeFunction}) yields
\begin{eqnarray}
\eh = 1-\rho(1-\e)\, .
\end{eqnarray}
It is easy to show that Eq.~(\ref{eq:StdMain}) implies 
the bound 
\begin{eqnarray}
h\ge \sup_{\e\in[0,1]} \phi(\e,1-\rho(1-\e)) \, ,\label{SupErasureLDPC}
\end{eqnarray}
where
\begin{eqnarray}
\phi(\e,\eh) = \Lambda'(1)\e(1-\eh)-
\frac{\Lambda'(1)}{P'(1)}[1-P(1-\e)]+\epsilon\, \Lambda(\eh)\, .
\label{TrialBec}
\end{eqnarray}
Notice that $\phi(\e,1-\rho(1-\e)) \equiv\psi(\e)$ is a smooth function of
$\e\in[0,1]$. Therefore  the $\sup$ in Eq. (\ref{SupErasureLDPC}) is
achieved either in $\e=0,1$ or for a  $\e\in (0,1)$ such that $\psi(\e)$
stationary. It is easy to realize that the stationarity condition is
equivalent to the equations
\begin{eqnarray}
\e = \epsilon\, \lambda(\eh)\, ,\;\;\;\;\;\;\;\;
\eh = 1-\rho(1-\e)\, .\label{FixedPointBEC}
\end{eqnarray}
The reader will recognize the fixed point conditions for BP decoding
\cite{Tornado0}.

\begin{figure}
\begin{tabular}{cc}
\epsfig{figure=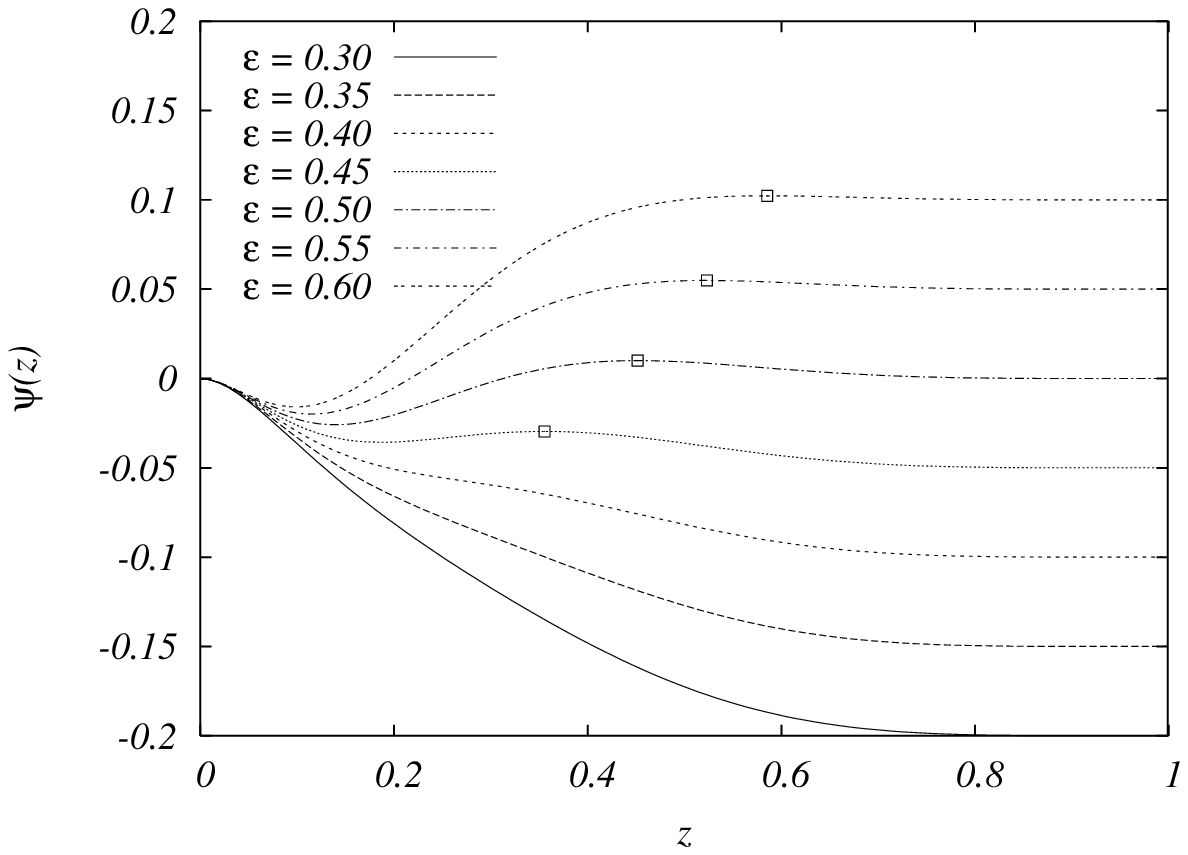,angle=0,width=0.45\linewidth}&
\epsfig{figure=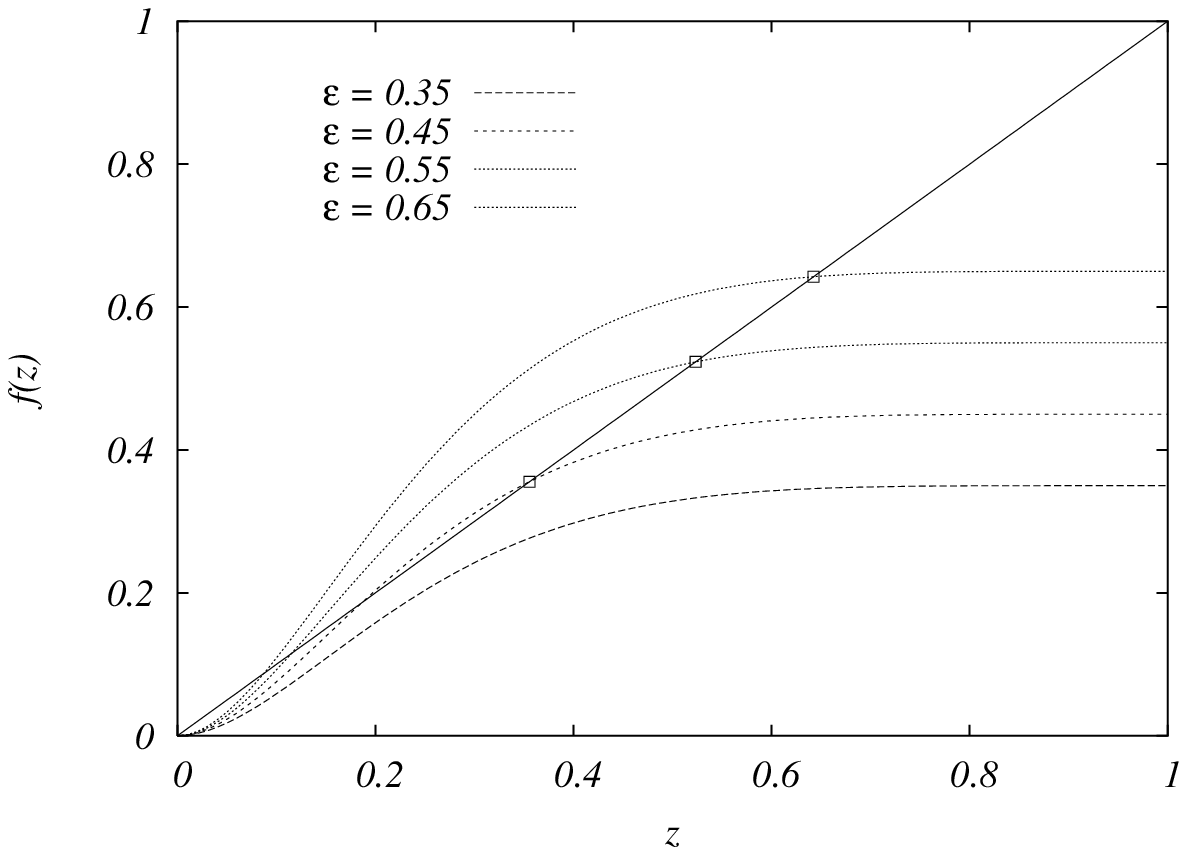,angle=0,width=0.45\linewidth}
\end{tabular}
\caption{\small Left frame: trial entropy $\phi_V$ on the erasure channel 
${\rm BEC}(\epsilon)$ as a function of the parameter $z$ characterizing the 
$V$-distribution, cf. Eq.~(\ref{TrialBec}). Here we consider the
$\PCS$ {\it ensemble} with $\Lambda(x) = x^3$ and 
$\Rho(x) = x^6$ (i.e. a regular $(3,6)$ Gallager code).
A square ($\Box$) marks the high bit-error-rate
local maximum $z_{\rm bad}(\epsilon)$. Right frame: graphical representation
of the equation $z=f(z)$ yielding the local extrema of the trial entropy,
cf. Eq.~(\ref{FixedPointBEC}).}
\label{PsiBecFig}
\end{figure}
Let us consider a specific example: the $(3,6)$ regular Gallager
code. We have $\Lambda(x) = x^3$ and $\Rho(x) = x^6$: $\Rho(x)$
is convex for any $x\in\R$ and therefore Theorem \ref{MainTheorem}
applies. The design rate is $\rdes = 1/2$. In Fig.~\ref{PsiBecFig}
we show the function $\psi(z)$ for several values of the erasure probability.
In the right frame we present the function 
$f(z) = \epsilon\lambda(1-\rho(1-\e))$ for some of these values.
At small $\epsilon$, the conditions (\ref{FixedPointBEC}) have a unique 
solution at $\e_{\rm good}(\epsilon)=0$, and $\psi(\e)$ has its unique 
local maximum there. The corresponding lower bound on the conditional entropy 
is $\psi(\e_{\rm good}(\epsilon)) = 0$. 
For $\epsilon > \epsilon_{\rm BP}\approx 0.4294398$
a secondary maximum $\e_{\rm bad}(\epsilon)$ appears. Density evolution
converges to $\e_{\rm bad}(\epsilon)$ and therefore this fixed point
control the bit error rate under BP decoding. For 
$\epsilon_{\rm BP}<\epsilon<\epsilon_{\rm MAP}\approx 0.4881508$, 
$\psi(\e_{\rm bad})<\psi(\e_{\rm good})$ and therefore this local 
maximum is irrelevant for MAP decoding. Above 
$\epsilon_{\rm MAP}$,  $\psi(\e_{\rm bad})>\psi(\e_{\rm good})$ and therefore
$\e_{\rm bad}(\epsilon)$ controls the properties of MAP decoding too.

\begin{figure}
\begin{tabular}{cc}
\epsfig{figure=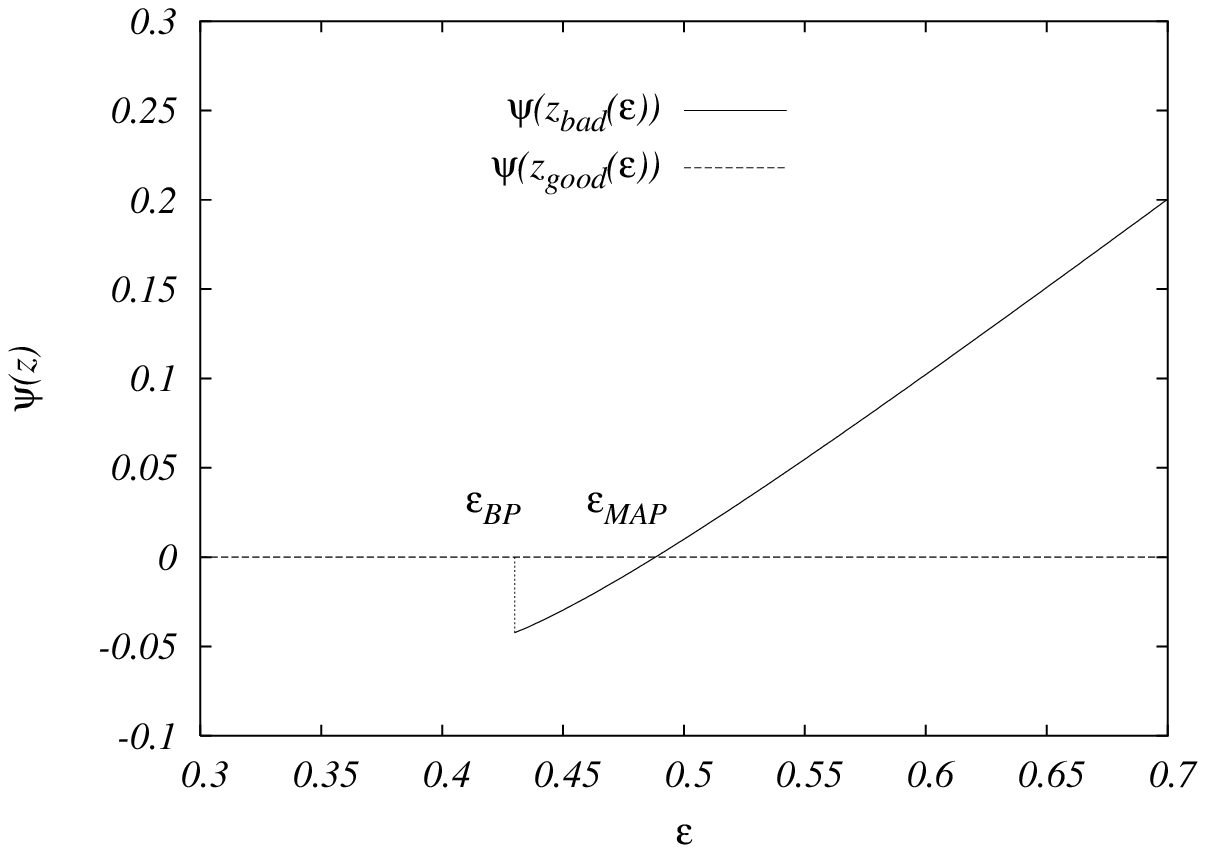,angle=0,width=0.45\linewidth}&
\epsfig{figure=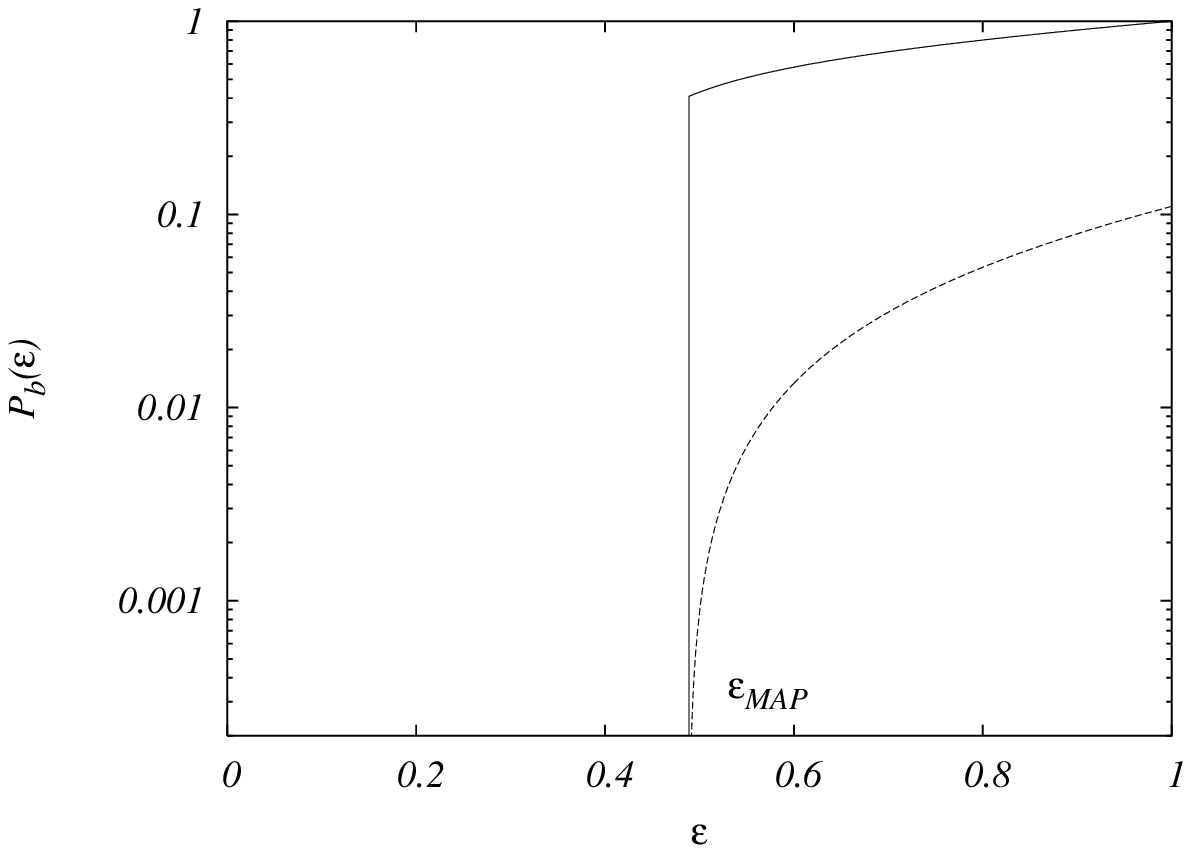,angle=0,width=0.45\linewidth}
\end{tabular}
\caption{\small Left frame: trial entropy evaluated at its local maxima
$z_{\rm good}(\epsilon)$ and $z_{\rm bad}(\epsilon)$,
as a function of the erasure probability $\epsilon$
(notice that $\psi(z_{\rm good}(\epsilon))=0$ identically). 
Right frame: bit error
rate under maximum likelihood decoding. The dashed line is the lower 
bound obtained from Fano inequality. The continuous curve is the 
conjectured exact result.}
\label{PsiEpsFig}
\end{figure}
In Fig.~\ref{PsiEpsFig}, left frame, we reproduce the function 
$\psi(\e_{\rm bad}(\epsilon))$ 
as a function of $\epsilon$. Fano inequality, cf. Lemma \ref{FanoLemma},
can be used for obtaining lower bounds on block and bit error rates
in terms of the quantity $\max\{\psi(\e_{\rm good}(\epsilon)),
\psi(\e_{\rm bad}(\epsilon)) \}$. The result for our running example is 
presented in Fig.~\ref{PsiEpsFig}, right frame.
It is evident that the result is not tight because of the sub-optimality 
of Fano inequality. For instance, in the $\epsilon\to 1$ limit, 
$\psi(\e_{\rm bad}(\epsilon))$ yields the lower bound 
$h_n\ge 1/2$. This result is easily understood: since no bit has been received,
all the $2^{n/2}$ codewords are equiprobable. On the other hand
Fano inequality yields a poor $P_{\rm b}(\epsilon=1)\ge 0.11003$.

A better (albeit non-rigorous) estimate is provided by the following recipe.
Notice that BP decoding yields (in the large blocklength limit)
\begin{eqnarray}
P^{\rm BP}_{\rm b}(\epsilon)=\left\{\begin{array}{ll}
\epsilon\Lambda(z_{\rm good}(\epsilon)) & \mbox{for $\epsilon<
\epsilon_{\rm BP}$}\, ,\\
\epsilon\Lambda(z_{\rm bad}(\epsilon))  & 
\mbox{for $\epsilon\ge \epsilon_{\rm BP}$}\, .
\end{array}\, \right.
\end{eqnarray}
Our prescription consists in taking 
\begin{eqnarray}
P^{\rm MAP}_{\rm b}(\epsilon)=\left\{\begin{array}{ll}
\epsilon\Lambda(z_{\rm good}(\epsilon)) & \mbox{for $\epsilon<
\epsilon_{\rm MAP}$}\, ,\\
\epsilon\Lambda(z_{\rm bad}(\epsilon))  & 
\mbox{for $\epsilon\ge \epsilon_{\rm MAP}$}\, .
\end{array}\, \right.
\end{eqnarray}
In other words, BP
is asymptotically optimal except in the interval 
$[\epsilon_{\rm BP},\epsilon_{\rm MAP}]$. Generalizations
and heuristic justification of this recipe will be provided in the next 
Sections. The resulting curve for our running example is reported in 
Fig.~\ref{PsiEpsFig}, right frame.

\begin{table}
\centerline{
\begin{tabular}{cccccc}
\hline
$k$ & $\gamma$ & $\epsilon_{\rm BP}$ & $\epsilon_{\rm MAP}$: New UB & 
Gallager UB &
Gallager LB\\
\hline
$2$ & $4$  & $1/3$ & $1/3$ & $\ast$ & $\ast$ \\
$3$ & $6$  & $0.4294398$ & $0.4881508$ & $0.4999118$ & $0.4833615$\\
$4$ & $8$  & $0.3834465$ & $0.4977409$ & $0.4999118$ & $0.4967571$\\
$5$ & $10$ & $0.3415500$ & $0.4994859$ & $0.4999997$ & $0.4992593$\\
$6$ & $12$ & $0.3074623$ & $0.4998757$ & $0.5000000$ & $0.4998207$\\
\hline
\end{tabular}}
\caption{\small Maximum a posteriori probability and belief propagation 
thresholds for several
ensembles of the form ${\rm LDPC}(n,\gamma,x^k)$ with $\gamma=(1-\rdes)k$
and $\rdes = 1/4$. For the MAP threshold we  compare several different
thresholds: `New UB' is the upper bound derived in this paper; 
`Gallager UB' is Gallager lower bound as generalized in 
Ref.~\cite{BurshteinUB};
`Gallager LB' is the upper bound derived using Gallager's 
technique, as applied in Ref.~\cite{Pishro}.}
\label{TableLDPCBEC}
\end{table}
The analysis of this simple  example
uncovers the existence of three distinct regimes: 
$(i)$ A low noise regime, $\epsilon<\epsilon_{\rm BP}$:
both BP and MAP decoding are effective in this case: the bit error
rate vanishes in the large blocklength limit;
$(ii)$ An intermediate noise regime, $\epsilon_{\rm BP}<
\epsilon<\epsilon_{\rm MAP}$. Only MAP decoding can  produce vanishing 
error rates here. $(iii)$ An high noise regime, $\epsilon_{\rm MAP}<\epsilon$.
The bit error rate under MAP decoding is bounded from below.
In Table \ref{TableLDPCBEC}
we report the values of $\epsilon_{\rm BP}$ and $\epsilon_{\rm MAP}$
for a few  ensembles $\PCS$ with $\Lambda(x) = x^l$, 
$\Rho(x) = x^k$ and $\rdes=1/2$.
As we shall discuss below, this pattern is quite general.

\underline{$\GMS$ ensemble}. It is interesting to look at the 
differences between LDGM and LDPC  ensembles within the BEC context.
The requirement (\ref{CnodeFunction}) implies
\begin{eqnarray}
\eh = 1-(1-\epsilon)\rho(1-\e)\, .\label{AdmissibilityLDGM}
\end{eqnarray}
Applying Theorem \ref{MainTheorem} we get the bound 
\begin{eqnarray}
h\ge \sup_{\e\in[0,1]} \phi(\e,1-(1-\epsilon)\rho(1-\e)) 
\, ,\label{SupErasureLDGM}
\end{eqnarray}
where, with a slight abuse of notation, we defined
\begin{eqnarray}
\phi(\e,\eh) = \Lambda'(1)\e(1-\eh)-
\frac{\Lambda'(1) (1-\epsilon)}{P'(1)}[1-P(1-\e)]+ \Lambda(\eh)\, .
\label{TrialBecLDGM}
\end{eqnarray}
As in the LDPC case we look at the stationarity condition of the function
$\psi(\e)\equiv \phi(\e,1-(1-\epsilon)\rho(1-\e))$. Elementary calculus 
yields the couple of equations 
\begin{eqnarray}
\e = \lambda(\eh)\, ,\;\;\;\;\;\;\;\;
\eh = 1-(1-\epsilon)\rho(1-\e)\, ,\label{FixedPointBECLDGM}
\end{eqnarray}
that, once again, coincide with the fixed point conditions for BP decoding.
These equations have a 
noise-independent solution $\e_{\rm bad}(\epsilon)=1$ (implying 
$\eh = 1$ because of Eq. (\ref{AdmissibilityLDGM})). Theorem
\ref{MainTheorem} yields $h\ge 1-\Cap(\epsilon)/\rdes$,
with $\Cap(\epsilon) = 1-\epsilon$ the channel capacity: 
we recover in this context the simple lower bound of Sec.~\ref{ShannonSection}.

\begin{figure}
\begin{tabular}{cc}
\epsfig{figure=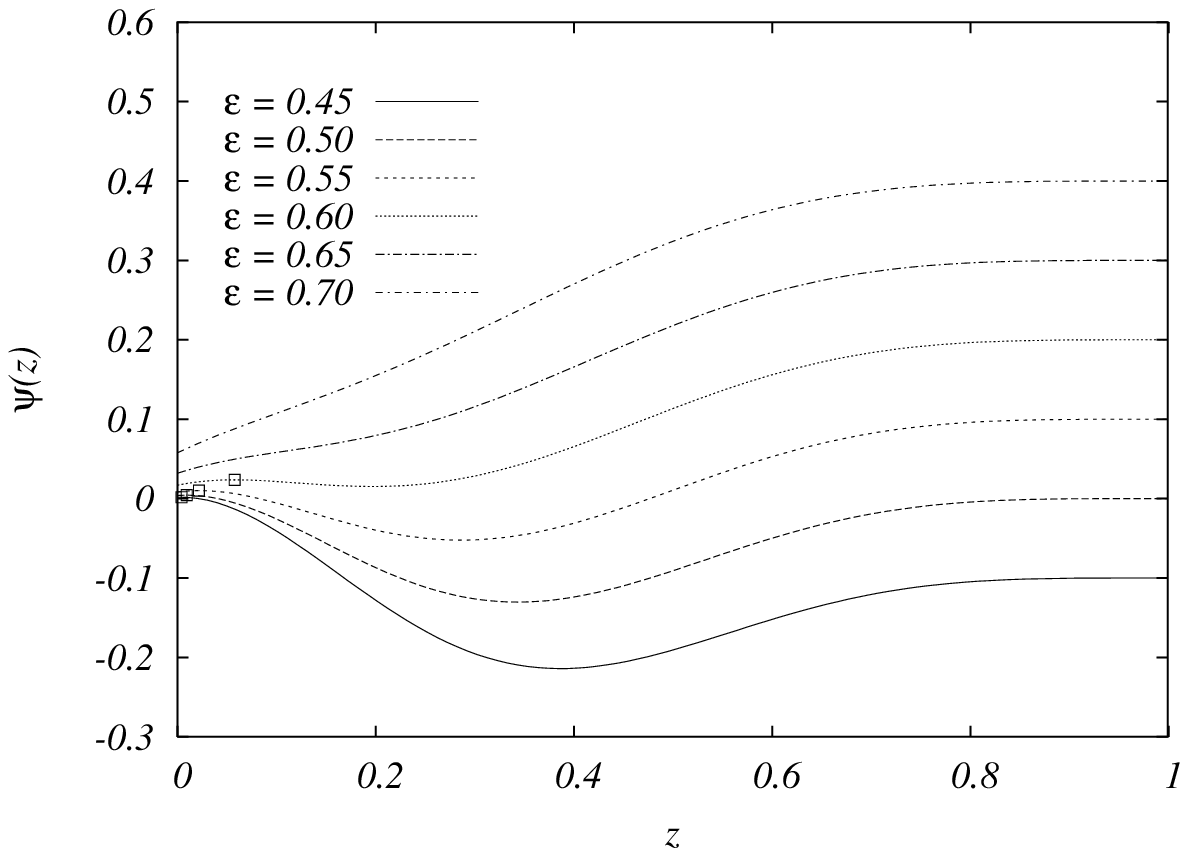,angle=0,width=0.45\linewidth}&
\epsfig{figure=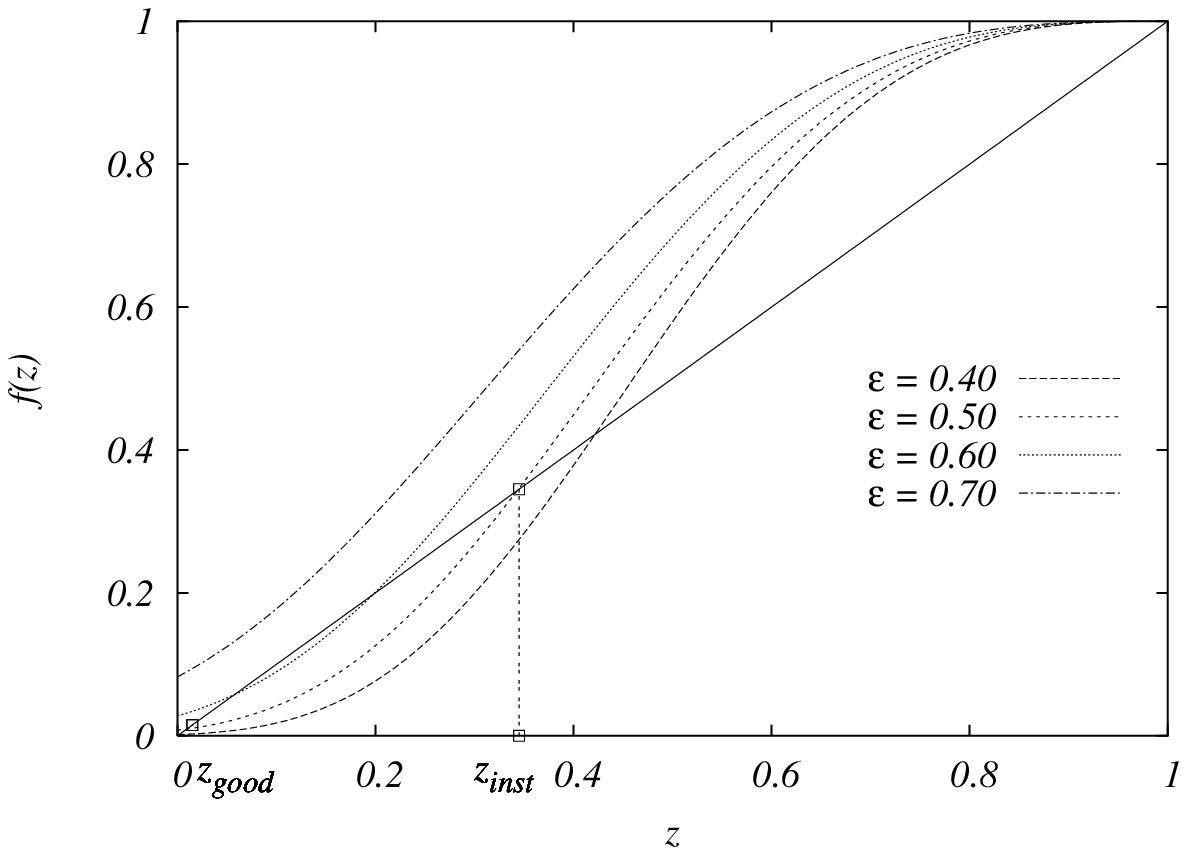,angle=0,width=0.45\linewidth}
\end{tabular}
\caption{\small Left frame: trial entropy for the $\GMS$ {\it ensemble} on the
BEC$(\epsilon)$ channel. Here $\Lambda(x) = x^8$ and $\Rho(x) = x^4$
(the design rate is $\rdes =1/2$).
A square ($\Box$) marks the small-$P_{\rm b}$ 
stationary point $z_{\rm good}(\epsilon)$. Right frame: graphical 
representation of the stationarity condition $z=f(z)$.}
\label{PsiBecFigGM}
\end{figure}
A better understanding of the peculiarities LDGM  ensembles
is obtained by looking at a particular example. Consider, for instance,
$\Lambda(x) = x^8$ and $\Rho(x) = x^4$ which corresponds to a design rate 
$\rdes = 1/2$. Theorem \ref{MainTheorem} applies because $\Rho(x)$
is convex on $\R$. In Fig.~\ref{PsiBecFigGM}, left frame, we plot the
function $\psi(\e)$ for several values of the erasure probability 
$\epsilon$. It is clear that $\e_{\rm bad} = 1$ is always a local maximum.
A second local maximum $\e_{\rm good}(\epsilon)$ appears
when the erasure probability becomes smaller than 
$\epsilon_{\rm BP}\approx 0.6165534$. The extremum at $\e_{\rm good}(\epsilon)$
becomes a global maximum for $\epsilon <\epsilon_{\rm MAP}$,
with $\epsilon_{\rm MAP}\approx 0.5022591$. In  
Fig.~\ref{PsiBecFigGM}, right frame,
we reproduce the function $f(z) = \lambda(1- (1-\epsilon)\rho(1-z))$
in terms of which the stationarity condition (\ref{FixedPointBECLDGM})
reads $z = f(z)$. We also mark the solutions $\e_{\rm good}(\epsilon)$
(corresponding to a local maximum of $\psi(z)$) and  
$\e_{\rm inst}(\epsilon)$ (corresponding to a local minimum of $\psi(z)$).

The interpretation of these results is straightforward. Maximum likelihood 
decoding is controlled by the stationary point $z_{\rm bad}=1$ for
$\epsilon>\epsilon_{\rm MAP}$. In this regime the lower bound
(\ref{TrialBecLDGM}) yields the same conditional entropy as for the random 
code ensemble.
We expect the bit error rate in this regime to be $P_{\rm b}(\epsilon)= 1/2$.
At low noise ($\epsilon<\epsilon_{\rm MAP}$)
the fixed point  $\e_{\rm good}(\epsilon)$ controls the 
MAP performances. Analogously to what argued in the previous 
Subsection, we expect this to imply to a bit error 
rate $P_{\rm b}(\epsilon)= \Lambda(1-(1-\epsilon)
\rho(1-\e_{\rm good}(\epsilon)))$.

As for BP decoding, it has a unique fixed point $\e_{\rm bad}=1$ for 
$\epsilon>\epsilon_{\rm BP}$. This corresponds to a high 
bit error rate $P_{\rm b}(\epsilon)= 1/2$. A second, locally stable, fixed
point appears at $\epsilon_{\rm BP}$.  If BP is initialized using
only erased messages (as is usually done), all the messages remain erased (BP 
does not know where to start from). The same remains true is a small
number of non-erased (correct) messages is introduced: density evolution
is still controlled by the  $\e_{\rm bad}$ fixed point. If however the
initial conditions contains a large enough fraction 
(namely, larger than $1-\e_{\rm inst}$) of correct messages,
the small-$P_{\rm b}$ fixed point $\e_{\rm good}$ is eventually reached.

Let us finally notice that the present results can be shown to be consistent 
with the ones of Refs.~\cite{UrbankeMeasson,UrbankeMeassonMontanari}.
%
%
\subsection{General channel: a simple minded bound}

The previous Section suggests a simple bound for the $\PCS$  ensemble 
on general BIOS channel. Take, as for the BEC case, $V=0$
with probability $\e$ and $=\infty$ with probability $1-\e$,
while $U=0$ with probability $\eh$ and $=\infty$ with probability $1-\eh$.
These conditions are consistent with the admissibility
requirement  (\ref{CnodeFunction}) if
\begin{eqnarray}
\eh = 1-\rho(1-\e)\, .
\end{eqnarray}
Plugging into Eq. (\ref{FreeEnergy}) we get a bound of the same form as for
the BEC, cf. Eq. (\ref{SupErasureLDPC}) with
\begin{eqnarray}
\phi(\e,\eh) = \Lambda'(1)\e(1-\eh)-
\frac{\Lambda'(1)}{P'(1)}[1-P(1-\e)]+[1-\Cap(Q)]\, \Lambda(\eh)\, .
\end{eqnarray}
Passing from the BEC to a general BIOS, amounts, under
this simple ansatz,  to substituting $1-\Cap(Q)$ to $\epsilon$.
%
%
\subsection{General channel: optimizing the bound}
\label{sec:BoundOpt}

We saw in Sec.~\ref{BECSection} that, for the BEC,  stationary points of the 
trial entropy function correspond to fixed points of the density-evolution 
equations. This fact is indeed quite general and holds for a general BIOS 
channel.

In order to discuss this point, it is useful to have a concrete
representation for  the random variables $U$, $V$ entering in the definition 
of the trial entropy (\ref{FreeEnergy}).  
A first possibility is to identify them with the distributions 
$\Ud(x)\equiv \Prob[U\le x]$ and $\Vd(x) \equiv \Prob[V\le x]$
as explained in \cite{RichardsonUrbankeBook}. The distributions
are right continuous, non decreasing functions such that
$\lim_{x\to -\infty}\Ad(x) =0$, and $\lim_{x\to +\infty}\Ad(x) \le 1$.
Viceversa, to any such function we may associate a well 
defined random variable. It is convenient to introduce
the densities $\ud(x)$ and $\vd(x)$ which are formal Radon-Nikodyn 
derivatives of $\Ud(x)$ and  $\Vd(x)$.
We also introduce the log-likelihood distributions associated to 
channel output, cf Eq. (\ref{LogLikelihoodDef}):
$\Jd(x) \equiv \Prob[J\le x]$ and $\Hd(x)  \equiv \Prob[h\le x]$.
The corresponding densities will be denoted by $\jd(x)$ and  
$\hd(x)$.

The admissibility condition (\ref{CnodeFunction}) translate of course into
a condition on the distributions $\Ud(x)$ and $\Vd(x)$.
Following once again \cite{RichardsonUrbankeBook}, we express 
this condition through `$G$-distributions'. More precisely
for any number $x\in(-\infty,+\infty]$, we define 
$\gamma(x) =(\gamma_1(x),\gamma_2(x))\in\{\pm 1\}\times [0,+\infty)$
by taking $\gamma_1(x) = \sign\, x$, and $\gamma_2(x) = -\log|\tanh x|$.
We define $\Gamma$ to be the change-of-measure operator associated
to the mapping $\gamma$. If $X$ is a random variable with 
distribution $\Ad$ and (formal) density $\ad$, $\Gamma(\ad)$ is
defined to be the density associated to $\gamma(X)$. Despite
the notation $\Gamma$ is defined on distributions, and only formally on 
densities. The action of $\Gamma$ is described in detail in
\cite{RichardsonUrbankeBook}. Among the other properties, it has a well 
defined inverse $\Gamma^{-1}$. We can now write the condition implied
by  (\ref{CnodeFunction}) in a compact form:
\begin{eqnarray}
\ud = \sum_{k=2}^{k_{\rm max}} \rho_k\; \Gamma^{-1}
\left[\Gamma(\jd)\otimes\Gamma(\vd)^{\otimes (k-1)}\right]\; \equiv
\; \rho_{\jd}(\vd)\, .
\end{eqnarray}

A second concrete representation is obtained by inverting the distributions
$\Ud(x)$, $\Vd(x)$. Of course this is not possible unless $\Ud(x)$,
$\Vd(x)$ are continuous and increasing. However, we can always define the
`inverse distributions':
\begin{eqnarray}
\Ui:(0,1)&\to & (-\infty,+\infty]\, \\
     \xi & \mapsto & \Ui(\xi)\equiv \min\{ x \mbox{ such that } 
\Ud(x)\ge\xi\}\, ,
\end{eqnarray}
with an analogous definition for $\Vi(\xi)$. We introduce analogously
the inverse distributions $\Ji(\xi)$ and $\Hi(\xi)$ for the 
log-likelihoods $J_a$ and $h_i$, cf. Eq.  (\ref{LogLikelihoodDef}).
Notice that, given a real valued random
variable $X$, its inverse distribution $\Ai(\xi)$  is non-decreasing
and left-continuous. We shall denote by $\invdistr$ the space of inverse 
distributions. Moreover, it has a simple practical
meaning: if one is able to sample $\xi$ uniformly in $(0,1)$, then
$\Ai(\xi)$ is distributed like $X$.
From this observation it follows that any inverse distribution 
$\Ai(\xi)$ uniquely determines its associated random variable.

We can now re-express the trial entropy (\ref{FreeEnergy}) 
as a functional over $\invdistr$, $\phi=\phi(\Ui,\Vi)$ 
using the above correspondence\footnote{Since throughout this
Section the degree sequences are kept fixed, we
shall drop the dependence of $\phi$ on $(\Lambda,\Rho)$, and 
(with an abuse of notation) replace it with its dependence upon $\Ui$
and $\Vi$.}. 
After some straightforward computations we get
\begin{eqnarray}
\phi & = & -\Lambda'(1)\int \!
\log_2[1+\tanh \Ui(\xi_1)\tanh\Vi(\xi_2)] \; d\xi_0\, d\xi_1\, +\\
&&+\sum_l\Lambda_l\, e^{-\gamma}\int\!
\log_2\left[\sum_{\sigma = \pm 1} 
(1+\sigma\tanh \Hi(\xi_0))\prod_{i=1}^l(1+\sigma\tanh \Ui(\xi_i)\right]\;
\prod_{i=0}^{l}d\xi_i\, + \nonumber\\
&& +\frac{\gamma}{\Rho'(1)}\sum_{k}\Rho_k\int\!
\log_2\left[1+\tanh\Ji(\xi_0)\prod_{i=1}^{k}\tanh\Vi(\xi_i)\right]
\prod_{i=0}^{l}d\xi_i\, - \Cap(\Qv) -\frac{\gamma}{\Rho'(1)}\Cap(\Qc)\, ,
 \nonumber
\end{eqnarray}
with all the integrals on the $\xi_i$'s being on the interval $(0,1)$.
This representation allows to easily derive the following result.
\begin{lemma}
Assume the supremum of the trial free energy $\phi_V$ over the space of 
admissible random variables is achieved for some couple $(U,V)$.
Then
\begin{eqnarray}
V \ed h+\sum_{i=1}^l U_i\, ,\label{VnodeFunction}
\end{eqnarray}
$l$ being a Poisson random variable of parameter $\gamma$ and $h$ 
distributed according to the definition  (\ref{LogLikelihoodDef}).
\end{lemma}
\proof
Look at $\phi$ as a functional $(\Ui,\Vi)\to \phi(\Ui,\Vi)$ of the inverse
distributions $\Ui$ and $\Vi$. The idea is to differentiate this functional
at its (assumed) maximum. Let $\Di:(0,1)\to(-\infty,+\infty]$ be 
left continuous and non decreasing. It is an easy calculus exercise 
to show that
\begin{eqnarray}
\left.\frac{d\phantom{\ve}}{d\ve}\phi(\Ui+\ve\Di,\Vi)\right|_{\ve=0} & = &
-\Lambda'(1)\int[1-\tanh^2\Ui(\xi)]\,\Di(\xi)\,{\cal F}_\xi(\Ui,\Vi)\, ,\\
\left.\frac{d\phantom{\ve}}{d\ve}\phi(\Ui,\Vi+\ve \Di)\right|_{\ve=0} & = &
-\Lambda'(1)\int[1-\tanh^2\Vi(\xi)]\,\Di(\xi)\,{\cal G}_\xi(\Ui,\Vi)\, ,
\end{eqnarray}
where
\begin{eqnarray}
{\cal F}_\xi(\Ui,\Vi)& = &\int\!\frac{\tanh \Vi(\xi_1)}{1+\tanh\Ui(\xi)
\tanh\Vi(\xi_1)}\; d\xi_1-\\
&&-\sum_{l}\lambda_l
\int\! \frac{\tanh [\Hi(\xi_0)+\sum_{i=1}^l\Ui(\xi_i)]}{1+\tanh\Ui(\xi)
\tanh[\Hi(\xi_0)+\sum_{i=1}^l\Ui(\xi_i)]}\;\prod_{i=0}^{l}d\xi_i\, ,
\nonumber\\
{\cal G}_\xi(\Ui,\Vi)& = & \int\!\frac{\tanh \Ui(\xi_1)}{1+\tanh\Ui(\xi_1)
\tanh\Vi(\xi)}\; d\xi_1\; -\\
&&-\;\sum_{k}\rho_k\int\!\frac{\tanh\Ji(\xi_0)
\prod_{i=1}^{k-1}\tanh\Vi(\xi_i)}{1+\tanh\Ji(\xi_0)
\prod_{i=1}^{k-1}\tanh\Vi(\xi_i)\tanh\Vi(\xi)}\;\prod_{i=0}^{k-1}d\xi_i\, .
\nonumber
\end{eqnarray}
Notice that ${\cal G}_\xi(\Ui,\Vi)$ vanishes because of the admissibility
condition (\ref{CnodeFunction}). It is then straightforward to show
that ${\cal F}_\xi(\Ui,\Vi)$ must vanish for any $\xi$ such that 
$\Ui(\xi)<\infty$, in order for $(\Ui,\Vi)$ to be a 
maximum. This in turns implies the thesis.
\endproof
%
%
\subsection{Numerical estimates and comparison with previous bounds}
\label{ComparisonSection}

The discussion in the previous Section suggests a natural possibility 
for evaluating numerically the lower bound in Theorem \ref{MainTheorem}.
Run density evolution \cite{RichardsonUrbanke} for $T$ iterations 
and then evaluate the trial entropy (\ref{FreeEnergy}) taking 
$U$ and $V$ to be random variables with the density of (respectively)
right-to-left or left-to-right messages. Notice that, in order 
for Eq.~(\ref{CnodeFunction}) to be satisfied, the right-to-left
density must be updated one last time before evaluating the trial entropy.
\begin{table}
\centerline{
\begin{tabular}{ccccccc}
\hline
$k$ & $l$ & $r_{\rm des}$ & New UB & Gallager UB & Gallager LB & Shannon limit\\
\hline
$4$ & $3$ & $1/4$ & $0.2101(1)$ & $0.2109164$ & $0.2050273$ & $0.2145018$\\
$5$ & $3$ & $2/5$ & $0.1384(1)$ & $0.1397479$ & $0.1298318$ & $0.1461024$\\
$6$ & $3$ & $1/2$ & $0.1010(2)$ & $0.1024544$ & $0.0914755$ & $0.1100279$\\
$6$ & $4$ & $1/3$ & $0.1726(1)$ & $0.1726268$ & $0.1709876$ & $0.1739524$\\
\hline
\end{tabular}}
\caption{\small Thresholds for regular LDPC codes over the binary symmetric
channel ${\rm BSC}(p)$ ($k$ and $l$ are, respectively, the check and variable 
node degrees, and $r_{\rm des}$ the design rate). The new upper bound
proved in this paper is evaluated numerically following the approach 
described in the text. The quoted error comes from Monte Carlo sampling of the
random variables $U$ and $V$. `Gallager UB' and `Gallager LB'
refer respectively to the upper and lower bounds obtained by 
Gallager~\cite{GallagerThesis}.}
\label{TableLDPCBSC}
\end{table}

This still leaves a lot of freedom. The first question is: how large $T$
(the number of iterations) should be? 
While it is difficult to provide a quantitative answer, 
in order to approach the supremum in Eqs.~(\ref{TheoremPoisson})
to (\ref{eq:StdMain}), one should get a good approximation of fixed point 
densities. Generically, this happens only as $T$ is let to infinity.

The next question is: how the densities should be initialized?
This question has a very simple answer in usual applications of density 
evolution: just initialize to the message density seen at the zeroth
step of message passing. This generally means $U,V$ identically equal to $0$.
Hereafter, we shall refer to this as the `$0$-initialization'
This answer is no longer complete in the present context. In fact any
initial condition, such that $U$ and  $V$ are symmetric random variables,
corresponds eventually to a valid lower bound of
the form in Eqs.~(\ref{TheoremPoisson}) to (\ref{eq:StdMain}).
At least one other simple initial condition consist in taking
$U=V=+\infty$ identically. In the case of standard ensembles with 
minimum left degree at least $2$ this is in fact a fixed point
and the corresponding trial entropy vanishes.
We shall refer to this as the `$\infty$-initialization'. 

Despite this freedom, Eqs.~(\ref{TheoremPoisson}) to (\ref{eq:StdMain})
always provide a lower bound, no matter how we implement the general strategy.
In Tab.~\ref{TableLDPCBSC} we report the numerically-evaluated upper
bound for a few regular ensembles over the binary symmetric channel. 
We implemented a sampled (Monte Carlo)
version of density evolution (with $10^4$ to $10^5$ sampling points)
and adopted the $0$-initialization. 
The trial entropy (\ref{FreeEnergy}) was averaged over $10^4$ iterations after
$10^4$ equilibration iterations. The threshold was estimated as the
smallest noise parameter such that the trial entropy is positive. 

In the same Table (and analogously in Table~\ref{TableLDPCBEC}
for the erasure channel), we compare our upper bounds with previous 
upper and lower bounds. In his thesis \cite{GallagerThesis}, 
Gallager used an estimate of the distance spectrum, together with 
a clever modification of the union bound in order to obtain lower
bounds. This technique was further generalized and refined
over the years, see for instance \cite{BurshteinLB,Bounds,Pishro} but it's fair
to say that there were no major modification over the initial result
for the simplest regular ensembles and channel models.
We should stress that a different technique, based on typical
pairs decoding was proposed in \cite{TypicalPairs}. Our evaluation of
Gallager bound wasn't numerically distinguishable from the results of 
\cite{TypicalPairs}.

As for upper bounds, Gallager approach is based on an 
information theoretic argument.
Also in this case, despite some improvements~\cite{BurshteinUB},
the main idea remained essentially unchanged. Moreover, the quantitative
estimates by Gallager remained essentially the state-of-the art
for the simplest regular ensembles and channel models considered here.

Despite the various estimates in Tables \ref{TableLDPCBEC}
and  \ref{TableLDPCBSC} are numerically close (which is partially
due to the proximity of capacity), the bound of Theorem~\ref{MainTheorem}
is clearly superior to previous upper bounds. 
%
%
\subsection{Relation with the Bethe free energy}
\label{BetheSection}

Until now we studied the average properties of the code {\it ensembles}
defined in Sec. \ref{DefinitionsSection}. Although the concentration 
result of Sec. \ref{ConcentrationSection} justify this approach,
it may be interesting to take a step backward and consider
the decoding problem for a single realization of the code and of the
channel. It is convenient to introduce the `Bethe free-energy'
\cite{Bethe}  $F_{\rm B}(\ub)$ associated to the probability distribution 
(\ref{ProbabilityDistribution}). We have
\begin{eqnarray}
\Fbe(\ub) & = &\Ube(\ub)-\Hbe(\ub)\, ,\label{eq:BetheFreeEnergy}
\end{eqnarray}
where
\begin{eqnarray}
\Ube(\ub) & = & -\sum_{a\in\Cnod}\sum_{\ux_a}b_a(\ux_{a})
\log_2\Qc(\hy_a|\ux_a)-
\sum_{i\in\Vnod}\sum_{x_i} b_i(x_i)\log_2 \Qv(y_i|x_i)\, ,\\
\Hbe(\ub) &= & -\sum_{a\in\Cnod}\sum_{\ux_a}b_a(\ux_a)
\log_2 b_a(\ux_a)+
\sum_{i\in\Vnod}(|\di|-1)\sum_{x_i} b_i(x_i)\log_2  b_i(x_i)\, ,
\end{eqnarray}
and we used the shorthands $\ux_{a}=(x_{i^a_1},\dots,x_{i_k^a})$
and $\Qc(\hy_a|\ux_a)=\Qc(\hy_a|x_{i^a_1}\oplus\cdots\oplus x_{i_k^a})$.
The parameters $\{ b_a(\ux_a):\; a\in\Cnod\}$ and 
$\{b_i(x_i): \; i\in\Vnod\}$  are probability 
distributions subject to the marginalization conditions
\begin{eqnarray}
\sum_{x_{j},\, j\in\da\backslash i} b_a(\ux_a) &=& b_i(x_i)
\;\;\;\;\;\;\;\;\;\; \forall i\in\da\, ,\label{Marg1}\\
\sum_{x_i} b_i(x_i) & = & 1\;\;\;\;\;\;\;\;\;\;\;\;\;\;\;\;\;\;\; \forall i\, .
\label{Marg2}
\end{eqnarray}
For LDGM codes  $\Fbe(\ub)$ is always finite. 
For LDPC codes, it takes values in $(-\infty,+\infty]$ and is
finite if $b_a(\ux_a) $ vanishes whenever 
$x_{i^a_1}\oplus\cdots\oplus x_{i_k^a}=0$ (as always we use the convention 
$0\log 0=0$). Moreover, as explained in \cite{Bethe}, its stationary points are
fixed points of BP decoding.

Following \cite{Bethe}, we consider the stationary points of
the Bethe free energy (\ref{eq:BetheFreeEnergy}) under the constraints 
(\ref{Marg1}), (\ref{Marg2}). This can be done by introducing
a set of Lagrange multipliers 
$\{\lambda_{ia}(x_i)\}$ for Eq. (\ref{Marg1}), the constraint
(\ref{Marg2}) being easily satisfied by normalizing the beliefs. 
One then consider the Lagrange function
\begin{eqnarray}
L_{\rm B}(b,\lambda) = F_{\rm B}(b) - \sum_{(ia)\in\Edges}\sum_{x_i}
\lambda_{ia}(x_i)\left[\sum_{x_j:\, j\in\da\backslash i}b_a(\ux_a)-
b_i(x_i)\right] \, .
\end{eqnarray}
We refer to \cite{Bethe} for further details of this computation.
Stationary points are eventually shown to have the form
\begin{eqnarray}
b_a(\ux_a) & = & \frac{1}{z_a}\, \Qc(\hy_a|\ux_a)
\prod_{j\in\da} P_{v_{j\to a}}(x_{j})\, ,\label{eq:BethePar1}\\
b_i(x_i) & = & \frac{1}{z_i}\, \Qv(y_i|x_i)\prod_{a\in\di}P_{u_{a\to i}}(x_i)
\, ,\label{eq:BethePar2}
\end{eqnarray}
with $P_{\cdot}(x)$ being defined as in Eq. (\ref{BinDef}).
The $z_a$'s and $z_i$'s are fixed by the normalization conditions 
$\sum_{\ux}b_a(\ux) = 1$ and $\sum_{x}b_i(x_i)$. 
The messages $\{v_{i\to a}\}$ are related to the Lagrange multipliers 
$\{\lambda_{ia}(x_i)\}$ by the relation
\begin{eqnarray}
P_{v_{i\to a}}(x_i) \propto \exp\{\lambda_{ia}(x_i)\}\, ,
\end{eqnarray}
while the $\{u_{a\to i}\}$ must satisfy the equation
\begin{eqnarray}
v_{i\to a} & = & h_i +\sum_{b\in\di\backslash a} u_{b\to i}\, ,
\label{eq:SaddleBethe1}
\end{eqnarray}
for any $i\in \Vnod$.
The marginalization constraint (\ref{Marg1}) is satisfied if the equation
\begin{eqnarray}
u_{a\to i} & = & \atanh\{ \tanh J_a\prod_{j\in\da\backslash i}
\tanh v_{j\to a}\}\, \label{eq:SaddleBethe2}
\end{eqnarray}
holds for any $a\in\Cnod$.

If we substitute the beliefs (\ref{eq:BethePar1}), 
(\ref{eq:BethePar2}) into Eq. (\ref{eq:BetheFreeEnergy})
we can express the Bethe free energy as a function of the messages
$\uu \equiv\{u_{a\to i}\}$ and $\uv \equiv\{v_{i\to a}\}$.
Using  Eqs. (\ref{eq:SaddleBethe1}), (\ref{eq:SaddleBethe2}), 
we get the following expression (with a slight abuse of notation we do not 
change the symbol denoting the free energy)
\begin{eqnarray}
F_{\rm B}(\uu,\uv) & = & \sum_{(ia)\in\Edges}\log_2 
\left[\sum_{x_i}P_{u_{a\to i}}(x_i)P_{v_{i\to a}}(x_i)\right]-
\sum_{i\in \Vnod}\log_2\left[\sum_{x_i}\Qv(y_i|x_i)\prod_{a\in\di}
P_{u_{a\to i}}(x_i)\right]- \nonumber\\
&&-\sum_{a\in\Cnod}\log_2\left[\sum_{\ux_a}\Qc(\hy_a|\ux_a)\prod_{i\in\da}
P_{v_{i\to a}}(x_i)\right]\, .
\end{eqnarray}
A simple comparison of this expression with Eq. (\ref{FreeEnergy})
yields the following interesting result.
\begin{propo}
Let $\Fbe(\uu,\uv)$ be the Bethe free energy for any of the code ensembles
$\GMP$, $\PCP$, $\GMS$, $\PCS$,  with the beliefs parameterized as in 
Eqs. (\ref{eq:BethePar1}) and (\ref{eq:BethePar2}), and assume that 
the messages are i.i.d. random variables $u_{i\to a}\ed U$ and 
$v_{a\to i}\ed V$. Then
\begin{eqnarray}
\lim_{n\to\infty} \frac{1}{n}\E\, \Fbe(\uu,\uv) = -\phi_V(\Lambda,\Rho)
+\kappa\sum_y Q(y|0)\log_2\, Q(y|0)\, ,\label{BetheTrial}
\end{eqnarray}
where the expectation $\E(\cdot)$ is taken both with respect to the
messages distribution and the code ensemble, and $\kappa = \gamma/\Rho'(1)$
(for {\rm LDGM} ensembles) or $\kappa = 1$ (for {\rm LDPC} ensembles).
For multi-Poisson ensembles $\GMM$ and $\PCM$, the same formula holds with 
$\Lambda$ being replaced by $\Lambda^{(\gamma)}$ on the right hand side.
\end{propo}
\prooft 
In order to compute the expectation on the left hand side
of Eq.~(\ref{BetheTrial}), let us proceed in two steps. In a first step,
we shall take the expectation with respect to the messages
$\{u_{i\to a},v_{a\to i}\}$, which 
in the Theorem statement are assumed to be i.i.d.'s, as
well as with respect to the channel output $\{ y_i,\hy_a\}$. Let us
denote by $\Vnod_l$ ($\Cnod_k$) the set of variable nodes 
(check nodes) of degree $l$ (degree $k$). By linearity of expectation, we get 
\begin{eqnarray}
\E_{u,v}F_{\rm B}(\uu,\uv) & = & |\Edges|\,\, \E_{u,v}\log_2 
\left[\sum_{x}P_{u}(x)P_{v}(x)\right]-
\sum_{l}|\Vnod_l|\,\, \E_y\E_{u}\log_2\left[\sum_{x}\Qv(y|x)\prod_{a=1}^l
P_{u_a}(x)\right]- \nonumber\\
&&-\sum_{k}|\Cnod_k|\,\, \E_v\E_{u}
\log_2\left[\sum_{\ux}\Qc(\hy|x_1\oplus\cdots\oplus x_k)\prod_{i=1}^k
P_{u_{i}}(x_i)\right]\, .\label{eq:ProofBetheFirst}
\end{eqnarray}
Now notice that the number of edges is equal to the number of variable
nodes times the average left degree: $|\Edges| = n\hLambda'(1)$.
The number of variable nodes of degree $l$ is, by definition
$|\Vnod_l| = n\hLambda_l$. Furthermore the total number of check nodes is
$n\hLambda'(1)/\hRho'(1)$, and therefore $|\Cnod_k| = 
(n\hLambda'(1)/\hRho'(1))\hRho_k$. Finally both for Poisson and standard 
ensembles, the expected degree profile converges to the 
design profile, see Sec.~\ref{PoissonDefinitionSection}. In other words
\begin{eqnarray}
\lim_{n\to\infty}\Eg\, \hLambda_l = \Lambda_l\, ,\;\;\;\;
\lim_{n\to\infty}\Eg\, \hRho_k = \Rho_k\, ,\;\;\;\;
\lim_{n\to\infty}\Eg\, \hLambda'(1) = \Lambda'(1)\, .
\end{eqnarray}
Therefore (\ref{BetheTrial}) is proved by taking the expectation with respect 
to the graph ensemble in Eq.~(\ref{eq:ProofBetheFirst}) and then taking 
the large blocklength 
limit\footnote{Notice that in the case of Poisson ensembles 
$\Lambda_l$ as no bounded support ($l$ can be arbitrarily large).
However the thesis follows from convergence of $\Eg\, \hLambda_l $ in total
variation distance.}.

Finally, for the multi-Poisson ensemble we gave just to notice that 
the expected left degree profile converges to $\Lambda^{(\gamma)}$
rather than to $\Lambda$, see App.~\ref{app:DegreeMultiPoisson}
\endproof

This result provides
an appealing interpretation for the trial entropy
entering in Theorem \ref{MainTheorem}. Apart from a simple rescaling,
it is asymptotically equal to the expected value of the Bethe free energy 
when the messages $\{u_{i\to a}\}$ and $\{v_{a\to i}\}$ are i.i.d.
random variables. Viceversa, Theorem \ref{MainTheorem} can be interpreted
as yielding a connection between the Bethe free energy, and the conditional 
entropy of the transmitted message.
%
%
\section{Generalizations and conclusion}
\label{GeneralizationSection}

We expect that the results derived in this paper can be extended in several
directions.

A first direction consists in proving the analogue of Theorem
\ref{MainTheorem} for more general code ensembles.
It is important to notice that the technique used in this paper (as well as in
\cite{FranzLeoneToninelli}) makes a crucial use of the convexity 
of $\Rho(x)$. Although the non-rigorous calculations
of \cite{Saad1,Saad2,Mio,NostroDynamical} suggest that the the result
will have the same form for a non-convex $\Rho(x)$, the proof is probably 
more difficult in this case. 

A second direction consists in proving that the bound of 
Theorem \ref{MainTheorem} is indeed tight. We precise this claim as follows.
\begin{conj}
Under the hypotheses of Theorem \ref{MainTheorem} we have
\begin{eqnarray}
\lim_{n\to\infty}h_n = \sup_{V}\, \phi_V\, ,\label{Conjecture}
\end{eqnarray}
where the $\sup$ has to be taken over the space of admissible random variables.
The degree sequences to be used as argument of $\phi_V$ are the
same as in  Theorem \ref{MainTheorem}.
\end{conj}
Once again, this claim is supported by \cite{Saad1,Saad2,Mio,NostroDynamical}.

Finally, in this paper we limited ourselves to estimating the
conditional entropy per channel use. 
As discussed in Section \ref{ConcentrationSection},
this implies only sub-optimal bounds on the bit error rate.
It would be therefore important to estimate directly this quantity
without passing through Fano inequality. The results of
\cite{Saad1,Saad2,Mio,NostroDynamical} suggest the following recipe for 
computing the bit error rate under symbol MAP decoding.
Determine the message densities maximizing the trial entropy,
cf. Eq. (\ref{TheoremPoisson}). Compute the density of a posteriori likelihoods
as in density evolution (this implies a convolution of all the densities 
incoming in a variable node). The bit error rate is simply given by the weight 
of negative log-likelihoods under this distribution. 

Finally, one may hope that the strong connection between message
passing techniques (density evolution) and MAP decoding (conditional
entropy) highlighted in the present approach may lead to a better understanding
of the former.
%
%
\section*{Acknowledgments}

I am grateful to Tom Richardson and R\"udi Urbanke for their encouragement,
and to Silvio Franz and Fabio Toninelli for many exchanges on the
interpolation method. 

Note: while this paper was was being revised, an alternative technique has 
been put forward \cite{Life} which allows to rederive part of the present
results. 
%
%
\appendix
%
%
\section{Coupling graph ensembles}
\label{CouplingApp}

In this Appendix we prove Lemma \ref{MultiLemma}. Instead of exhibiting 
directly a coupling between a standard graph $\graph_{\rm s}\ed 
(n,\Lambda,\Rho)$ and a multi-Poisson graph 
$\graph_{\rm mP}\ed (n,\Lambda,\Rho,\gamma)$, we shall proceed
in two step. More precisely, we shall exhibit two couplings
$(\graph_{\rm s},\graph_{*})$
and $(\graph_{*},\graph_{\rm mP})$ where
the distribution of $\graph_{*}\ed(n,\Lambda,\Rho,\gamma)_* $ 
is defined  below
(as in Sec.~\ref{MultiPoissonDefinitionSection}, we let 
$t_{\rm max}\equiv \lfloor\Lambda'(1)/\gamma\rfloor-1$ be the number of 
rounds).

In order to generate a random element in $(n,\Lambda,\Rho,\gamma)_*$,
proceed as for the multi-Poisson ensemble (see Definition 
\ref{def:MultiPoisson}) but the following modification. During 
stage $t$, for each check node $a=(\hat{a},k,t)\in\Cnod_t$, and for each 
$r = 1,\dots, k$, $i^a_r$ is chosen randomly in $\Vnod$ with distribution
$w_i(t,a,r) = (\soc_i(t)-\Delta_i(t,a,r))/[\sum_{j}  
(\soc_i(t)-\Delta_i(t,a,r))]$,
where $\Delta_i(t,a,r)$ is the number of times $i$ has been already 
chosen during stage $t$. In other words, unlike in the multi-Poisson
ensemble, we keep track faithfully of the number of free sockets.

Let us now describe how the two-step coupling works.
\begin{itemize}
\item[] {\bf From $\graph_{\rm mP}$ to $\graph_*$:}
Consider round $t$. Let $\soc^{\rm mP}_i(t)$ and $\soc^{*}_i(t)$ be the 
number of free sockets respectively for $\graph_{\rm mP}$ and
 $\graph_{\rm *}$. Choose the variable nodes $(i^a_r)_{\rm mP}$
and $(i^a_r)_{*}$ in the two graphs by coupling optimally 
(see discussion below) the  distributions 
$w^{\rm mP}_i(t) = [\soc^{\rm mP}_i(t)]_+/
(\sum_j [\soc^{\rm mP}_j(t)]_+)$ and $w^{*}_i(t;a,r) 
= (\soc^*_i(t)-\Delta_i(t;a,r))/[\sum_{j}  (\soc^*_i(t)-\Delta_i(t;a,r))]$. 
If $(i^a_r)_{*}\neq (i^a_r)_{\rm mP}$, we say that a `discrepancy'
has occurred.
We claim that, if $\gamma<1$, then the total number of discrepancies is 
smaller than $An\gamma^b$ w.h.p.
($C$ and $b$ being  $n$-- and $\gamma$--independent constants). 
The proof of this claim 
is provided below. Of course the number of rewirings necessary to pass from 
$\graph_{\rm mP}$ to $\graph_*$ is bounded by twice
the number of discrepancies.
\item[] {\bf From $\graph_{*}$ to $\graph_{\rm s}$:} Notice that 
$\graph_{*}$ is generated in the same way as $\graph_{\rm s}$ 
(see discussion at the end of Sec.~\ref{StandardDefinitionSection}) 
but for the fact that it contains a random number of check nodes. 
In fact, the total number of check nodes of degree $k$ (call it $m_k$) in a 
$(n,\Lambda,\Rho,\gamma)_*$ graph is a Poisson random variable with mean 
$\om_k = t_{\rm max}n\gamma\Rho_k/\Rho'(1)$. Denoting by
$m^{({\rm s})}_k= n\Lambda'(1)\Rho_k/\Rho'(1)$ the number of check nodes in a 
standard $(n,\Lambda,\Rho)$ graph, it is easy to see that 
$m^{({\rm s})}_k-2Bn\gamma< \om_k 
\le  m^{({\rm s})}_k - Bn\gamma$ for some positive 
($n$ and $\gamma$ independent) constant $B$. 
By elementary properties of Poisson random variables, one obtains
$m^{({\rm s})}_k-3Bn\gamma< m_k \le  m^{({\rm s})}_k - (1/2)Bn\gamma$ 
for each $k\in\{2,\dots,k_{\rm max}\}$ with probability 
greater than $1-2e^{-C\gamma^2n}$ for some constant $C$.

We therefore obtain the desired coupling as follows: first generate 
$\graph_{*}$. If $m_k>m^{({\rm s})}_k$ for any $k$, then generate
an independent graph $\graph_{\rm s}$. In the opposite case, generate
$\graph_{\rm s}$ by adding 
$m^{({\rm s})}_k-m_k$ check nodes for each $k$ and connect them to 
variable nodes as described at the end of Sec.~\ref{StandardDefinitionSection}.
Because of the above argument, the number of rewirings (check nodes added) 
is smaller than $A'n\gamma$ with high probability.
\end{itemize}

We are now left with the task of proving the claim in the first step. 
Before accomplishing this task, it is worth recalling an elementary
fact 
which is useful in this proof~\cite{Aldous}. Given two distributions 
$\{w^{(1)}_i\}$ and $\{w^{(2)}_i\}$ over $i\in [n]$, their total
variation distance is defined as 
\begin{eqnarray}
||w^{(1)}-w^{(2)}|| = \frac{1}{2}\sum_{i=1}^n |w^{(1)}_i-w^{(2)}_i|\, .
\end{eqnarray}
Furthermore, if $i_1$ and $i_2$ are distributed according to 
$w^{(1)}$ and $w^{(2)}$, there exist coupling between them (i.e. a joint 
distribution which has $w^{(1)}_\cdot$, $w^{(2)}_\cdot$ as marginals) 
such that $||w^{(1)}-w^{(2)}|| = \Prob(i_1\neq i_2)$. Such coupling
is `optimal' in the sense that, for any coupling we have
$||w^{(1)}-w^{(2)}|| \le \Prob(i_1\neq i_2)$.
 
The proof of the claim is obtained by recursively estimating
the number of discrepancies between $\graph_{\rm mP}$ and $\graph_{*}$.
Suppose that we have terminated the first $t$ rounds 
(denoted as $0,\dots, t-1$ in Definition~\ref{def:MultiPoisson}) 
in the generation
of the couple $(\graph_{\rm mP},\graph_*)$ and no more than 
$C_tn\gamma$ discrepancies occurred so far (with $C_t$ $n$-independent). 
This hypothesis
trivially holds for $t=0$. We will determine an
$n$-independent constant $C_{t+1}$, such that, at the end
of the $t$-th  step there will be, w.h.p., less than $C_{t+1}n\gamma$ 
discrepancies.
By iterating this argument, we deduce
that $\graph_{\rm mP}$ and $\graph_*$ have less than 
$C_{t_{\rm max}}n\gamma$ discrepancies with high probability, and 
will be able to obtain the estimate $C_{t_{\rm max}}\le 
A+B\gamma^{-\rho}$ with $0<\rho< 1$, $A,B>0$ three 
$\gamma$-independent constants. This implies Lemma \ref{MultiLemma}
with $b=1-\rho$.

During the round $t$, $(i^a_r)_*$ and $(i^a_r)_{\rm mP}$ are taken from
the distributions $w^{\rm mP}_i(t)$ and $w^{*}_i(t;a,r)$ described above.
Let us start by noticing that, with high probability
\begin{eqnarray}
\sum_{i=1}^n|[\soc_i^{\rm mP}(t)]_+-\soc_i^*(t)+\Delta_i(t;a,r)| & \le &
\sum_{i=1}^n|\soc_i^{\rm mP}(t)-\soc_i^*(t)| + \sum_{i=1}^n \Delta_i(t;a,r)\le
\nonumber\\
&\le & C_tn\gamma^2 t + 2n\gamma\, .
\end{eqnarray}
The first inequality follows from the fact that $\soc_i^*(t)$
and $\Delta_i(t;a,r)$ are non-negative. 
The second one, from the induction hypothesis 
together with the observation that $\sum_{i=1}^n \Delta_i(t;a,r)$ is
smaller than the total number of variable node choices during round 
$t$, which in turn is a Poisson random variable with mean $n\gamma$.

Next, we observe that, w.h.p. 
\begin{eqnarray}
\sum_{i=1}^{n} [\soc_i^{\rm mP}(t)]_+\ge n(\Lambda'(1)-\gamma t-\gamma)\, .
\end{eqnarray}
In fact, at $t=0$ the sum on the left-hand side has value $n\Lambda'(1)$,
and after each round it decreases at most by the number of 
left sockets which are occupied during that round. This is a Poisson 
random variable of mean $n\gamma$.

Now we can estimate the variation distance between the distributions
of $(i^a_r)_{\rm mP}$ and $(i^a_r)_*$:
\begin{eqnarray}
||w^{\rm mP}_{\cdot}(t)-w^{*}_{\cdot}(t;a,r)|| &\le & 
\frac{\sum_{i=1}^n|[\soc_i^{\rm mP}(t)]_+-\soc_i^*(t)+\Delta_i(t;a,r)| }{\sum_{i=1}^n
 [\soc_i^{\rm mP}(t)]_+} \le \nonumber \\
&\le &\frac{C_t\gamma^2 t +2\gamma}{\Lambda'(1)-\gamma(t+1)}
\le \\
& \le & \frac{C_t\Lambda'(1)+2}{t_{\rm max}-t}\, ,
\label{eq:DistanceEstimate}
\end{eqnarray}
where the second inequality follows from $t<t_{\rm max}$ and 
$t_{\rm max}\le \Lambda'(1)/\gamma-1$. During round $t$, about $n\gamma$
couples $(i^a_r)_{\rm mP}$ and $(i^a_r)_*$ are chosen and they
differ with probability $||w^{\rm mP}_{\cdot}(t)-w^{*}_{\cdot}(t;a,r)|| $
(because we coupled $w^{\rm mP}_{\cdot}(t)$ and $w^{*}_{\cdot}(t;a,r)$
optimally).
The total number of discrepancies is therefore smaller
than $2n\gamma ||w^{\rm mP}_{\cdot}(t)-w^{*}_{\cdot}(t;a,r)|| $
with high probability. Unhappily this estimate worsen as $t$
approaches $t_{\rm max}$ because of the denominator in 
(\ref{eq:DistanceEstimate}). This problem is overcome as follows. Fix
$t_* \equiv t_{\rm max}-\lfloor t_{\rm max}^{\rho}\rfloor$,
where $0<\rho<1$ is the solution of $\rho = 2\Lambda'(1)(1-\rho)$, and use
the estimate (\ref{eq:DistanceEstimate}) only for $t< t_*$. For 
$t_{*}\le t< t_{\rm max}$ we just use the fact that during each round no 
more than $n\gamma$ discrepancies can be introduced.
In other words
\begin{eqnarray}
C_{t+1} =\left\{\begin{array}{ll}
 C_t+ 2\Lambda'(1)\, (C_t+\lambda)/(t_{\rm max}-t)
 & \mbox{  if $t< t_*$,}\\
C_t+1 & \mbox{  if $t_*\le t< t_{\rm max}$,}
\end{array}\right.
\end{eqnarray}
where we introduced the constant $\lambda\equiv 2/\Lambda'(1)>0$. 
This recursion is easily summed up, yielding 
\begin{eqnarray}
\log(C_{t_*}+\lambda)-\log(\lambda) & = &\sum_{t=0}^{t_{*}-1}
\log\left(1+ \frac{2\Lambda'(1)}{t_{\rm max}-t}\right) \le  
\sum_{t=0}^{t_{*}-1}\frac{2\Lambda'(1)}{t_{\rm max}-t}  \le\\
& \le & 2\Lambda'(1) 
\sum_{t=\lfloor t_{\rm max}^{\rho}\rfloor+1}^{t_{\rm max}}\frac{1}{t} 
\le 2\Lambda'(1)\, \log\left(\frac{t_{\rm max}e}{\lfloor t_{\rm max}^{\rho}\rfloor}\right)\, .
\end{eqnarray}
Using the definition of $t_{\rm max}$ and the
relation $2\Lambda'(1)(1-\rho)=\rho$, we get 
$C_{t_*}\le A+B\gamma^{-\rho}$ with $A$ and $B$ two $\gamma$ independent 
constants. Finally $C_{t_{\rm max}}\le C_{t_*} + 
\lfloor t_{\rm max}^{\rho}\rfloor\le A'+B'\gamma^{-\rho}$.  
%
%
\section{Degree distribution for multi-Poisson ensembles}
\label{app:DegreeMultiPoisson}

In this Appendix we provide an asymptotic characterization of 
the variable-node degree profile for the multi-Poisson ensemble
$(n,\Lambda,\Rho,\gamma)$. We shall start by defining a construction
which yields the degree profile in the large blocklength limit and then 
prove convergence to this construction.

Let $\Omega^{(t)}_{l,\soc}$ be a sequence of distributions 
over $l\in\{2,\dots ,l_{\rm max}\}$ and $\soc\in \Z$ indexed by 
$t\in\{0,\dots,t_{\rm max}\}$ and defined as follows. Introduce the 
kernel
\begin{eqnarray}
W_t(\Delta|\soc) = e^{-\lambda(\soc)}\;\frac{\lambda(\soc)^{\Delta}}{\Delta!}\, ,\;\;\;\;\;
\lambda(\soc) \equiv\frac{\gamma [\soc]_+}{\sum_{\soc,l}
\Omega^{(t)}_{l,soc}\, [\soc]_+}\, .
\label{eq:KernelDefinition}
\end{eqnarray}
Next define recursively $\Omega^{(t)}_{l,\soc}$ as
\begin{eqnarray}
\Omega^{(t+1)}_{l,\soc} = \sum_{\soc'\ge \soc}\Omega^{(t)}_{l,\soc'}
\, W(\soc'-\soc|\soc')\, ,\;\;\;\;\;\;\;\;\;
\Omega^{(0)}_{l,\soc} = \Lambda_l\, \ind_{\soc=0}\, ,
\end{eqnarray}
where $\ind_{A}$ is the indicator function of the event $A$. 
Notice that the sum in the denominator
of Eq.~(\ref{eq:KernelDefinition}) is always well-defined.
In fact from the definition follows that $\Omega^{(t)}_{l,\soc}=0$
if $\soc>l_{\rm max}$.
Finally, we define the asymptotic degree profile to be
\begin{eqnarray}
\Lambda^{(\gamma)}_l = \sum_{l'}\Omega^{(t_{\rm max})}_{l',l'-l}\, .
\end{eqnarray}

The following result implies that $\{\Lambda^{(\gamma)}_l\}$ is
in fact the correct asymptotic degree profile.
\begin{lemma}
Let $\{\hLambda_l\, :\, l = 2,\dots, l_{\rm max}\}$ be the variable nodes
degree profile for a random Tanner graph from the $(n,\Lambda,\Rho,\gamma)$
multi-Poisson ensemble. Denote by $||\mu-\nu||$ the total variation distance
between distributions $\mu$ and $\nu$ (see previous Section).
If $\{\Lambda^{(\gamma)}_l\}$ is defined as above, then there
exist $A,B>0$, such that
\begin{eqnarray}
(I).\;\;\;\;\;\;\; && \lim_{n\to\infty}\E\, \hLambda_l = \Lambda^{(\gamma)}_l\, ,
\\
(II).\;\;\;\;\;\;\; &&\lim_{n\to\infty}||\E\, \hLambda- \Lambda^{(\gamma)}||=0\, ,\\
(III).\;\;\;\;\;\;\; && \Prob\left\{ | \hLambda_l-\Lambda^{(\gamma)}_l|\ge
\ve\right\}\le A\, e^{-Bn\ve^2}\, ,
\end{eqnarray}
for any positive $\ve$.
\end{lemma}
\prooft\, Notice that $(III)$  obviously implies
$(I)$. Moreover (see proof of Corollary \ref{coro:Degree}), if a sequence of 
distributions over the integers $\mu^{(n)}$ 
converges pointwise to a (normalized) distribution $\mu^{(\infty)}$,
then $|| \mu^{(n)}- \mu^{(\infty)}||\to 0$. Therefore $(I)$ implies
$(III)$. 

We are left with the task of proving $(III)$. We shall in fact prove the 
following stronger statement.
Consider a multi-Poisson graph generated as in Definition
(\ref{def:MultiPoisson}).
Let $\hOmega^{(t)}_{l,\soc}$ be the fraction of variable nodes $i\in\Vnod$
such that  $i\in\Vnod_l$ (or, equivalently, $\soc_i(0) = l$)
and  $\soc_i(t) = \soc$. We claim that $\hOmega^{(t)}_{l,\soc}$ is well
approximated by the sequence $\Omega^{(t)}_{l,\soc}$ defined above. 
More precisely, there exist constants $A,B$ (which may depend on the ensemble
parameters as well as on $t$) such that 
\begin{eqnarray}
\Prob\left\{|\hOmega_{l,\soc}^{(t)}-\Omega_{l,\soc}^{(t)}|\ge
\ve\right\}\le A\, e^{-Bn\ve^2}\, .\label{eq:ConcetrationMPThesis}
\end{eqnarray}
Recall that the degree of a variable node $i\in\Vnod_l$  is 
$l-\soc_i(t_{\rm max})$,
we have
\begin{eqnarray}
\hLambda_l = \sum_{l'}\hOmega^{(t_{\rm max})}_{l',l'-l}\, .\label{eq:FiniteSum}
\end{eqnarray}
The thesis therefore follows from Eq.~(\ref{eq:ConcetrationMPThesis}) together 
with the observation that the sum (\ref{eq:FiniteSum})
contains a finite number of terms.

The claim is proved by induction on $t$. It is obviously true 
for $t=0$. Assume now that the claim is true up to stage $t$, and
consider the distribution of $\hOmega^{(t+1)}_{l,\soc}$ conditional
to $\hOmega^{(t)}_{l,\soc}$. By the induction hypothesis we can 
assume that $|\hOmega_{l,\soc}^{(t)}-\Omega^{(t)}_{l,\soc}| \le \ve$.
Furthermore, since $\hOmega_{l,\soc}^{(t)}=0$ whenever $\soc>l_{\rm max}$,
we can also assume 
\begin{eqnarray}
|\sum_d\hOmega_{l,\soc}^{(t)}[\soc]_+-
\sum_\soc\Omega^{(t)}_{l,\soc}[\soc]_+| \le \ve\, .
\end{eqnarray}
We shall neglect the exponentially rare cases in which these conditions do not 
hold.

The total number of variable nodes
sampled during stage $t+1$ is $\Delta_{\rm tot} = \sum_k km^{(t)}_k$
where $m^{(t)}_k$ is a Poisson random variable with mean 
$n\gamma\Rho_k/\Rho'(1)$. We can therefore assume that 
$|\Delta_{\rm tot}-n\gamma|\le n\ve $ and neglect the rare cases 
in which this is false. Next, consider a variable $i$, such that
$\soc_i(t) = \soc$. The probability that this is chosen when selecting
one of the neighbors of a function node $a$ is
\begin{eqnarray}
w_i(t) = \frac{[\soc_i(t)]_+}{\sum_j [\soc_j(t)]_+} =
\frac{[\soc]_+}{\sum_{l,\soc} \hOmega^{(t)}_{l,\soc} [\soc]_+} \equiv w(\soc)\, .
\end{eqnarray}
The probability that, during stage $t$, this variable node is selected 
$\Delta_i(t) = \Delta$ times
(conditional to the total number $\Delta_{\rm tot}$) 
is therefore
\begin{eqnarray}
\Prob[\Delta_i(t) =\Delta| \soc_i(t) = \soc] 
& =& {\Delta_{\rm tot}\choose \Delta} \, 
w(\soc)^{\Delta}(1-w(\soc))^{\Delta_{\rm tot}-\Delta} =\label{eq:DistributionDelta}\\
& = & W_t(\Delta|\soc)+O(1/n)+O(\ve)\, .\nonumber
\end{eqnarray}
Therefore, the fraction of variable nodes such that $\soc_i(0)=l$, 
$\soc_i(t) =\soc$ and
$\Delta_i(t)=\Delta$ (to be denoted by $\hOmega^{(t)}_{\Delta,l,\soc}$)
is concentrated around $[W_t(\Delta|\soc)+O(\ve)]\hOmega^{(t)}_{l,\soc}$. 
Using the induction hypothesis, this implies that 
\begin{eqnarray}
\Prob\left\{|\hOmega^{(t)}_{\Delta,l,\soc}-W_t(\Delta|\soc)\Omega^{(t)}_{l,\soc}|\ge\ve
\right\}\le A\, e^{-nB\ve^2}\, .\label{eq:ConcetrationMPProof}
\end{eqnarray}
Next we notice that $\soc_i(t) =\soc$ and $\Delta_i(t)=\Delta$ implies
$\soc_i(t+1)=\soc-\Delta$. Therefore $\hOmega^{(t+1)}_{l,\soc} = 
\sum_{\soc'\ge \soc} \hOmega^{(t)}_{\soc'-\soc,l,\soc'}$. 
Since a finite number of terms
enter in this sum, Eq.~(\ref{eq:ConcetrationMPProof}) implies
\begin{eqnarray}
\Prob\left\{\left|\hOmega^{(t+1)}_{l,\soc}-
\sum_{\soc'\ge \soc}W_t(\soc'-\soc|\soc')\Omega^{(t)}_{l,\soc'}\right|\le \ve\right\}
\le A\, e^{-nB\ve^2}\, ,\label{eq:ConcetrationMPProofSecond}
\end{eqnarray}
for some, eventually different, constants $A$ and $B$. Notice that
the sum in the above expression is exactly $\Omega^{(t+1)}_{l,\soc}$.
Therefore, thesis $(III)$ is proved.
\endproof
%
%
\section{Derivative of the conditional entropy: Poisson ensemble}
\label{StraightforwardApp}

In this Appendix we  compute derivative of the conditional entropy 
with respect to the interpolation parameter, cf. Eq. (\ref{EntropyDerivative}).
The crucial observation is the following. Let $n$ a poissonian random variable
with parameter $\lambda>0$, and $f:{\mathbb N} \to {\mathbb R}$ any function on
the positive integers, then:
\begin{eqnarray}
\frac{d\phantom{\lambda}}{d\lambda}\E f(n) = \E[f(n+1)-f(n)]\, .
\label{eq:DerivativePoisson}
\end{eqnarray}
Consider now the expression (\ref{EntropyInterpolation}) for the
interpolating conditional entropy. This depends upon $t$ through the 
distributions of the $m_k$'s (i.e. the number of right nodes of degree $k$
which is a poissonian variable with parameter $nt\Rho_k/\Rho'(1)$), 
and the distribution of the $l_i$'s (i.e. 
the number of repetitions for the variable $x_i$, which is a random 
variable with parameter $\gamma-t$).
When differentiating with respect to $t$ we get therefore a sum of several 
contributions.
For the sake of clarity, let us compute in detail one of such terms.
In order to single out a term, assume that the parameter entering in 
the distribution of $m_k$ is $t_k$ and is distict from 
$t$ (i.e. the number of right nodes of degree $k$
is a poissonian variable with parameter $nt_k\Rho_k/\Rho'(1)$).
Let  $\partf=\partf(m_k)$ be the 
normalization defined in Eq. (\ref{InterProbabilityDistribution})
for a graph with $m_k$ check nodes of degree $k$.
Applying the above formula we get
\begin{eqnarray}
\frac{dh_n}{dt_k}(t,t_k) 
&=& \frac{\Rho_k}{\Rho'(1)}\Et\log_2\left\{\partf(m_k+1)/
\partf(m_k)\right\}\, .
\end{eqnarray}
The symbol $\Et$ includes expectation with respect to $m_k$,
the choice of $m_k+1$ check nodes of degree $k$, 
as well as with respect to the corresponding received message.
We can however single out expectation with respect to the 
last of these nodes and use the fact that $ \partf(m_k)$ does not depend on it.
Denote $\partf_{\cal C}(i_1\dots i_k;\hy)$ the  normalization constant 
in Eq.~(\ref{InterProbabilityDistribution})  when a factor  
$\Qc(\hy|x_{i_1}\oplus\cdots\oplus x_{i_k})$ is is multiplied to the
probability distribution.
Then we have
\begin{eqnarray}
\frac{dh_n}{dt_k}(t,t_k) &=& \frac{\Rho_k}{\Rho'(1)}
\, \frac{1}{n^k}\,
\sum_{i_1\dots i_k} 
\E_{\hy}\Et \log_2 \left\{
\partf_{\cal C}(i_1\dots i_k;\hy)/\partf\right\}\, .
\end{eqnarray}
The same calculation can be repeated for check each degree $k$
as well as for the dependency upon $t$ of the distribution of the $l_i$'s. 
We introduce the notation $\partf_{\rm eff}(i;z)$ for the normalization
constant in Eq. (\ref{InterProbabilityDistribution}) when a factor  
$\Qeff(z|x_i)$ is multiplied.
With these definitions we have (here we set aagain $t_k=t$)
\begin{eqnarray}
\frac{dh_n}{dt}(t) &=& \sum_k \frac{\Rho_k}{\Rho'(1)}\, \frac{1}{n^k}\,
\sum_{i_1\dots i_k} \E_{\hy}\Et \log_2 \left\{
\partf_{\cal C}(i_1\dots i_k;\hy)/\partf\right\}-\frac{1}{n}\sum_{i\in{\cal V}}
\E_z\Et\log_2\left\{\partf_{\rm eff}(i;z)/\partf\right\}-\nonumber\\
&&-\frac{1}{\Rho'(1)}\sum_y \Qc(y|0)\log_2 \Qc(y|0)+
\sum_z \Qeff(z|0)\log_2 \Qeff(z|0)\, .
\end{eqnarray}
The expression (\ref{EntropyDerivative}) is recovered by noticing that 
$\partf_{\cal C}(i_1\dots i_k;\hy)/\partf = 
\<\Qc(y|x_{i_1}\oplus\cdots\oplus x_{i_k})\>_t$ and
$\partf_{\rm eff}(i;z)/\partf = \<\Qeff(z|x_i)\>$.
%
%
\section{Positivity of $R_n(t)$: Poisson ensemble}
\label{PositivityApp}

In this Appendix we show that, under the hypotheses of Theorem
\ref{MainTheorem}, the remainder $R_n(t)$ in Eq. (\ref{RemainderDef}) 
is positive. This completes the proof of the Theorem. We start 
by writing the remainder in the form
\begin{eqnarray}
R_{n}(t) = R_{a,n}(t) - R_{b,n}(t) - \Rh_{a,n}(t) +\Rh_{b,n}(t) \, ,
\label{SumRemainder}
\end{eqnarray}
where (throughout this Appendix entropies will be measured in nats:
this clearly does not affect the sign of $R_{n}(t)$)
\begin{eqnarray}
R_{a,n}(t) & = & \frac{1}{\Rho'(1)}\sum_k \Rho_k\,  \frac{1}{n^k}
\sum_{i_1\dots i_k} \E_{\hy}\Et\log
\left\<\frac{2\, \Qc(\hy|x_{i_1}\oplus\cdots\oplus  x_{i_k})}
{\Qc(\hy|0)+\Qc(\hy|1)}\right\>_t\, ,\\
R_{b,n}(t) & = & \frac{1}{n}\sum_i \E_z\Et \log \left\<
\frac{2\,\Qeff(z|x_i)}{\Qeff(z|0)+\Qeff(z|1)}\right\>_t\, .
\end{eqnarray}
Analogous definitions are understood for $\Rh_{a,n}(t)$ 
and $\Rh_{b,n}(t)$ with the $\<\cdot\>_t$ average being substituted by
an average over $P_{v_1}(x_1)\cdots P_{v_n}(x_n)$ as in the passage from
Eq. (\ref{EntropyDerivative}) to Eq. (\ref{EntropyDerivativeApprox}). 
The code average $\Et$ has to be of course substituted by an average 
over $V$ variables $\E_v$.

We shall treat each of the four terms  $R_{a,n}(t)\dots \Rh_{b,n}(t)$ 
separately and put everything together at the end. Let us start from the first 
term. Using the  definitions (\ref{LogLikelihoodDef}) and
(\ref{Aposteriori}) we get
\begin{eqnarray}
R_{a,n}(t) & = & \frac{1}{\Rho'(1)}\sum_k \Rho_k\,  \frac{1}{n^k}
\sum_{i_1\dots i_k} \E_{J}\Et\log[1+\tanh J\,\tanh \ell_{i_1\dots i_k}]\, .
\end{eqnarray}
Here we did not write explicitly the dependence of the log-likelihood
$\ell_{i_1\dots i_k}$ for the sum $x_{i_1}\oplus\cdots\oplus  x_{i_k}$
upon the received message $(\uy,\uhy,\uz)$ and the code realization.
We notice now that  $J$ and $\ell_{i_1\dots i_k}$ are two independent 
symmetric random variables. We can therefore apply the observations
of Sec. \ref{VariablesSection} (and in particular Lemma \ref{Tricks}
to get
\begin{eqnarray}
R_{a,n}(t) & = & \sum_{m=1}^{\infty}c_{m} \, R_{a,n}(t;m)\, ,\label{Series}
\end{eqnarray}
where 
\begin{eqnarray}
c_{m} \equiv \left(\frac{1}{2m-1}-\frac{1}{2m}\right) 
\, >\,  0   \, ,
\end{eqnarray}
and
\begin{eqnarray}
R_{a,n}(t;m) \equiv \frac{1}{\Rho'(1)}\, \E_J[(\tanh J)^{2m}]\,
\sum_k\Rho_k \, \frac{1}{n^k}
\sum_{i_1\dots i_k}\Et[(\tanh\ell_{i_1\dots i_k})^{2m}]\, .\label{SeriesTerm}
\end{eqnarray}
It is now convenient to introduce the `spin' variables\footnote{This
name comes from the statistical mechanics analogy \cite{Sourlas1}.} 
$\sigma_i$, 
$i\in \Vnod$ as follows
\begin{eqnarray}
\sigma_i = \left\{\begin{array}{ll} 
+1 & \mbox{if $x_i = 0$}\, ,\\
-1 & \mbox{if $x_i = 1$}\, .
\end{array}
\right.
\end{eqnarray}
Notice that $\tanh\ell_{i_1\dots i_k} = \<\sigma_{i_1}\cdots 
\sigma_{i_k}\>_t$. We can also write the $2m$-th power of 
$\tanh\ell_{i_1\dots i_k}$ introducing $2m$ i.i.d. copies 
$\us^{(1)},\dots,\us^{(2m)}$. Using the notation introduced in 
Eq. (\ref{ReplicaDef}) we get 
\begin{eqnarray}
(\tanh\ell_{i_1\dots i_k})^{2m} = \<(\sigma^{(1)}_{i_1}\cdots
\sigma^{(1)}_{i_k})\cdots(\sigma^{(2m)}_{i_1}\cdots\sigma^{(2m)}_{i_k})
\>_{t,*}\, .
\end{eqnarray}
We replace this formula in Eq. (\ref{SeriesTerm}), and we are finally able
to carry on the sums over $i_1\dots i_k$ and $k$. The final result is 
remarkably compact
\begin{eqnarray}
R_{a,n}(t;m) = \frac{1}{\Rho'(1)}\, \E_J[(\tanh J)^{2m}]\,
\Et\<P(Q_{2m})\>_{t,*}\, ,
\end{eqnarray}
where we defined the `multi-overlaps'
\begin{eqnarray}
Q_{2m}(\us^{(1)},\dots,\us^{(2m)}) = \frac{1}{n}\sum_{i=1}^n
\sigma^{(1)}_i\cdots\sigma^{(2m)}_i\, .
\end{eqnarray}
Notice that $Q_{2m}\in [-1,1]$.

The same procedure can be repeated for  $R_{b,n}(t)$. We get 
$R_{b,n}(t) = \sum_{m} c_m R_{b,n}(t;m)$, with
\begin{eqnarray}
R_{b,n}(t;m) & = &\E_u[(\tanh u)^{2m}]\, \frac{1}{n}\sum_{i}\Et 
[(\tanh\ell_i)^{2m}] = \\
& = & \E_u[(\tanh u)^{2m}]\, \frac{1}{n}\sum_{i}\Et 
[\<\sigma_i\>_t^{2m}] =\\
& = & \E_u[(\tanh u)^{2m}]\, \frac{1}{n}\sum_{i}\Et 
[\<\sigma_i^{(1)}\cdots\sigma_i^{(2m)}\>_{t,*}]=\\
& = & \E_u[(\tanh u)^{2m}]\, \Et\<Q_{2m}\>\, .\label{Rb}
\end{eqnarray}

Let us now consider $\Rh_{a,n}(t)$. Since the probability distribution 
for the bits $x_i$'s is factorized, the averages can now be easily computed.
We get
\begin{eqnarray}
\Rh_{a,n}(t) & =  & \frac{1}{\Rho'(1)}\sum_k \Rho_k\, \E_J\E_v
\log[1+\tanh J\tanh v_1\cdots\tanh v_k]\, .
\end{eqnarray}
Notice that in fact the right-hand side is independent both
of $n$ and $t$
Once again we observe that $J$ and the $v_i$'s are independent symmetric
random variables. Using the properties exposed in Sec. \ref{VariablesSection}
we obtain $\Rh_{a,n}(t) = \sum_m c_m \Rh_{a,n}(t;m)$, where 
\begin{eqnarray}
\Rh_{a,n}(t;m) & = & \frac{1}{\Rho'(1)}\,\E_J[(\tanh J)^{2m}]\sum_k \Rho_k\, 
\left\{ \E_v[(\tanh v)^{2m}]\right\}^k =\\
 & = & \frac{1}{\Rho'(1)}\,\E_J[(\tanh J)^{2m}]\, \Rho(q_{2m})\, ,
\end{eqnarray}
where we defined $q_{2m} \equiv \E_v[(\tanh v)^{2m}]\in[-1,1]$.

Finally, the same procedure is applied to $\Rh_{b,n}(t)$. We obtain
$\Rh_{b,n}(t) = \sum_{m}c_m\Rh_{b,n}(t;m)$ with 
\begin{eqnarray}
\Rh_{b,n}(t;m) = \E_u[(\tanh u)^{2m}]\, q_{2m}\, .\label{Rhb}
\end{eqnarray}
The next step consists in noticing that, because $U$ and $V$ are 
admissible, we can apply Eq. (\ref{CnodeFunction}) to get 
\begin{eqnarray}
\E_u[(\tanh u)^{2m}] = \E_J [(\tanh J)^{2m}]\sum_{k}\rho_h \left\{
\E_v [(\tanh v)^{2m}]\right\}^{k-1} =   
\frac{1}{\Rho'(1)}\, \E_J [(\tanh J)^{2m}]\,\Rho'(q_{2m})\, .
\end{eqnarray}
This identity allows us to rewrite Eqs. (\ref{Rb}) and (\ref{Rhb}) in
the form
\begin{eqnarray}
R_{b,n}(t;m) & = &\frac{1}{\Rho'(1)}\, \E_J [(\tanh J)^{2m}]\,\Rho'(q_{2m})\, 
\Et\<Q_{2m}\>\, ,\\
\Rh_{b,n}(t;m) & = &\frac{1}{\Rho'(1)}\, \E_J [(\tanh J)^{2m}]\,
\Rho'(q_{2m})\, q_{2m}\, .
\end{eqnarray}

All  the series obtained are absolutely convergent because $c_{m}\sim m^{-2}$
as $m\to \infty$ and $|R_{\cdot,n}(t;m)|\le 1$. We can therefore
obtain $R_n(t)$ by performing the sum in Eq. (\ref{SumRemainder}) term by term.
We get 
\begin{eqnarray}
R_n(t) = \frac{1}{\Rho'(1)}\sum_{m=1}^{\infty}c_m\, \E_J[(\tanh J)^{2m}]\, 
\E_t[\<f(Q_{2m},q_{2m})\>_{t,*}]\, ,
\end{eqnarray}
where 
\begin{eqnarray}
f(Q,q) \equiv  P(Q)-P'(q)Q-P(q)+P'(q)q\, .\label{eq:ConvexfDef}
\end{eqnarray}
Since we assumed $P(x)$ to be convex for $x\in[-1,1]$, $f(Q,q)\ge 0$
for any $Q,q\in [-1,1]$. This completes the proof.
%
%
\section{Derivative of the conditional entropy: multi-Poisson ensemble}
\label{app:multi-Derivative}

Throughout this Section $t_*\in\{0,\dots,t_{\rm max}-1\}$ and $s\in[0,\gamma]$
are fixed.
Let us start by noticing that the expected conditional entropy with respect 
to the multi-Poisson has the structure (here we use the shorthand $\uY$ 
for the received message, which in our formalism is in fact $(\uY,\uhY)$): 
\begin{eqnarray}
h_n = \frac{1}{n}\E_C H_n(\uX|\uY) = \frac{1}{n}\Ey
\E_{t_{*}-1}\E_{t_*|t_{*}-1}\E_{t_{\rm max}-1|t_{*}} 
H_n(\uX|\uY)\, .
\end{eqnarray}
Here we denoted by $\E_{t_2|t_1}$, with $t_2\ge t_1$ the expectation 
with respect to the rounds $t_1+1,\dots,t_2$ in the code construction,
and by $\E_{t_1}$ the unconditional expectation over the first $t_1$ rounds. 
Notice that parameter $s$ enters uniquely in the state $\E_{t_*|t_{*}-1}$,
and more precisely in the mean of the Poisson variables $\{m^{(t_*)}_k\}$
and $\{l_i(t_*)\}$.  We can therefore apply Eq.~(\ref{eq:DerivativePoisson})
to the expression (\ref{EntropyInterpolationMP}). 
We get
\begin{eqnarray}
\frac{dh_n}{ds} &=& \sum_k \frac{\Rho_k}{\Rho'(1)}\,
\sum_{i_1\dots i_k} \,
\Ey\E_{t_*}w_{i_1}\cdots w_{i_k}
\Big\{  \underbrace{\E^{\{i_1\dots i_k\}}_{t_{\rm max}-1|t_{*}}  
 \log_2 \partf_{\cal C}(i_1\dots i_k;\hy)-
 \E_{t_{\rm max}-1|t_{*}}  \log_2 \partf}_{(i)} \Big\}\, \nonumber\\
&&-\sum_{i\in{\cal V}}\,
\E_z\E_{t_*}w_i\Big\{\underbrace{\E^{\{i\}}_{t_{\rm max}-1|t_{*}}
\log_2\partf_{\rm eff}(i;z)
-\E_{t_{\rm max}-1|t_{*}}\log_2\partf}_{(ii)}\Big\}-\nonumber\\
&&-\frac{1}{\Rho'(1)}\sum_y \Qc(y|0)\log_2 \Qc(y|0)+
\sum_z \Qeff(z|0)\log_2 \Qeff(z|0)\, ,\label{eq:DerivativeMultiFirst}
\end{eqnarray}
where we used the shorthand $w_i=w_i(t_*)$.
The definition of the modified partition functions  
$\partf_{\cal C}(i_1\dots i_k;\hy)$ and $\partf_{\rm eff}(i;z)$ is the same 
as in App.~\ref{StraightforwardApp}. The resulting expression
is here more complicated because the expectation over the 
stages $t_*+1,\dots,t_{\rm max}-1$ is not independent 
of the graph realization after stage $t_*$. For instance, 
if an extra check node is added during round $t_*$ (as a result of
Eq.~(\ref{eq:DerivativePoisson})), the following check nodes are
going to be added with a slightly different distribution in rounds
$t_*+1,\dots,t_{\rm max}-1$. This fact is taken into account by defining 
the state $\E^{\{i_1\dots i_k\}}_{t_{\rm max}-1|t_{*}}$ as follows.
At the end of round $t_*$, set
$\soc_i(t_*+1) = \soc_i(t_*)-\Delta_i(t_*)-l_i(t_*)-\nu_i$ with  
$\nu_i$ equal to the number of times $i$ appears in  
$\{i_1,\dots,i_k\}$.
Then proceed as for the interpolating ensemble introduced in 
Sec.~\ref{sec:MultiProof} for rounds $t_*+1,\dots,t_{\rm max}-1$.

We now decompose the underbraced terms
in (\ref{eq:DerivativeMultiFirst}) as follows:
\begin{eqnarray}
\Big\{\; (i)\;\Big\} & = &
\Big\{\E^{\{i_1\dots i_k\}}_{t_{\rm max}-1|t_{*}}  
 \log_2 \partf_{\cal C}(i_1\dots i_k;\hy)-
\E^{\{i_1\dots i_k\}}_{t_{\rm max}-1|t_{*}} 
 \log_2 \partf\Big\}+\,\nonumber \\
&&\hspace{-0.3cm}+\Big\{
\underbrace{\E^{\{i_1\dots i_k\}}_{t_{\rm max}-1|t_{*}} 
\log_2 \partf-
 \E_{t_{\rm max}-1|t_{*}}  \log_2 \partf}_{F(i_1\dots i_k)}\Big\}\, ,
\end{eqnarray}
and analogously for terms of type $(ii)$. It is now a matter of simple 
algebra to obtain Eq.~(\ref{EntropyDerivativeMP}), where (dropping the 
dependence of $\varphi$ upon $t_*$ and $s$)
\begin{eqnarray}
\varphi(n) \equiv \E_{y,z}\E_{t_*} \left\{\sum_k \frac{\Rho_k}{\Rho'(1)}\,
\sum_{i_1\dots i_k} w_{i_1}\cdots w_{i_k} \,
 F(i_1\dots i_k)-\sum_{i\in{\cal V}} w_i\, F(i)\right\}\, .
\label{eq:SmallMulti}
\end{eqnarray}

We are now faced with the task of proving that $\varphi(n)$ is bounded as 
claimed in Sec.~\ref{sec:MultiProof}. Denote the quantity 
in parenthesis as  $\phb(n)$.
Notice that $|\phb(n)|\le 2n$: in fact
$\phb(n)$ is the difference among the entropies of two $n$-bits
distributions. 

First, we shall show that $\phb(n)\le C\sqrt{(\log n)^3/n}$,
under the  hypothesis that $\sum_{i} \soc_i(t_*+1)\ge An$
for some positive constant $A$. Notice that the condition
holds with high probability at least $1-2e^{-Bn}$, for some $B>0$. 
The thesis follows from the inequality
(hereafter we set $\soc_i \equiv\soc_i(t_*+1)$)
\begin{eqnarray}
|\varphi(n)|\le \Prob\left[\sum_{i} \soc_i\ge An\right] C\sqrt{
\frac{(\log n)^3}{n}}+ 
\Prob\left[\sum_{i} \soc_i < An\right]\, 2n \le C'\sqrt{\frac{(\log n)^3}{n}}
\end{eqnarray}

We start by two simple observations which hold under the above
condition.
\begin{enumerate}
\item There exist a constant $F_0>0$ such that 
$|F(i_1\dots i_k)|\le F_0$. $F_0$ is understood to depend on the ensemble 
parameters as well as on $k$, but not on $n$, $t_*$ or $s$. 
This fact
is proved by noticing that $F(i_1\dots i_k)$ is the difference between
the  expectation values of $\log\partf$ in two different 
ensembles. These ensembles can be coupled as $\graph_*$ and 
$\graph_{\rm mP}$ in App.~\ref{CouplingApp}. Each time a new variable node
must be chosen in the two graphs, choose it by coupling optimally the 
corresponding $w_i$ distributions. The number of discrepancies obtained 
in this way is bounded: there is probability $O(1/n)$ of discrepancy 
(here the condition on $\sum_{i} \soc_i$ is used) at each 
step and less than $n\Lambda'(1)$ steps. 
Finally the variation in $\log\partf$ produced
by a single rewiring is smaller than $2$ in absolute value.
\item Let $i_1,\dots, i_k$ be i.i.d. with distribution  $\{w_i\}$.
There exist a constant $w_0$ such that  the probability that any two of them
coincide is smaller than $w_0/n$. This is proved by noticing 
that $w_i = [\soc_i(t_*)]_+/\sum_i[\soc_i(t_*)]_+\le l_{\rm max}/An$ 
because of the above condition. Therefore, the probability of having 
coinciding indices is smaller than $k(k-1) l_{\rm max}/An$.
\end{enumerate}
In a nutshell, these two remarks imply that terms with coincident indices
give a contribution bounded by $C/n$ in Eq.~(\ref{eq:SmallMulti}).
Moreover, the first of these observation implies $|\phi(n)|\le C_1$
uniformly in $t_*$ and $s$, as claimed in Sec.~\ref{sec:MultiProof}.

We next rewrite the function $F(i_1\dots i_k)$ by singling out the
stage $t_*+1$ in the code construction
\begin{eqnarray}
F(i_1\dots i_k) & = & \E^{\{i_1\dots i_k\}}_{t_{*}+1|t_{*}}[
\E_{t_{\rm max}-1|t_{*}+1}\log_2\partf] - \E_{t_{*}+1|t_{*}}
[\E_{t_{\rm max}-1|t_{*}+1}\log_2\partf] \equiv\nonumber\\
& \equiv & \E^{\{i_1\dots i_k\}}_{t_{*}+1|t_{*}}\Psi(j_1\dots j_m) 
- \E_{t_{*}+1|t_{*}}\Psi(j_1\dots j_m) \, .\label{eq:FSimplified}
\end{eqnarray}
Here $\Psi(\cdots)$ denotes the quantity in square brackets in the previous 
line, and we made explicit its dependency upon the variable nodes
chosen during stage $t_*+1$. Notice that $j_1\dots j_m$
are i.i.d.'s with distributions
\begin{eqnarray}
\hw^{\{i_1\dots i_k\}}_j =\frac{[\soc_j-\nu_j]_+}{\sum_{i}[\soc_i-\nu_i]_+}
\, ,\;\;\;\;\;\;
\hw_j =\frac{[\soc_j]_+}{\sum_{i}[\soc_i]_+}\, ,
\end{eqnarray}
where $\nu_j$ is (as above) the number of times $j$ appears
in $\{i_1\dots i_k\}$. Notice that $\hw_j$ is not the same as $w_j$,
the former being computed in terms of the $\{\soc_i(t_*+1)\}$ while the
latter depends upon   $\{\soc_i(t_*)\}$. The only property of
$\Psi(j_1\dots j_m)$ we shall need hereafter is that it concentrates
exponentially when $j_1\dots j_m$ are distributed according to 
$\hw_j$. More precisely
\begin{eqnarray}
\Prob\left[|\Psi-\E\Psi|\ge n\ve\right]\le Ae^{-nB\ve^2}\, ,
\label{eq:PsiConcentration}
\end{eqnarray}
for some positive constants $A$ and $B$. This result is obtained by 
repeating the proof of Theorem~\ref{TheoremConcentration} for the quantity
$\Psi(j_1\dots j_m)$.

Now, we use the general fact that, given two distributions $p(s)$ and 
$q(s)$, such that $p(s)=0$ only if $q(s)=0$, we can write
\begin{eqnarray} 
\E_q\,f(s) = \E_p\,\left[ X(s)\, f(s)\right]
\, ,\;\;\;\;\;\; X(s) \equiv \frac{1}{z}\, \frac{q(s)}{p(s)}\, ,
\;\; z\equiv \E_p\,\left[ \frac{q(s)}{p(s)}\right]\, .
\end{eqnarray} 
Applying this general relation to Eq.~(\ref{eq:FSimplified}), we get
\begin{eqnarray} 
F(i_1\dots i_k) & = & \E\left[ X_{\{i_1\dots i_k\} }(j_1\dots j_m) ; 
\Psi (j_1\dots j_m)\right]\, ,\label{FeqCorrel}\\
X_{\{i_1\dots i_k\} } & \equiv & \left(1-\frac{[k]}{L}\right)^{-m}\prod_{a=1}^m
\frac{[\soc_{j_a}-\nu_{j_a}]_+}{[\soc_{j_a}]_+}\, ,\label{eq:XMulti}
\end{eqnarray} 
where we denoted $L\equiv \sum_i [\soc_i]_+$ and 
$[k]\equiv \sum_{i}([\soc_i]_+ -[\soc_i-\nu_i]_+)$. 
Furthermore, we assumed $d_{i}>0$. We denote by $\Vnod_+$ the set of 
variable nodes satisfying this condition. 
Notice that 
$0\le [k]\le k$. Moreover $0\le X_{\{i_1\dots i_k\} } \le C$ 
for some constant $C$ (recall that $m=O(n)$) and $\E X_{\{i_1\dots i_k\} }=1$.
In view of the remarks 1. and 2. above, we focus here that 
$i_1,\dots, i_k$ are distinct. Under this assumption
\begin{eqnarray} 
X_{\{i_1\dots i_k\} } & = & \left(1-\frac{k}{L}\right)^{-m}
\left(1-\frac{1}{l_{i_1}}\right)^{\sum_a\ind_{j_a=i_1}}\cdots \left(1-\frac{1}{l_{i_r}}\right)^{\sum_a\ind_{j_a=i_1}}=\nonumber\\
& = & (1+\delta(n))\, X_{i_1}\cdots X_{i_k}\, ,\label{eq:XMultiAppr}\\
X_i &\equiv& \left(1-\frac{1}{L}\right)^{-m}\left(1-\frac{1}{l_{i}}
\right)^{\sum_a\ind_{j_a=i}}\, ,\label{eq:XOne}
\end{eqnarray} 
where $\ind_A$ is the indicator function for the event $A$ 
and $\delta(n)$ is a non random function, which can be
bounded $|\delta(n)|\le\delta_0/n$. Inserting into Eq.~(\ref{FeqCorrel}),
and using observation 1 above,
we get 
\begin{eqnarray} 
F(i_1\dots i_k) & = & (1+\delta(n))\E\left[ X_{i_1}\cdots X_{i_k}; 
\Psi\right]= \E\left[ X_{i_1}\cdots X_{i_k}; 
\Psi\right] + O(1/n)\, .
\end{eqnarray} 

We can now plug this result in Eq.~(\ref{eq:SmallMulti}), to get
\begin{eqnarray} 
\phb(n) & = & \sum_k \frac{\Rho_k}{\Rho'(1)}\,
\sum_{i_1\dots i_k} w_{i_1}\cdots w_{i_k} \,
 \E\left[ X_{i_1}\cdots X_{i_k}; 
\Psi\right]-\sum_{i\in{\cal V}} w_i\, \E\left[ X_i; 
\Psi\right] +O(1/n) =\nonumber\\
& = & \frac{1}{\Rho'(1)}\,  \E\left\{[\Rho(X)-\Rho'(1)X];\Psi\right\} + O(1/n)
=\\
& = & \frac{1}{\Rho'(1)}\,  \E\left\{f(X,\E X)\,
(\Psi-\E\Psi)\right\} + O(1/n)\, .\label{eq:ExpressionRemainder}
\end{eqnarray} 
Here $f(X,x)$ is defined as in Eq.~(\ref{eq:ConvexfDef}) and we introduced the 
site average $X \equiv \sum_{i=1}^nw_iX_i$.
Furthermore we 
used the fact that terms with coincident indices induce an error of order 
$O(1/n)$, and that $\E X=1$. Since $f(X,x)$ is convex positive with
$f(x,x)=0$, we have
\begin{eqnarray} 
|\phb(n)| \le C \, \E\left\{(X-\E X)^2\,(\Psi-\E\Psi)\right\} + O(1/n)\, .
\end{eqnarray} 
Finally, we notice that $X$ satisfies a concentration law of the form
\begin{eqnarray}
\Prob\left[|X-1|\ge \ve\right]\le Ae^{-nB\ve^2}\, .
\label{eq:XConcentration}
\end{eqnarray}
The proof is, once again, the same as for Theorem~\ref{TheoremConcentration}.

Using the expression (\ref{eq:ExpressionRemainder}) together with
Eqs.~(\ref{eq:PsiConcentration}), (\ref{eq:XConcentration}), and the 
fact that $\Psi\le n$ we finally get
\begin{eqnarray} 
|\phb(n)| \le C_2 \,n \ve^3 + C_2n n e^{-nB\ve^2} + O(1/n)\le 
C\sqrt{\frac{(\log n)^3}{n}}\, .
\end{eqnarray} 
where the last inequality follows by choosing $\ve = \alpha\sqrt{\log n/n}$.
%
%
\section{Positivity of $R_n(t_*;s)$: multi-Poisson ensemble}
\label{MultiPositivityApp}

We start as for the Poisson ensemble, by writing the remainder in the form
\begin{eqnarray}
R_{n}(t_*,s) = R_{a,n}(t_*,s) - R_{b,n}(t_*,s) - \Rh_{a,n}(t_*,s) +
\Rh_{b,n}(t_*,s) \, ,
\label{SumRemainderMP}
\end{eqnarray}
where (to lighten formulae, entropies will be measured in nats in this 
Appendix)
\begin{eqnarray}
R_{a,n}(t_*,s) & = & \sum_k \frac{\Rho_k}{\Rho'(1)}
\sum_{i_1\dots i_k} \E_{\hy}\E^{\{i_1\dots i_k\}}
w_{i_1}\cdots w_{i_k} \log
\left\<\frac{2\, \Qc(\hy|x_{i_1}\oplus\cdots\oplus  x_{i_k})}
{\Qc(\hy|0)+\Qc(\hy|1)}\right\>\, ,\\
R_{b,n}(t_*,s) & = & \sum_i \E_z\E^{\{i\} } w_i\log \left\<
\frac{2\,\Qeff(z|x_i)}{\Qeff(z|0)+\Qeff(z|1)}\right\>\, .
\end{eqnarray}
Analogous expressions hold for $\Rh_{a,n}(t_*,s)$, $\Rh_{b,n}(t_*,s)$
with the conditional measure $\<\, \cdot\,\>$ substituted by the product 
measure $P_{v_1}(x_1)\cdots P_{v_n}(x_n)$.
Here we set $w_i=w_i(t_*)$ as in the previous Section, and we use the 
notation $\E^{\{i_1\dots i_k\}}$ introduced there. 

The treatment of the four terms $R_{a,n}(\cdot),R_{b,n}(\cdot),
\Rh_{a,n}(\cdot),\Rh_{b,n}(\cdot)$ parallels closely the calculations
in App.~\ref{PositivityApp}. Here we limit ourself to discussing 
$R_{a,n}(\cdot)$: this should be more than sufficient for understanding 
the necessary changes with respect to the Poisson case.
As in that case, we use (\ref{LogLikelihoodDef}) and
(\ref{Aposteriori}) to write
\begin{eqnarray}
R_{a,n}(t_*,s) & = & \sum_k \frac{\Rho_k}{\Rho'(1)}
\sum_{i_1\dots i_k} \E_{J}\E^{\{i_1\dots i_k\}}
w_{i_1}\cdots w_{i_k} \log
\left[1+\tanh J\tanh\ell_{i_1\dots i_k}\right]=\label{eq:RMPfirst}\\
 & = & \sum_k \frac{\Rho_k}{\Rho'(1)}
\sum_{i_1\dots i_k} \E_{J}\E\left\{w_{i_1}\dots w_{i_k}X_{\{i_1\dots i_k\}}
\log
\left[1+\tanh J\tanh\ell_{i_1\dots i_k}\right]\right\}=\label{eq:RMPsecond}\\
 & = & \sum_k \frac{\Rho_k}{\Rho'(1)}
\sum_{i_1\dots i_k} \E_{J}\E\left\{w_{i_1}X_{i_k}\dots w_{i_k}X_{i_k}
\log
\left[1+\tanh J\tanh\ell_{i_1\dots i_k}\right]\right\}+O(1/n)\, .\nonumber
\end{eqnarray}
In passing from Eq.~(\ref{eq:RMPfirst}) to Eq.~(\ref{eq:RMPsecond}), we 
used the general identity $\E^{\{i_1\dots i_k\}}[A] =
\E[X_{\{i_1\dots i_k\}}A]$, where $X_{\{i_1\dots i_k\}}$ is defined in 
Eq.~(\ref{eq:XMulti}). We then used Eq.~(\ref{eq:XMultiAppr})
to approximate $X_{\{i_1\dots i_k\}}$ with an error of order $O(1/n)$.

Now we can Taylor expand the logarithm as in App.~\ref{PositivityApp}.
We obtain 
\begin{eqnarray}
R_{a,n}(t_*,s) & = & \sum_{m=1}^{\infty}c_{m} \, R_{a,n}(t_*,s;m)\, ,
\end{eqnarray}
where $c_{m} = 1/(2m-1)-1/2m$ and
\begin{eqnarray}
R_{a,n}(t_*,s;m) \equiv  \frac{1}{\Rho'(1)}\, \E_J[(\tanh J)^{2m}]\,
\Et\<P(Q_{2m})\>_*\, . \, .
\end{eqnarray}
The unique difference with respect to Poisson ensemble is in the definition 
of `multi-overlaps'. Now in fact we have (with an abuse we use the same 
notation as for the simple Poisson ensemble):
\begin{eqnarray}
Q_{2m}(\us^{(1)},\dots,\us^{(2m)}) = \sum_{i=1}^n w_i X_i
\sigma^{(1)}_i\cdots\sigma^{(2m)}_i\, .
\end{eqnarray}
Notice that we no longer have $Q_{2m}\in [-1,+1]$ because of the terms $X_i$.
However Eq.~(\ref{eq:XOne}) implies $|X_i|\le \exp(m/L)$.
Recall that $m$ is the number of variable nodes chosen during
stage $t_*+1$, which is exponentially concentrated around its expectation
$n\gamma$. On the other hand  $L=\sum_{i}[d_i]_+\ge \sum_{i}d_i$,
and the last quantity is exponentially concentrated around 
$n\gamma(t_{\rm max}-t_*)\ge n\gamma$. Therefore, for any $\delta>0$,
we have $|X_i|\le e(1+\ve)$ for any $i\in [n]$ with high probability.
As a consequence $|Q_{2m}|\le e(1+\ve)$ with high probability.

The other terms in Eq.~(\ref{SumRemainderMP}) are treated analogously. 
We finally get 
\begin{eqnarray}
R_{n}(t_*,s) = \frac{1}{\Rho'(1)}
\sum_{m=1}^{\infty}c_m\, \E_J[(\tanh J)^{2m}]\, 
\E_t[\<f(Q_{2m},q_{2m})\>]\, ,
\end{eqnarray}
with $f(X,x)$ defined as in (\ref{eq:ConvexfDef}). Positivity follows 
from the assumption that $\Rho(x)$ is convex in $[-e(1+\ve),e(1+\ve)]$.
%
%

\vspace{1cm}

\end{document}